\documentclass[12pt] {article}
\usepackage{psfrag}
\usepackage{graphicx}
\usepackage{latexsym,amsfonts}
\usepackage{amssymb}
\begin{document}
\setcounter{page}{1}
\def\theequation{\arabic{section}.\arabic{equation}}
\def\theequation{\thesection.\arabic{equation}}
\setcounter{section}{0}

\title{On kaonic deuterium.\\ Quantum field theoretic and relativistic
 covariant approach}

\author{A. N. Ivanov\,\thanks{E--mail: ivanov@kph.tuwien.ac.at, Tel.:
+43--1--58801--14261, Fax: +43--1--58801--14299}~\thanks{Permanent
Address: State Polytechnic University, Department of Nuclear Physics,
195251 St. Petersburg, Russian Federation}\,,
M. Cargnelli\,\thanks{E--mail: michael.cargnelli@oeaw.ac.at}\,,
M. Faber\,\thanks{E--mail: faber@kph.tuwien.ac.at, Tel.:
+43--1--58801--14261, Fax: +43--1--58801--14299}\,, H.
Fuhrmann\,\thanks{E--mail: hermann.fuhrmann@oeaw.ac.at}\,,\\
V. A. Ivanova\,\thanks{E--mail: viola@kph.tuwien.ac.at, State
Polytechnic University, Department of Nuclear Physics, 195251
St. Petersburg, Russian Federation}\,, J. Marton\,\thanks{E--mail:
johann.marton@oeaw.ac.at}\,, N. I. Troitskaya\,\thanks{State
Polytechnic University, Department of Nuclear Physics, 195251
St. Petersburg, Russian Federation}~~, J. Zmeskal\,\thanks{E--mail:
johann.zmeskal@oeaw.ac.at}}

\date{\today}

\maketitle

\vspace{-0.5in}
\begin{center}
{\it Atominstitut der \"Osterreichischen Universit\"aten,
Arbeitsbereich Kernphysik und Nukleare Astrophysik, Technische
Universit\"at Wien, \\ Wiedner Hauptstr. 8-10, A-1040 Wien,
\"Osterreich \\ und\\ Institut f\"ur Mittelenergiephysik
\"Osterreichische Akademie der Wissenschaften,\\
Boltzmanngasse 3, A-1090, Wien, \"Osterreich}
\end{center}

\begin{center}
\begin{abstract}
We study kaonic deuterium, the bound $K^- d$ state $A_{K d}$.  Within
a quantum field theoretic and relativistic covariant approach we
derive the energy level displacement of the ground state of kaonic
deuterium in terms of the amplitude of $K^-d$ scattering for arbitrary
relative momenta.  Near threshold our formula reduces to the
well--known DGBT formula. The S--wave amplitude of $K^-d$ scattering
near threshold is defined by the resonances $\Lambda(1405)$,
$\Sigma(1750)$ and a smooth elastic background, and the inelastic
channels $K^- d \to NY$ and $K^- d \to NY\pi$, where $Y =
\Sigma^{\pm}, \Sigma^0$ and $\Lambda^0$, where the final--state
interactions play an important role. The Ericson--Weise formula for
the S--wave scattering length of $K^-d$ scattering is derived. The
total width of the energy level of the ground state of kaonic
deuterium is estimated using the theoretical predictions of the
partial widths of the two--body decays $A_{Kd} \to NY$ and
experimental data on the rates of the $NY$--pair production in the
reactions $K^-d \to NY$. We obtain $\Gamma_{1s} = (630 \pm 100)\,{\rm
eV}$. For the shift of the energy level of the ground state of kaonic
deuterium we predict $\epsilon_{1s} = (353 \pm 60)\,{\rm eV}$.
\end{abstract}


\end{center}

\newpage

\section{Introduction}
\setcounter{equation}{0}

Kaonic deuterium $A_{K d}$ is an analogy of hydrogen with an electron
and the proton replaced by the $K^-$ meson and the deuteron,
respectively. The relative stability of kaonic deuterium is fully due
to Coulomb forces \cite{SD54}--\cite{CD03}.  The Bohr radius of kaonic
deuterium, is
\begin{eqnarray}\label{label1.1}
a_B = \frac{1}{\mu\,\alpha} = \frac{1}{\alpha}\,\Big(\frac{1}{m_K}
+ \frac{1}{m_d}\Big) = 69\,{\rm fm},
\end{eqnarray}
where $\mu = m_Km_d/( m_K + m_d) = 391\,{\rm MeV}$ is a
reduced mass of the $K^-d$ system, calculated at $m_K = 494\,{\rm
MeV}$ and $m_p = 1876\,{\rm MeV}$ \cite{DG00}, and $\alpha = e^2/\hbar
c = 1/137.036$ is the fine--structure constant \cite{DG00}. Below we
use the units $\hbar = c = 1$, then $\alpha = e^2 = 1/137.036$. Since
the Bohr radius of kaonic hydrogen is much greater than the range of
strong low--energy interactions $R_{\rm str} \sim 1/m_{\pi^-} =
1.42\,{\rm fm}$ and the radius of the deuteron $r_d = 4.32\,{\rm fm}$
\cite{MN79}, the strong low--energy interactions can be taken into
account perturbatively \cite{SD54}--\cite{CD03}.

According to Deser, Goldberger, Baumann and Thirring \cite{SD54} the
energy level displacement of the ground state of kaonic deuterium can
be defined in terms of the S--wave amplitude $f^{\,K^-d}_0(Q)$ of
low--energy $K^-d$ scattering as follows
\begin{eqnarray}\label{label1.2}
-\epsilon_{1s} + i\,\frac{\Gamma_{1s}}{2} =
 \frac{2\pi}{\mu}\,f^{\,K^-d}_0(0)\,|\Psi_{1s}(0)|^2,
\end{eqnarray}
where $\Psi_{1s}(0) = 1/\sqrt{\pi a^3_B}$ is the wave function of the
ground state of kaonic hydrogen at the origin and $f^{\,K^-d}_0(0)$ is
the amplitude of $K^-d$ scattering in the S--wave state, calculated at
zero relative momentum $Q = 0$ of the $K^-d$ pair. The DGBT formula
can be rewritten in the equivalent form
\begin{eqnarray}\label{label1.3}
-\epsilon_{1s} + i\,\frac{\Gamma_{1s}}{2} = 2\,\alpha^3 \mu^2\,
  f^{\,K^-d}_0(0),
\end{eqnarray}
where $2\,\alpha^3 \mu^2 = 602\,{\rm eV\,fm^{-1}}$ and
$f^{\,K^-d}_0(0)$ is measured in ${\rm fm}$. The formula
(\ref{label1.3}) is used by experimentalists for the analysis of
experimental data on the energy level displacement of the ground state
of kaonic deuterium \cite{DEAR1}.

For non--zero relative momentum $Q$ the amplitude $f^{\,K^-d}_0(Q)$ is
defined by
\begin{eqnarray}\label{label1.4}
f^{\,K^-d}_0(Q) = \frac{1}{2iQ}\,\Big(\eta^{K^-d}_0(Q)\,
e^{\textstyle\,2i\delta^{K^-d}_0(Q)} - 1\Big),
\end{eqnarray}
where $\eta^{K^-d}_0(Q)$ and $\delta^{K^-d}_0(Q)$ are the inelasticity
and the phase shift of the reaction $K^- + d \to K^- + d$,
respectively. At relative momentum zero, $Q = 0$, the inelasticity and
the phase shift are equal to $\eta^{K^-d}_0(0) = 1$ and
$\delta^{K^-d}_0(0) = 0$. For $Q \to 0$ the phase shift behaves as
$\delta^{K^-d}_0(Q) = a^{K^-d}_0\,Q + O(Q^2)$, where $a^{K^-d}_0$ is
the S--wave scattering length of $K^-d$ scattering.

The real part of $f^{\,K^-d}_0(0)$ is related to $a^{K^-d}_0$ as
\begin{eqnarray}\label{label1.5}
{\cal R}e\,f^{\,K^-d}_0(0) = a^{K^-d}_0.
\end{eqnarray}
Due to the optical theorem the imaginary part of the amplitude
$f^{\,K^-d}_0 (0)$ can be expressed in terms of the total cross section
$\sigma^{K^-d}_0(Q)$ for $K^-d$ scattering in the S--wave state
\begin{eqnarray}\label{label1.6}
{\cal I}m\,f^{\,K^-d }_0(0) = \lim_{Q \to
0}\frac{Q}{4\pi}\,\sigma^{K^-d}_0(Q) = \frac{1}{2}\lim_{Q \to
0}\frac{1}{Q}\,(1 - \eta^{K^-d}_0(Q)\cos 2\delta^{K^-d}_0(Q)).
\end{eqnarray}
The r.h.s. of (\ref{label1.6}) can be transcribed into the form
\begin{eqnarray}\label{label1.7}
{\cal I}m\,f^{\,K^-d}_0(0) =
-\frac{1}{2}\,\frac{d\eta^{K^-d}_0(Q)}{dQ}\Big|_{Q = 0}.
\end{eqnarray}
Hence, according to the DGBT formula the energy level displacement of
the ground state of kaonic hydrogen is defined by
\begin{eqnarray}\label{label1.8}
\epsilon_{1s} &=&- 2\,\alpha^3 \mu^2\, {\cal R}e\,f^{\,K^-d}_0(0) = -
 2\,\alpha^3 \mu^2\, a^{K^-d}_0,\nonumber\\ \Gamma_{1s} &=& ~~4
 \,\alpha^3 \mu^2\,{\cal I}m\,f^{\,K^-d}_0(0) = - 2 \,\alpha^3 \mu^2\,
 \frac{d\eta^{K^-d}_0(Q)}{dQ}\Big|_{Q = 0}.
\end{eqnarray}
These are general expressions describing the energy level displacement
of the ground state of kaonic deuterium.

The paper is organised as follows. In Section 2 we give the wave
function of the ground state of kaonic deuterium in the momentum and
the particle number representations. We derive the general formula for
the energy level displacement of the ground state of kaonic deuterium
in terms of the amplitude of $K^-d$ scattering for arbitrary relative
momenta of the $K^-d$ pair and define the S--wave amplitude
$f^{K^-d}_0(0)$ of elastic $K^-d$ scattering in terms of the S--wave
amplitudes of elastic $K^-p$, $K^-n$ and $K^-(pn)_{{^3}{\rm S}_1}$
scattering, where the $np$ pair couples in the ${^3}{\rm S}_1$ state
with isospin zero. In Section 3 we compute the S--wave amplitude of
elastic $K^-n$ scattering near threshold. In Section 4 we derive the
Ericson--Weise formula for the S--wave scattering length of $K^-d$
scattering. In Section 5 we adduce the general formula for the S--wave
amplitude of elastic $K^-pn$ scattering, saturated by the intermediate
two--baryon states $NY = n\Lambda^0$, $n \Sigma^0$ and $p
\Sigma^-$. In Sections 6, 7 and 8 we compute the amplitudes of the
reactions $K^-(pn)_{{^3}{\rm S}_1} \to NY \to K^-(pn)_{{^3}{\rm S}_1}$
with the $NY$ pair in the ${^3}{\rm P}_1$ and ${^1}{\rm P}_1$ states
for $NY = n\Lambda^0$, $n \Sigma^0$ and $p \Sigma^-$ pairs,
respectively. We compute the contribution of these reactions to the
energy level displacement of the ground state of kaonic hydrogen. In
Section 9 we compare our results with experimental data on the rates
of the reactions $K^-d \to n\Lambda^0$, $K^-d \to n\Sigma^0$ and $K^-d
\to p \Sigma^-$, other theoretical approaches and estimate the
expected value of the total width and the shift of the ground state of
kaonic deuterium, which are $\Gamma_{1s} = (630 \pm 100)\,{\rm eV}$
and $\epsilon_{1s} = (325 \pm 60)\,{\rm eV}$. In the Conclusion we
discuss the obtained results. In Appendix A we show that the wave
function of the kaonic deuterium, which we use for the calculation of
the energy level displacement of the ground state of kaonic deuterium,
describes the bound state of the $np$ pair in the ${^3}{\rm S}_1$
state with isospin zero and the bound $K^-d$ system in the ground
state. In Appendix B we compute the contribution of the elastic
background to the S--wave amplitude of elastic $K^-n$ scattering near
threshold. The calculation is carried out within the Effective quark
model with chiral $U(3)\times U(3)$ symmetry. In Appendix C we compute
in the momentum representation the spinorial wave functions of the
$np$ pair in the ${^3}{\rm S}_1$ state and the $NY$ pair in the
${^3}{\rm P}_1$ and ${^1}{\rm P}_1$ states. In Appendix D we compute
the phenomenological Lagrangians of the low--energy $PBB$, $SBB$ and
$SPP$ interactions within Gell--Mann's unitary scheme of strong
low--energy interactions. The masses of scalar mesons are taken
infinite that corresponds to the use of a non--linear realization of
chiral symmetry \cite{SG69} accepted for the analysis of strong
low--energy interactions within Chiral Perturbation Theory (ChPT) by
Gasser and Leutwyler \cite{JG99}. In Appendixes E and F we compute the
amplitudes of the $NY$ rescattering in the ${^3}{\rm P}_1$ and
${^1}{\rm P}_1$ states, respectively. Our procedure is equivalent to
the procedure developed by Anisovich {\it et al.} \cite{VA93} within
the dispersion relation technique.

\section{Energy level displacement and the wave function of kaonic 
deuterium in the ground state}
\setcounter{equation}{0}

\subsection{Energy level displacement of the ground state of kaonic
deuterium. General formula}

According to \cite{IV1}--\cite{IV3} the energy level displacement of
the ground state of kaonic  deuterium is defined by
\begin{eqnarray}\label{label2.1}
- \epsilon_{1s} + i\,\frac{\Gamma_{1s}}{2} &=&
\frac{1}{4m_Km_d}\,\frac{1}{3}\sum_{\lambda_d = 0,\pm
1}\int\frac{d^3k}{(2\pi)^3} \int\frac{d^3q}{(2\pi)^3}\,
\sqrt{\frac{m_Km_d}{E_K(\vec{k}\,)E_d(\vec{k}\,)}}\,
\sqrt{\frac{m_Km_d}{E_K(\vec{q}\,) E_d(\vec{q}\,)}}\nonumber\\
&\times&\Phi^{\dagger}_{1s}(\vec{k}\,)\,
M(K^-(\vec{q}\,)d(-\vec{q},\lambda_d) \to
K^-(\vec{k}\,)d(-\vec{k},\lambda_d))\, \Phi_{1s}(\vec{q}\,),
\end{eqnarray}
where we have averaged over polarisations of the deuteron $\lambda_d =
0, \pm 1$, $\Phi^{\dagger}_{1s}(\vec{k}\,)$ and $\Phi_{1s}(\vec{q}\,)$
are the wave functions of kaonic deuterium in the ground state in the
momentum representation and $M(K^-(\vec{q}\,)d(-\vec{q},\lambda_d) \to
K^-(\vec{k}\,)d(-\vec{k},\lambda_d))$ is the amplitude of elastic
$K^-d$ scattering\,\footnote{Of course, the energy level displacement
of the ground state of kaonic deuterium does not depend on the
polarisation of the deuteron and the formula (\ref{label2.1}) is
valid for a fixed $\lambda_d$.}. Due to the wave functions
$\Phi^{\dagger}_{1s}(\vec{k}\,)$ and $\Phi_{1s}(\vec{q}\,)$ the main
contributions to the integrals over $\vec{k}$ and $\vec{q}$ come from
the regions of 3--momenta $k \sim 1/a_B$ and $q \sim 1/a_B$, where
$1/a_B \simeq \alpha \mu \simeq 3\,{\rm MeV}$. Since typical momenta
in the integrand are much less than the masses of coupled particles,
$m_d \gg m_K \gg 1/a_B$, the amplitude of $K^-d$ scattering can be
defined for low--energy momenta only\,\footnote{It is obvious that due
to the formula (\ref{label2.1}) a knowledge of the amplitude of $K^-d$
scattering for all relative momenta from zero to infinity should give
a possibility to calculate the energy level displacement of the ground
state of kaonic deuterium without any low--energy approximation.}.

In the low--energy limit $k, q \to 0$ the relation (\ref{label2.1})
can be transcribed into the form \cite{IV1}--\cite{IV3}
\begin{eqnarray}\label{label2.2}
- \epsilon_{1s} + i\,\frac{\Gamma_{1s}}{2} =
\frac{2\pi}{\mu}\,f^{\,K^-d}_0(0)\,|\Psi_{1s}(0)|^2\,(1 + \delta^{(\rm
sm)}_{1s}),
\end{eqnarray}
where $\delta^{(\rm sm)}_{1s}$ is equal to \cite{IV1}
\begin{eqnarray}\label{label2.3}
\delta^{(\rm sm)}_{1s} =
-\,\alpha\,\frac{\mu}{~m_K}\,\frac{8}{\sqrt{\pi}}\,
\frac{\Gamma(3/4)}{\Gamma(1/4)} \simeq - 10^{-2}.
\end{eqnarray}
The correction $\delta^{(\rm sm)}_{1s}$ is universal and related to
the smearing of the wave function of exotic atom in the ground state
around the origin $r = 0$ \cite{IV1}.

For the analysis of the energy level displacement of the ground state
of kaonic deuterium in terms of the $K^-N$ and $K^- NN$ interactions
we suggest the following.  According to \cite{IV1}--\cite{IV3} the
energy level displacement of the ground state of kaonic deuterium can
be defined as
\begin{eqnarray}\label{label2.4}
-\,\epsilon_{1s} + i\,\frac{\Gamma_{1s}}{2} = \lim_{T,V\to
\infty}\frac{\langle A^{(1s)}_{K^-
d}(\vec{P},\lambda_d)|\mathbb{T}|A^{(1s)}_{K^-
d}(\vec{P},\lambda_d)\rangle}{2 E^{(1s)}_A(\vec{P}\,)VT}\Big|_{\vec{P}
= 0},
\end{eqnarray}
where $E^{(1s)}_A(\vec{P}\,) = \sqrt{\vec{P}^{\,2} + (M^{(1s)}_A)^2}$
is the total energy of kaonic deuterium in the ground
state\,\footnote{Here $M^{(1s)}_A = m_K + m_d + E_{1s}$ and
$E_{1s} = - \alpha^2 \mu/2 = - 10.41\,{\rm keV}$ are the mass and the
binding energy of kaonic deuterium in the ground state.} , $TV$ is a
4--dimensional volume defined by $(2\pi)^4\delta^{(4)}(0) = TV$
\cite{SS61} and $\mathbb{T}$ is the $T$--matrix obeying the unitary
condition \cite{SS61}
\begin{eqnarray}\label{label2.5}
\mathbb{T} - \mathbb{T}^{\dagger} = i\,\mathbb{T}^{\dagger}\mathbb{T}.
\end{eqnarray}
Then, $|A^{(1s)}_{\pi d}(\vec{P},\lambda_d)\rangle$ is the ground
state wave function of kaonic deuterium in the momentum and particle
number representation.

\subsection{Wave function of kaonic deuterium in the ground state}

The wave function $|A^{(1s)}_{\pi d}(\vec{P},\lambda_d)\rangle$ of
kaonic  deuterium in the ground state we determine as
\begin{eqnarray}\label{label2.6}
\hspace{-0.3in}&&|A^{(1s)}_{K^- d}(\vec{P},\lambda_d = \pm 1)\rangle
=\nonumber\\
\hspace{-0.3in}&&= \frac{1}{(2\pi)^3}\int \frac{d^3k_K}{\sqrt{2
E_K(\vec{k}_K)}}\frac{d^3k_d}{\sqrt{2
E_d(\vec{k}_d)}}\,\sqrt{2 E^{(1s)}_A(\vec{k}_K +
\vec{k}_d)}\,\delta^{(3)}(\vec{P} - \vec{k}_K -
\vec{k}_d)\,\Phi_{1s}(\vec{k}_K)\nonumber\\
\hspace{-0.3in}&&\times\,\frac{1}{(2\pi)^3}\int \frac{d^3k_p}{\sqrt{2
E_p(\vec{k}_p)}}\frac{d^3k_n}{\sqrt{2 E_n(\vec{k}_n)}}\,\sqrt{2
E_d(\vec{k}_p + \vec{k}_n)}\,\delta^{(3)}(\vec{k}_d - \vec{k}_p -
\vec{k}_n)\,\Phi_{d}\Big(\frac{\vec{k}_p -
\vec{k}_n}{2}\Big)\nonumber\\
\hspace{-0.3in}&&\times\,c^{\dagger}_{K^-}(\vec{k}_K)
a^{\dagger}_p(\vec{k}_p,\pm 1/2) a^{\dagger}_n(\vec{k}_n, \pm
1/2)|0\rangle,\nonumber\\
\hspace{-0.3in}&&|A^{(1s)}_{K^- d}(\vec{P},\lambda_d = 0)\rangle
=\nonumber\\
\hspace{-0.3in}&&= \frac{1}{(2\pi)^3}\int \frac{d^3k_K}{\sqrt{2
E_K(\vec{k}_K)}}\frac{d^3k_d}{\sqrt{2
E_d(\vec{k}_d)}}\,\sqrt{2 E^{(1s)}_A(\vec{k}_K +
\vec{k}_d)}\,\delta^{(3)}(\vec{P} - \vec{k}_K -
\vec{k}_d)\,\Phi_{1s}(\vec{k}_K)\nonumber\\
\hspace{-0.3in}&&\times\,\frac{1}{(2\pi)^3}\int \frac{d^3k_p}{\sqrt{2
E_p(\vec{k}_p)}}\frac{d^3k_n}{\sqrt{2 E_n(\vec{k}_n)}}\,\sqrt{2
E_d(\vec{k}_p + \vec{k}_n)}\,\delta^{(3)}(\vec{k}_d - \vec{k}_p -
\vec{k}_n)\,\Phi_{d}\Big(\frac{\vec{k}_p -
\vec{k}_n}{2}\Big)\nonumber\\
\hspace{-0.3in}&&\times\, c^{\dagger}_{K^-}(\vec{k}_K)
\frac{1}{\sqrt{2}}\,[a^{\dagger}_p(\vec{k}_p,+ 1/2)
a^{\dagger}_n(\vec{k}_n, - 1/2) + a^{\dagger}_p(\vec{k}_p,- 1/2)
a^{\dagger}_n(\vec{k}_n, + 1/2)]|0\rangle,
\end{eqnarray}
where $\Phi_d(\vec{k}\,)$ is the wave function of the deuteron as a
bound $np$ state with a total isospin zero, $ I = 0$, and a total spin
one, $S = 1$. It is normalised to unity
\cite{IV1}
\begin{eqnarray}\label{label2.7}
\int \frac{d^3k}{(2\pi)^3}\,|\Phi_d(\vec{k}\,)|^2 = 1.
\end{eqnarray}
The operators $c^{\dagger}_{K^-}(\vec{k}_K)$,
$a^{\dagger}_p(\vec{k}_p,\sigma_p)$ and
$a^{\dagger}_n(\vec{k}_n,\sigma_n)$ create the $K^-$--meson, the
proton and the neutron and obey standard canonical and relativistic
covariant commutation (for the $K^-$--meson) and anti--commutation
(for the proton and the neutron) relations. In the Appendix A we show
that the wave function (\ref{label2.6}) describes the $np$ pair in the
bound ${^3}{\rm S}_1$ state with a total isospin zero, $ I = 0$. One
can show \cite{IV1} that the wave function (\ref{label2.6}) is
normalised as

\begin{eqnarray}\label{label2.8}
\langle A^{(1s)}_{\pi d}(\vec{P}\,',\lambda\,'_d )|A^{(1s)}_{\pi
d}(\vec{P},\lambda_d)\rangle = (2\pi)^3 2
E^{(1s)}_A(\vec{P}\,)\,\delta^{(3)}(\vec{P}\,' -
\vec{P}\,)\,\delta_{\lambda\,'_d \lambda_d}.
\end{eqnarray}
Using the wave function (\ref{label2.6}) the energy level displacement
of the ground state of kaonic deuterium can be transcribed into the
form
\begin{eqnarray}\label{label2.9}
\hspace{-0.3in}&&- \epsilon_{1s} + i\,\frac{\Gamma_{1s}}{2} = \int
\frac{d^3k
d^3K}{(2\pi)^6}\,\frac{\Phi^*_{1s}(\vec{k}\,)\Phi^*_d(\vec{K} +
\vec{k}/2)}{\sqrt{2 E_K(\vec{k}\,)2 E_p(\vec{K}\,)}}\int
\frac{d^3qd^3Q}{(2\pi)^6}\,\frac{\Phi_{1s}(\vec{q}\,)\Phi_d(\vec{Q} +
\vec{q}/2)}{\sqrt{2 E_K(\vec{q}\,)2 E_p(\vec{Q}\,)}}\nonumber\\
\hspace{-0.3in}&&\times\,(2\pi)^3\,\delta^{(3)}(\vec{K} + \vec{k} -
\vec{Q} - \vec{q})\;M(K^-(\vec{q}\,)p(\vec{Q},\sigma_p) \to
K^-(\vec{k}\,)p(\vec{K},\sigma_p))\nonumber\\
\hspace{-0.3in}&&+ \int
\frac{d^3kd^3K}{(2\pi)^6}\,\frac{\Phi^*_{1s}(\vec{k}\,)\Phi^*_d(\vec{K}
+ \vec{k}/2)}{\sqrt{2 E_K(\vec{k}\,) 2 E_n(\vec{K}\,)}}\int
\frac{d^3qd^3Q}{(2\pi)^6}\,\frac{\Phi_{1s}(\vec{q}\,)\Phi_d(\vec{Q} +
\vec{q}/2)}{\sqrt{2 E_K(\vec{q}\,) 2 E_n(\vec{Q}\,)}}\nonumber\\
\hspace{-0.3in}&&\times\,(2\pi)^3\,\delta^{(3)}(\vec{K} + \vec{k} -
\vec{Q} - \vec{q})\;M(K^-(\vec{q}\,) n(\vec{Q},\sigma_n) \to
K^-(\vec{k}\,)n(\vec{K},\sigma_n))\nonumber\\
\hspace{-0.3in}&&+ \int
\frac{d^3kd^3K}{(2\pi)^6}\,\frac{\Phi^*_{1s}(\vec{k}\,)\Phi^*_d(\vec{K}
+ \vec{k}/2)}{\sqrt{2 E_K(\vec{k}\,) 2 E_p(\vec{K}\,) 2
E_n(\vec{K} + \vec{k}\,)}}\int \frac{d^3q
d^3Q}{(2\pi)^6}\,\frac{\Phi_{1s}(\vec{q}\,)\Phi_d(\vec{Q} +
\vec{q}/2)}{\sqrt{2 E_K(\vec{q}\,) 2 E_p(\vec{Q}\,)2 E_n(\vec{Q} +
\vec{q}\,)}}\nonumber\\
\hspace{-0.3in}&&\times\,M(K^-(\vec{q}\,) p(\vec{Q},\sigma_p) n(-
\vec{Q} - \vec{q},\sigma_n) \to K^-(\vec{k}\,)
p(\vec{K},\sigma_p)n(- \vec{K} - \vec{k},\sigma_n))\nonumber\\
\hspace{-0.3in}&&+ \int
\frac{d^3kd^3K}{(2\pi)^6}\,\frac{\Phi^*_{1s}(\vec{k}\,)\Phi^*_d(\vec{K}
+ \vec{k}/2)}{\sqrt{2 E_p(\vec{K}\,) 2 E_n(\vec{K} + \vec{k}\,)}}\int
\frac{d^3q d^3Q}{(2\pi)^6}\,\frac{\Phi_{1s}(\vec{q}\,)\Phi_d(\vec{Q} +
\vec{q}/2)}{\sqrt{2 E_p(\vec{Q}\,)2 E_n(\vec{Q} + \vec{q}\,)}}
\nonumber\\
\hspace{-0.3in}&&\times\,(2\pi)^3\,\delta^{(3)}(\vec{k} - \vec{q}\,)
\,M(p(\vec{Q},\sigma_p) n(- \vec{Q} - \vec{q},\sigma_n) \to
p(\vec{K},\sigma_p)n(- \vec{K} - \vec{k},\sigma_n)).
\end{eqnarray}
The r.h.s. of (\ref{label2.9}) is expressed in terms of the amplitudes
of the reactions $K^-p \to K^-p$, $K^-n \to K^- n$ and $K^- (n
p)_{{^3}{\rm S}_1} \to K^- (n p)_{{^3}{\rm S}_1}$, where the $np$ pair
couples in the ${^3}{\rm S}_1$ with isospin zero.  In principle, the
amplitudes of the reactions $K^-p \to K^-p$, $K^-n \to K^- n$ and $K^-
(n p)_{{^3}{\rm S}_1} \to K^- (n p)_{{^3}{\rm S}_1}$ should contain
all corrections caused by QCD isospin--breaking and electromagnetic
interactions and all inelastic channels induced by both strong, QCD
isospin--breaking and electromagnetic interactions.

We would like to accentuate that the contribution of the last term in
(\ref{label2.9}), describing the transition $n p \to n p$, should be
dropped, since it corresponds to a disconnected Feynman diagram of
elastic low--energy $K^-d$ scattering.

In order to show this we represent the amplitude of elastic
low--energy $K^-d$ scattering as
\begin{eqnarray}\label{label2.10}
\hspace{-0.3in}&&M(K^-(\vec{q}\,)d(-\vec{q},\lambda_d) \to
K^-(\vec{k}\,)d(-\vec{k},\lambda_d)) = \lim_{T,V\to
\infty}\frac{\langle K^-(\vec{k}\,)d(-\vec{k},\lambda_d)|\mathbb{T}
|K^-(\vec{q}\,)d(-\vec{q},\lambda_d)\rangle}{VT}.\nonumber\\
\hspace{-0.3in}&&
\end{eqnarray}
The wave function of the state
$|K^-(\vec{k}_K)d(\vec{k}_d,\lambda_d)\rangle$ we define as
\cite{SS61}
\begin{eqnarray}\label{label2.11}
|K^-(\vec{k}_K)d(\vec{k}_d,\lambda_d)\rangle =
 c^{\dagger}_{K^-}(\vec{k}_K)|d(\vec{k}_d,\lambda_d)\rangle,
\end{eqnarray}
where $|d(\vec{k}_d,\lambda_d)\rangle$ is the wave function of the
deuteron in the ground state, which we take in the form
\cite{IV1}--\cite{IV3}
\begin{eqnarray}\label{label2.12}
\hspace{-0.3in}&&|d(\vec{k}_d, \lambda_d = \pm 1)\rangle =
\frac{1}{(2\pi)^3}\int \frac{d^3k_p}{\sqrt{2
E_p(\vec{k}_p)}}\frac{d^3k_n}{\sqrt{2 E_n(\vec{k}_n)}}\,\sqrt{2
E_d(\vec{k}_p + \vec{k}_n)}\,\delta^{(3)}(\vec{k}_d - \vec{k}_p -
\vec{k}_n)\nonumber\\
\hspace{-0.3in}&&\times\,\Phi_{d}\Big(\frac{\vec{k}_p -
\vec{k}_n}{2}\Big)\,
a^{\dagger}_p(\vec{k}_p, \pm 1/2) a^{\dagger}_n(\vec{k}_n, \pm
1/2)|0\rangle,\nonumber\\
\hspace{-0.3in}&&|d(\vec{k}_d,\lambda_d = 0)\rangle =
\frac{1}{(2\pi)^3}\int \frac{d^3k_p}{\sqrt{2
E_p(\vec{k}_p)}}\frac{d^3k_n}{\sqrt{2 E_n(\vec{k}_n)}}\,\sqrt{2
E_d(\vec{k}_p + \vec{k}_n)}\,\delta^{(3)}(\vec{k}_d - \vec{k}_p -
\vec{k}_n)\nonumber\\
\hspace{-0.3in}&&\times\,\Phi_{d}\Big(\frac{\vec{k}_p -
\vec{k}_n}{2}\Big)\, \frac{1}{\sqrt{2}}\,[a^{\dagger}_p(\vec{k}_p, +
1/2)) a^{\dagger}_n(\vec{k}_n, - 1/2) + a^{\dagger}_p(\vec{k}_p, -
1/2)) a^{\dagger}_n(\vec{k}_n, + 1/2)]|0\rangle,
\end{eqnarray}
normalised as
\begin{eqnarray}\label{label2.13}
\langle d(\vec{k}\,'_d,\lambda\,'_d|d(\vec{k}_d,\lambda_d)\rangle =
(2\pi)^3 2E_d(\vec{k}_d)\,\delta^{(3)}(\vec{k}\,'_d -
\vec{k}_d)\,\delta_{\lambda\,'_d \lambda_d}.
\end{eqnarray}
Following the procedure expounded in the Appendix A one can show that
the wave function (\ref{label2.12}) describes the $np$ pair in the
bound ${^3}{\rm S}_1$ state with isospin zero, $I = 0$.

Using (\ref{label2.12}) for the calculation of the matrix elements of
the $T$--matrix in (\ref{label2.10}) we get
\begin{eqnarray}\label{label2.14}
\hspace{-0.3in}&&M(K^-(\vec{q}\,)d(-\vec{q},\lambda_d) \to
K^-(\vec{k}\,)d(-\vec{k},\lambda_d)) = \nonumber\\
\hspace{-0.3in}&&= \sqrt{2 E_d(\vec{k}\,) 2
E_d(\vec{q}\,)}\int\frac{d^3K}{(2\pi)^3}\frac{d^3Q}{(2\pi)^3}
\frac{\Phi^*_d(\vec{K} + \vec{k}/2)}{\sqrt{2 E_p(\vec{K}\,)}}\frac{
\Phi_d(\vec{Q} +\vec{q}/2)}{\sqrt{2 E_p(\vec{Q}\,)}}\nonumber\\
\hspace{-0.3in}&&\times\,(2\pi)^3\,\delta^{(3)}(\vec{K} + \vec{k} -
\vec{Q} - \vec{q}\,)\,M(K^-(\vec{q}\,)p(\vec{Q},\sigma_p) \to
K^-(\vec{k}\,)p(\vec{K},\sigma_p)) \nonumber\\
\hspace{-0.3in}&&+ \sqrt{2 E_d(\vec{k}\,) 2
E_d(\vec{q}\,)}\int\frac{d^3K}{(2\pi)^3}\frac{d^3Q}{(2\pi)^3} \frac{
\Phi^*_d(\vec{K} + \vec{k}/2}{\sqrt{2
E_n(\vec{K}\,)}}\frac{\Phi_d(\vec{Q} + \vec{q}/2)}{\sqrt{2
E_n(\vec{Q}\,)}}\nonumber\\
\hspace{-0.3in}&&\times\,(2\pi)^3\,\delta^{(3)}(\vec{K} + \vec{k} -
\vec{Q} - \vec{q}\,)\,M(K^-(\vec{q}\,)n(\vec{Q},\sigma_n) \to
K^-(\vec{k}\,)n(\vec{K},\sigma_n))\nonumber\\
\hspace{-0.3in}&&+ \sqrt{2 E_d(\vec{k}\,) 2
E_d(\vec{q}\,)}\int\frac{d^3K}{(2\pi)^3}\frac{d^3Q}{(2\pi)^3} \frac{
\Phi^*_d(\vec{K} + \vec{k}/2)}{\sqrt{2 E_p(\vec{K}\,)2 E_n(\vec{k}
+ \vec{K}\,)}}\frac{\Phi_d(\vec{Q} + \vec{q}/2)}{\sqrt{2
E_p(\vec{Q}\,)2
E_n(\vec{q} + \vec{Q} \,)}}\nonumber\\
\hspace{-0.3in}&&\times\,M(K^-(\vec{q}\,)p(\vec{Q},\sigma_p) n(-
\vec{q} - \vec{Q},\sigma_n) \to K^-(\vec{k}\,)p(\vec{K},\sigma_p)
n(- \vec{k} - \vec{K},\sigma_n))\nonumber\\
\hspace{-0.3in}&&+ \,(2\pi)^3\,2
E_K(\vec{k}\,)\,\delta^{(3)}(\vec{k} - \vec{q}\,)\,2
E_d(\vec{k}\,)\int\frac{d^3K}{(2\pi)^3}\frac{d^3Q}{(2\pi)^3}\,
\Phi^*_d\Big(\vec{K} + \frac{1}{2}\,\vec{k}\Big)\Phi_d\Big(\vec{Q} +
\frac{1}{2}\,\vec{k}\Big)\nonumber\\
\hspace{-0.3in}&&\times\, \frac{M(p(\vec{Q},\sigma_p) n(- \vec{k} -
\vec{Q},\sigma_n) \to p(\vec{K},\sigma_p) n(- \vec{k} -
\vec{K},\sigma_n))}{\sqrt{2 E_p(\vec{K}\,)2 E_n(\vec{k}\,)2
E_p(\vec{Q}\,)2 E_n(\vec{k} + \vec{Q} \,)}}.
\end{eqnarray}
The amplitude of low--energy elastic $K^-d$ scattering
(\ref{label2.14}) can be represented by Feynman diagrams depicted in
Fig.1. 
\begin{figure}
\centering
\psfrag{pm}{$K^-$}
\psfrag{p}{$p$}
\psfrag{n}{$n$}
\psfrag{d}{$d$} 
\includegraphics[height=0.75\textheight]{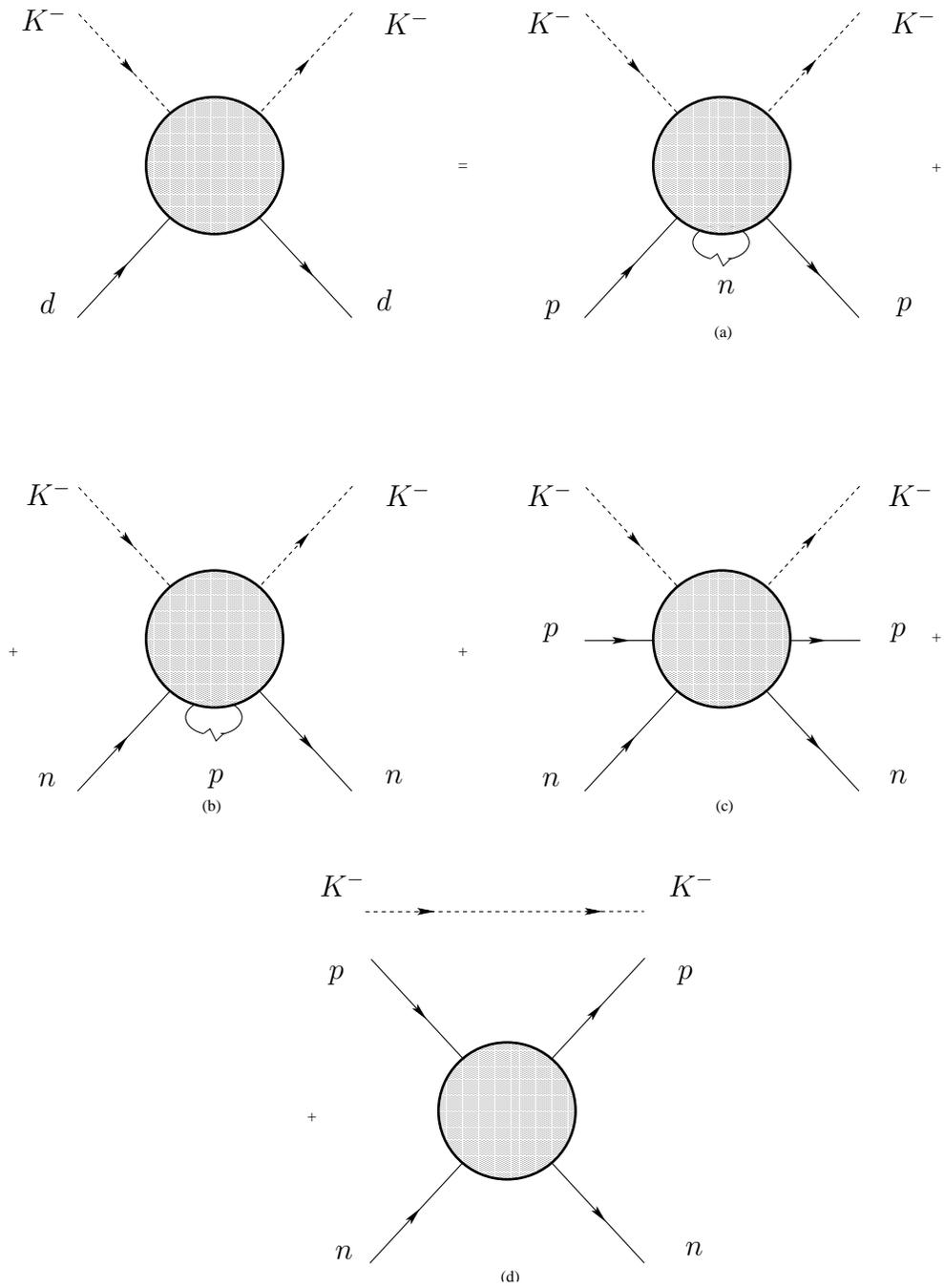}
\caption{Feynman diagrams of the reactions, describing the
amplitude of low--energy elastic $K^-d$ scattering}
\end{figure}
It is seen that the last term is described by the disconnected diagram
and, therefore, it does not contribute to the amplitude of $K^-d$
scattering. Dropping this term the amplitude of low--energy elastic
$K^-d$ scattering reads
\begin{eqnarray}\label{label2.15}
\hspace{-0.3in}&&M(K^-(\vec{q}\,)d(-\vec{q},\lambda_d) \to
K^-(\vec{k}\,)d(-\vec{k},\lambda_d)) = \nonumber\\
\hspace{-0.3in}&&= \sqrt{2 E_d(\vec{k}\,) 2
E_d(\vec{q}\,)}\int\frac{d^3K}{(2\pi)^3}\frac{d^3Q}{(2\pi)^3}
\frac{\Phi^*_d(\vec{K} + \vec{k}/2)}{\sqrt{2 E_p(\vec{K}\,)}}\frac{
\Phi_d(\vec{Q} +\vec{q}/2)}{\sqrt{2 E_p(\vec{Q}\,)}}\nonumber\\
\hspace{-0.3in}&&\times\,(2\pi)^3\,\delta^{(3)}(\vec{K} + \vec{k} -
\vec{Q} - \vec{q}\,)\,M(K^-(\vec{q}\,)p(\vec{Q},\sigma_p) \to
K^-(\vec{k}\,)p(\vec{K},\sigma_p)) \nonumber\\
\hspace{-0.3in}&&+ \sqrt{2 E_d(\vec{k}\,) 2
E_d(\vec{q}\,)}\int\frac{d^3K}{(2\pi)^3}\frac{d^3Q}{(2\pi)^3} \frac{
\Phi^*_d(\vec{K} + \vec{k}/2}{\sqrt{2
E_n(\vec{K}\,)}}\frac{\Phi_d(\vec{Q} + \vec{q}/2)}{\sqrt{2
E_n(\vec{Q}\,)}}\nonumber\\
\hspace{-0.3in}&&\times\,(2\pi)^3\,\delta^{(3)}(\vec{K} + \vec{k} -
\vec{Q} - \vec{q}\,)\,M(K^-(\vec{q}\,)n(\vec{Q},\sigma_n) \to
K^-(\vec{k}\,)n(\vec{K},\sigma_n))\nonumber\\
\hspace{-0.3in}&&+ \sqrt{2 E_d(\vec{k}\,) 2
E_d(\vec{q}\,)}\int\frac{d^3K}{(2\pi)^3}\frac{d^3Q}{(2\pi)^3} \frac{
\Phi^*_d(\vec{K} + \vec{k}/2)}{\sqrt{2 E_p(\vec{K}\,)2 E_n(\vec{k}
+ \vec{K}\,)}}\frac{\Phi_d(\vec{Q} + \vec{q}/2)}{\sqrt{2
E_p(\vec{Q}\,)2
E_n(\vec{q} + \vec{Q} \,)}}\nonumber\\
\hspace{-0.3in}&&\times\,M(K^-(\vec{q}\,)p(\vec{Q},\sigma_p) n(-
\vec{q} - \vec{Q},\sigma_n) \to K^-(\vec{k}\,)p(\vec{K},\sigma_p)
n(- \vec{k} - \vec{K},\sigma_n)).
\end{eqnarray}
Now we able to define the S--wave amplitude of elastic $K^-d$
scattering near threshold.

\subsection{S--wave amplitude of elastic $K^-d$ scattering near 
threshold}

From (\ref{label2.15}) the S--wave amplitude of $K^-d$ scattering near
threshold can be defined by
\begin{eqnarray}\label{label2.16}
f^{\,K^-d}_0(0) &=& \frac{1}{8\pi}\,\frac{1}{1 +
m_K/m_d}\int\frac{d^3K}{(2\pi)^3}
\frac{|\Phi_d(\vec{K})|^2}{E_p(\vec{K}\,)}M(K^-(\vec{0}\,)p(\vec{K},
\sigma_p) \to K^-(\vec{0}\,)p(\vec{K},\sigma_p)) \nonumber\\
&+&\frac{1}{8\pi}\,\frac{1}{1 + m_K/m_d}\int\frac{d^3K}{(2\pi)^3}
\frac{|\Phi_d(\vec{K})|^2}{E_n(\vec{K}\,)}M(K^-(\vec{0}\,)
n(\vec{K},\sigma_n) \to K^-(\vec{0}\,) n(\vec{K},\sigma_n))\nonumber\\
&+& \frac{1}{16\pi}\,\frac{1}{1 +
m_K/m_d}\int\frac{d^3K}{(2\pi)^3}\frac{d^3Q}{(2\pi)^3} \frac{
\Phi^*_d(\vec{K} )}{\sqrt{E_p(\vec{K}\,) E_n(
\vec{K}\,)}}\frac{\Phi_d(\vec{Q})}{\sqrt{ E_p(\vec{Q}\,) E_n( \vec{Q}
\,)}}\nonumber\\ &\times&\frac{1}{3}\sum_{(\sigma_p,\sigma_n; {^3}{\rm
S}_1)} M(K^-(\vec{0}\,)p(\vec{Q},\sigma_p) n(- \vec{Q},\sigma_n) \to
K^-(\vec{0}\,)p(\vec{K},\sigma_p) n( - \vec{K},\sigma_n)),\nonumber\\
&&
\end{eqnarray}
where we have averaged over the polarisations of the $np$ pair in the
${^3}{\rm S}_1$ state.

In terms of the S--wave amplitudes $f^{\,K^-p}_0(K)$ and $f^{\,K^-n}_0(K)$
of $K^-p$ and $K^-n$ scattering, respectively, the S--wave amplitude
$f^{\,K^-d}_0(0)$ reads
\begin{eqnarray}\label{label2.17}
\hspace{-0.3in}f^{\,K^-d}_0(0) &=& \frac{1}{1 +
m_K/m_d}\int\frac{d^3K}{(2\pi)^3}\,\Big(1 +
\frac{m_K}{E_p(\vec{K}\,)}\Big)\,
f^{\,K^-p}_0(K)\,|\Phi_d(\vec{K}\,)|^2 \nonumber\\
\hspace{-0.3in}&+& \frac{1}{1 +
m_K/m_d}\int\frac{d^3K}{(2\pi)^3}\,\Big(1 +
\frac{m_K}{E_n(\vec{K}\,)}\Big)\,
f^{\,K^-n}_0(K)\,|\Phi_d(\vec{K}\,)|^2\nonumber\\
\hspace{-0.3in}&+& \frac{1}{16\pi}\,\frac{1}{1 +
m_K/m_d}\int\frac{d^3K}{(2\pi)^3}\frac{d^3Q}{(2\pi)^3} \frac{
\Phi^*_d(\vec{K} )}{\sqrt{E_p(\vec{K}\,) E_n(
\vec{K}\,)}}\frac{\Phi_d(\vec{Q})}{\sqrt{ E_p(\vec{Q}\,) E_n( \vec{Q}
\,)}}\nonumber\\
\hspace{-0.3in}&\times&\frac{1}{3}\sum_{(\sigma_p,\sigma_n; {^3}{\rm
S}_1)} M(K^-(\vec{0}\,)p(\vec{Q},\sigma_p) n(- \vec{Q},\sigma_n) \to
K^-(\vec{0}\,)p(\vec{K},\sigma_p) n( - \vec{K},\sigma_n)).\nonumber\\
\hspace{-0.3in}&&
\end{eqnarray}
The real part ${\cal R}e\,f^{\,K^-d}_0(0)$ of the S--wave amplitude of
$K^-d$ scattering is defined by
\begin{eqnarray}\label{label2.18}
{\cal R}e\,f^{\,K^-d}_0(0) = \frac{1 + m_K/m_N}{1 +
m_K/m_d}\,\Big({\cal R}e\,f^{\,K^-p}_0(0) + {\cal
R}e\,f^{\,K^-n}_0(0)\Big) + {\cal R}e\,\tilde{f}^{\,K^-d}_0(0),
\end{eqnarray}
where we have set $m_n = m_p = m_N$ and denoted
\begin{eqnarray}\label{label2.19}
\hspace{-0.3in}&&{\cal R}e\,\tilde{f}^{\,K^-d}_0(0) =
\frac{1}{16\pi}\,\frac{1}{1 +
m_K/m_d}\int\frac{d^3K}{(2\pi)^3}\frac{d^3Q}{(2\pi)^3} \frac{
\Phi^*_d(\vec{K} )}{\sqrt{E_p(\vec{K}\,) E_n(
\vec{K}\,)}}\frac{\Phi_d(\vec{Q})}{\sqrt{ E_p(\vec{Q}\,) E_n( \vec{Q}
\,)}}\nonumber\\
\hspace{-0.3in}&&\times\,\frac{1}{3}\sum_{(\sigma_p,\sigma_n; {^3}{\rm
S}_1)} {\cal R}e\,M(K^-(\vec{0}\,)p(\vec{Q},\sigma_p) n(-
\vec{Q},\sigma_n) \to K^-(\vec{0}\,)p(\vec{K},\sigma_p) n( -
\vec{K},\sigma_n)).
\end{eqnarray}
The imaginary part ${\cal I}m\,f^{\,K^-d}_0(0)$ of the S--wave
amplitude of $K^-d$ scattering is determined by the imaginary of the
amplitude $\tilde{f}^{\,K^-d}_0(0)$ only. This gives
\begin{eqnarray}\label{label2.20}
\hspace{-0.3in}&&{\cal I}m\,{f}^{\,K^-d}_0(0) =
\frac{1}{16\pi}\,\frac{1}{1 +
m_K/m_d}\int\frac{d^3K}{(2\pi)^3}\frac{d^3Q}{(2\pi)^3} \frac{
\Phi^*_d(\vec{K} )}{\sqrt{E_p(\vec{K}\,) E_n(
\vec{K}\,)}}\frac{\Phi_d(\vec{Q})}{\sqrt{ E_p(\vec{Q}\,) E_n( \vec{Q}
\,)}}\nonumber\\\hspace{-0.3in}&&\times\,\frac{1}{3}\sum_{(\sigma_p,\sigma_n;
{^3}{\rm S}_1)} {\cal I}m\,M(K^-(\vec{0}\,)p(\vec{Q},\sigma_p) n(-
\vec{Q},\sigma_n) \to K^-(\vec{0}\,)p(\vec{K},\sigma_p) n( -
\vec{K},\sigma_n)).
\end{eqnarray}
We accentuate that the decomposition of the real part of the S--wave
amplitude of $K^-d$ scattering, given by (\ref{label2.18}), agrees
well with that suggested by Ericson and Weise for the S--wave
scattering length of $\pi^-d$ scattering \cite{TE88} and by Barrett
and Deloff for the S--wave scattering length of $K^-d$ scattering
\cite{RB99}.

Thus, for the calculation of the energy level displacement of the
ground state of kaonic deuterium we have to compute the amplitudes of
$K^-p$, $K^-n$ and $K^- (pn)_{{^3}{\rm S}_1}$ scattering near
thresholds, $f^{\,K^-p}_0(0)$, $f^{\,K^-n}_0(0)$ and
$\tilde{f}^{\,K^-d}_0(0)$, respectively. Since the amplitude
$f^{\,K^-p}_0(0)$ has been computed in \cite{IV3}, it is left to
compute the real part of the amplitude $f^{\,K^-n}_0(0)$ and the
amplitude $\tilde{f}^{\,K^-d}(0)$.

\section{S--wave amplitude of $K^-n$ scattering near threshold}
\setcounter{equation}{0}

The calculation of the S--wave amplitude $f^{\,K^-n}_0(Q)$ of $K^-n$
scattering near threshold we carry out following \cite{IV3}. The
amplitude of low--energy $K^-n$ scattering we represent in the form
\begin{eqnarray}\label{label3.1}
f^{\,K^-n}_0(Q) &=& \frac{1}{2iQ}\,\Big(\eta^{K^-n}_0(Q)\,
e^{\textstyle\,2i\delta^{K^-n}_0(Q)} - 1\Big) =\nonumber\\ &=&
\frac{1}{2iQ}\,\Big( e^{\textstyle\,2i\delta^{K^-n}_B(Q)} - 1\Big) +
e^{\textstyle\,2i\delta^{K^-n}_B(Q)}f^{\,K^-n}_0(Q)_R,
\end{eqnarray}
where $\eta^{K^-n}_0(Q)$ and $\delta^{K^-n}_0(Q)$ are the inelasticity
and the phase shift of low--energy $K^-n$ scattering, which we
describe in terms of $\delta^{K^-n}_0(Q)_B$, the phase shift of an
elastic background of low--energy $K^-n$ scattering, and
$f^{\,K^-n}_0(Q)_R$, the contribution of the resonances. In the
low--energy limit $\delta^{K^-n}_0(Q)_B = A^{K^-n}_B Q$, where
$A^{K^-n}_B$ is a real parameter, and
\begin{eqnarray}\label{label3.2}
f^{\,K^-n}_0(0) = A^{K^-n}_B + f^{\,K^-n}_0(0)_R,
\end{eqnarray}
Since the state $|K^-n\rangle$ has the isospin one, $I = 1$, we assume
\cite{IV3} that $f^{\,K^-n}_0(Q)_R$ is defined by the contribution of
the $\Sigma(1750)$ resonance with isospin $I = 1$ and strangeness $S =
- 1$, the component of the $SU(3)_{\rm flavour}$ octet
\cite{DG00a}\,\footnote{ For simplicity we denote $\Sigma(1750)$ as
$\Sigma^-_2$.}.  The effective Lagrangian of the
$\Sigma(1750)\bar{K}N$--interaction reads \cite{IV3}
\begin{eqnarray}\label{label3.3}
{\cal L}_{\Sigma^-_2BP}(x) &=& f_2\,\bar{\Sigma}^-_2(x) (\Sigma^-(x)
\pi^0(x) - \Sigma^0(x) \pi^-(x)) +
\frac{g_2}{\sqrt{3}}\,\bar{\Sigma}^-_2(x)\Lambda^0(x)\pi^-(x)\nonumber\\
&&- \frac{1}{\sqrt{2}}\,(g_2 - f_2)\,\bar{\Sigma}^-_2(x) n(x) K^-(x) +
{\rm h.c.},
\end{eqnarray}
where $f_2$ and $g_2$ are the phenomenological coupling constants
\cite{IV3}. The value $f_2 = - g_2/3$ has been fixed from the
experimental data on the cross sections for inelastic reactions $K^-p
\to \Sigma^{\pm}\pi^{\mp}$, $K^-p \to \Sigma^0\pi^0$ and $K^-p \to
\Lambda^0\pi^0$. The value $g_2 = 1.123$ has been calculated from the
fit of the width of the $\Sigma(1750)$--resonance equal to
$\Gamma_{\Sigma_2} = 50\,{\rm MeV}$ for mass $m_{\Sigma_2} =
1750\,{\rm MeV}$ \cite{DG00a}.

\subsection{Real part of $f^{\,K^-n}_0(0)_R$}

According to \cite{IV3} the real part ${\cal R}e\,f^{\,K^-n}_0(0)_R$
of the amplitude $f^{\,K^-n}_0(0)_R$ is equal to
\begin{eqnarray}\label{label3.4}
{\cal R}e\,f^{\,K^-n}_0(0)_R =
\frac{1}{4\pi}\,\frac{\mu}{~m_K}\,\frac{8}{9}\,
\frac{g^2_2}{m_{\Sigma_2} - m_K - m_N} = 0.037\,{\rm fm}.
\end{eqnarray}
The numerical value is obtained for $m_K = 494\,{\rm MeV}$ and $m_N =
940\,{\rm MeV}$.

Now we should proceed to computing the contribution of the smooth
elastic background of low--energy $K^-n$ scattering.

\subsection{Elastic background of low--energy $K^-n$ scattering}

Using the results obtained in \cite{IV3} one can show that the smooth
elastic background $A^{K^-n}_B$ of low--energy elastic $K^-n$
scattering does not contain the contribution of exotic four--quark
states $a_0(980)$ and $f_0(980)$ and, as has been pointed out in
\cite{IV3}, can be fully determined within the soft--kaon theorem and
current algebra approach \cite{SW66}--\cite{EK98}. The result reads
\cite{IV3} 
\begin{eqnarray}\label{label3.5}
A^{K^-n}_B = \frac{1}{8\pi}\, \frac{1}{F^2_K}\,
\frac{m_Km_N}{m_K + m_N} = (0.200 \pm 0.024)\,{\rm fm},
\end{eqnarray}
where $F_K = 113\,{\rm MeV}$ is the PCAC constant of the
$K$--mesons \cite{DG00} and $\pm 0.024\,{\rm fm}$ is an uncertainty of
the current algebra approach \cite{IV3}.

As has been shown in \cite{IV3} the smooth elastic background of
low--energy $K^-n$ scattering can be also defined by the lowest quark
box--diagram depicted in Fig.2, calculated with the Effective quark
model with chiral $U(3)\times U(3)$ symmetry \cite{AI99}--\cite{AI92}.
\begin{figure}
  \centering
  \includegraphics[width=0.35\textwidth]{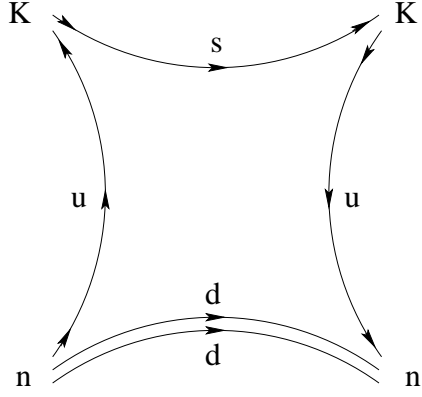}
\caption{The quark diagram describing a smooth elastic background of
low--energy elastic $K^-n$ scattering in the Effective quark model
with chiral $U(3)\times U(3)$ symmetry.}
\end{figure}
Using the reduction technique \cite{IZ80} the amplitude of elastic
low--energy $K^-n$ scattering we define as
\begin{eqnarray}\label{label3.6}
\hspace{-0.3in}&&(2\pi)^4 i\,\delta^{(4)}(q\,' + p\,' - q - p)\,M(K^-
n \to K^- n) =\nonumber\\
\hspace{-0.3in}&& = \lim_{p\,'^{\,2},\, p^2 \to m^2_N,\, q\,'^{\,2},
\,q^2 \to m^2_K}\int d^4x_1 d^4x_2 d^4x_3 d^4x_4\,e^{\textstyle
i\,q\,' \cdot x_1 + ip\,'\cdot x_2 - ip\cdot x_3 - i q\cdot
x_4}\nonumber\\
\hspace{-0.3in}&&\times\,(\Box_1 + m^2_K)(\Box_4 +
m^2_K)\,\bar{u}(p\,',\sigma\,'_n)\,\overrightarrow{(i\gamma_{\nu}
\,\partial^{\nu}_2 - m_N)}\langle 0|{\rm T}(K^-(x_1) n(x_2)
\bar{n}(x_3) K^+(x_4))|0\rangle\nonumber\\
\hspace{-0.3in}&&\times\,\overleftarrow{(-
i\gamma_{\mu}\,\partial^{\mu}_3 - m_N)}\,u(p,\sigma_n),
\end{eqnarray}
where $n(x)$ and $u(p,\sigma_n)$ are the interpolating field operator
and the Dirac bispinor of the neutron, and $K^{\pm}(x)$ are the
interpolating fields of the $K^{\mp}$--mesons.

In order to describe the r.h.s. of Eq.(\ref{label3.6}) at the quark
level we follow \cite{AI99} and use the equations of motion
\begin{eqnarray}\label{label3.7}
\overrightarrow{(i\gamma_{\nu}\,\partial^{\nu}_2 - m_N)}\,p(x_1) &=&
\frac{g_{\rm B}}{\sqrt{2}}\,\eta_n(x_2),\nonumber\\
\bar{p}(x_3)\overleftarrow{(- i\gamma_{\mu}\partial^{\mu}_3 - m_N)}
&=& \frac{g_{\rm B}}{\sqrt{2}}\,\bar{\eta}_n(x_3),
\end{eqnarray}
where $\eta_n(x_2)$ and $\bar{\eta}_n(x_3)$ are the three--quark
current densities \cite{AI99}
\begin{eqnarray}\label{label3.8}
\eta_n(x_2) &=& -
\,\varepsilon^{ijk}\,[\bar{d^c}_i(x_2)
\gamma^{\mu}d_j(x_2)]\gamma_{\mu}\gamma^5 u_k(x_2),\nonumber\\ 
\bar{\eta}_n(x_3) &=&+\,\varepsilon^{ijk}\,\bar{u}_i(x_3)\gamma^{\mu}
\gamma^5[\bar{d}_j(x_3)\gamma_{\mu}d^c_k(x_3)]
\end{eqnarray}
where $i,j$ and $k$ are colour indices and $\bar{\psi^{\,c}}(x) =
\psi(x)^T C$ and $C = - C^T = - C^{\dagger} = - C^{-1} $ is the charge
conjugate matrix, $T$ denotes transposition, and $g_{\rm B}$ is the
phenomenological coupling constant of the low--lying baryon octet
$B_8(x)$ coupled to the three--quark current densities \cite{AI99}
\begin{eqnarray}\label{label3.9}
\hspace{-0.5in}{\cal L}^{(\rm B)}_{\rm int}(x) = \frac{g_{\rm
B}}{\sqrt{2}}\,\bar{B}_8(x)\eta_8(x) + {\rm h.c.}
\end{eqnarray}
The coupling constant $g_{\rm B}$ is equal to $g_{\rm B} = 1.34\times
10^{-4}\,{\rm MeV}^{-2}$ \cite{AI99}.

For the interpolating field operators of the $K^{\pm}$--mesons we use
the following equations of motion \cite{AI99}
\begin{eqnarray}\label{label3.10}
(\Box_1 + m^2_K)K^-(x_1) &=&
\frac{g_K}{\sqrt{2}}\,\bar{u}(x_1)i\gamma^5 s(x_1),\nonumber\\ (\Box_4
+ m^2_K)K^+(x_4) &=& \frac{g_K}{\sqrt{2}}\,\bar{s}(x_4)i\gamma^5
u(x_4),
\end{eqnarray}
where $g_K = (m + m_s)/\sqrt{2}F_K$, $m = 330 \,{\rm MeV}$ and $m_s =
465\,{\rm MeV}$ are the masses of the constituent $u$, $d$ and $s$
quarks, respectively \cite{AI99,AI92} (see also \cite{AI80}).

The amplitude of low--energy elastic  $K^-p$ scattering is defined by
\begin{eqnarray}\label{label3.11}
\hspace{-0.3in}&&M(K^- n \to K^- n) = -\,i\,\frac{1}{4}\,g^2_{\rm
B}\,g^2_K\int d^4x_1 d^4x_2 d^4x_3\,e^{\textstyle i\,q\,' \cdot x_1 +
ip\,'\cdot x_2 - ip\cdot x_3}\nonumber\\
\hspace{-0.3in}&&\times\,\bar{u}(p\,',\sigma\,'_n)\langle 0|{\rm
T}(\bar{u}(x_1)i\gamma^5 s(x_1)\eta_n(x_2) \bar{\eta}_n(x_3)
\bar{s}(0)i\gamma^5 u(0))|0\rangle\,u(p,\sigma_n).
\end{eqnarray}
where the external momenta $q\,'$, $p\,'$, $q$ and $p$ should be kept
on--mass shell $q\,'^{\,2} = q^2 = m^2_K$ and $p\,'^{\,2} = p^2 =
m^2_N$. 

In Appendix B we have computed the vacuum expectation value. The
parameter $A^{K^-n}_B$, defining the smooth elastic background of
low--energy $K^-n$ scattering, is equal to
\begin{eqnarray}\label{label3.12}
A^{K^-n}_B = \frac{M(K^- n \to K^- n)}{8\pi (m_K + m_N)} = (0.221\pm
0.024)\,{\rm fm}.
\end{eqnarray}
The value of the smooth elastic background of low--energy $K^-n$
scattering, calculated with the Effective quark model with chiral
$U(3)\times U(3)$ symmetry, agrees well with that calculated within
the soft--kaon theorem and current algebra approach (\ref{label3.5}).

\subsection{Real part of the S--wave amplitude of $K^-n$ scattering
near threshold and the S--wave scattering lengths of $K^-N$ scattering
with isospin $I = 0$ and $I = 1$}

Using the results obtained in Subsections 3.1 and 3.2 we can compute
the real part of the amplitude $f^{\,K^-n}_0(0)$ of $K^-n$ scattering
near threshold
\begin{eqnarray}\label{label3.13}
{\cal R}e\,f^{\,K^-n}_0(0) = (0.258 \pm 0.024)\,{\rm fm}.
\end{eqnarray}
Since the $K^-n$ couples in the state with isospin $I = 1$, the real
part of the S--wave amplitude $f^{\,K^-n}_0(0)$ defines the S-wave
scattering length $a^1_0$ of $\bar{K}N$ scattering with isospin $I=1$:
$a^1_0 = (0.258 \pm 0.024)\,{\rm fm}$. Using the S--wave amplitude of
$K^-p$ scattering, calculated in \cite{IV3}, the S--wave scattering
length $a^0_0$ of $\bar{K}N$ scattering is equal to: $a^0_0 =
(-\,1.221 \pm 0.072)\,{\rm fm}$. The values
\begin{eqnarray}\label{label3.14}
a^0_0 &=& -\,1.221 \pm 0.072\,{\rm fm},\nonumber\\ a^1_0 &=& +\,0.258
\pm 0.024 \,{\rm fm}
\end{eqnarray}
we will use for the numerical calculation of the Ericson--Weise
contribution \cite{TE88} to the S--wave scattering length of $K^-d$
scattering.

\section{Amplitude of $K^- (p n)_{{^3}{\rm S}_1}$ scattering near threshold}
\setcounter{equation}{0}

The amplitude of $K^- (p n)_{{^3}{\rm S}_1}$ scattering, $K^- (p
n)_{{^3}{\rm S}_1} \to K^- (p n)_{{^3}{\rm S}_1}$, can be represented
in the form of two main contributions: (i) the amplitude, defining the
S--wave scattering length of $K^-d$ scattering in the Ericson--Weise
form, described by the Feynman diagrams depicted in Fig.3 and caused
by one--kaon exchanges, and (ii) the amplitude, defined by the
inelastic two--body $K^- (p n)_{{^3}{\rm S}_1}\to NY$ and three--body
$K^- (p n)_{{^3}{\rm S}_1} \to NY\pi$ channels, where $Y = \Lambda^0,
\Sigma^0$ or $\Sigma^{\pm}$ hyperons.
\begin{figure}
\centering \psfrag{pm}{$K^-$} \psfrag{p0}{$\bar{K}^0$} \psfrag{p}{$p$}
\psfrag{n}{$n$} \includegraphics[height=0.24\textheight]{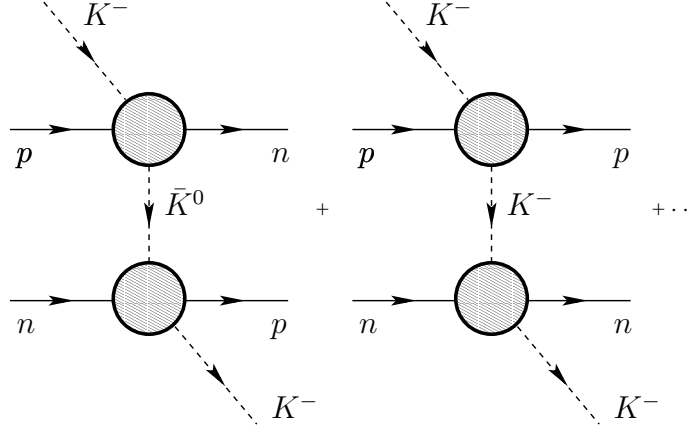}
\caption{Feynman diagrams of the amplitude of the low--energy reaction
$K^- (p n)_{{^3}{\rm S}_1} \to K^- (p n)_{{^3}{\rm S}_1}$ defining the
Ericson--Weise contribution to the S--wave scattering length of $K^-d$
scattering}
\end{figure}

This gives the following analytical representation of the amplitude of
the reaction $K^- (p n)_{{^3}{\rm S}_1} \to K^- (p n)_{{^3}{\rm S}_1}$:
\begin{eqnarray}\label{label4.1}
\hspace{-0.3in}&&\frac{1}{3}\sum_{(\sigma_p,\sigma_n; {^3}{\rm
S}_1)}M(K^-(\vec{0}\,) p(\vec{Q},\sigma_p) n(- \vec{Q},\sigma_n) \to
K^-(\vec{0}\,) p(\vec{K},\sigma_p) n( - \vec{K},\sigma_n)) =
\nonumber\\ \hspace{-0.3in}&&= \frac{1}{3}\sum_{(\sigma_p,\sigma_n;
{^3}{\rm S}_1)}M(K^-(\vec{0}\,) p(\vec{Q},\sigma_p) n(-
\vec{Q},\sigma_n) \to K^-(\vec{0}\,) p(\vec{K},\sigma_p) n( -
\vec{K},\sigma_n))_{\rm EW}\nonumber\\ \hspace{-0.3in}&&+
\frac{1}{3}\sum_{(\sigma_p,\sigma_n; {^3}{\rm
S}_1)}\tilde{M}(K^-(\vec{0}\,) p(\vec{Q},\sigma_p) n(-
\vec{Q},\sigma_n) \to K^-(\vec{0}\,) p(\vec{K},\sigma_p) n( -
\vec{K},\sigma_n)),
\end{eqnarray}
where $M(K^- p n \to K^- p n )_{\rm EW}$ reproduces the Ericson--Weise
formula of the S--wave scattering length of $K^-d$ scattering, defined
by the one--kaon exchanges, whereas $\tilde{M}(K^- p n \to K^- p n)$
is the amplitude of the reaction $K^- (p n)_{{^3}{\rm S}_1} \to K^- (p
n)_{{^3}{\rm S}_1}$, saturated by the inelastic two--body $K^- (p
n)_{{^3}{\rm S}_1} \to NY$ with $NY = n \Lambda^0, n \Sigma^0, p
\Sigma^-$ and three--body $K^- (p n)_{{^3}{\rm S}_1} \to NY\pi$ with
$NY\pi = n\Lambda^0\pi^0, p\Lambda^0\pi^-, n\Sigma^0\pi^0,
n\Sigma^+\pi^-, p\Sigma^0\pi^-, p\Sigma^-\pi^0, n \Sigma^- \pi^+$
reactions.

\subsection{The Ericson--Weise formula for $a^{K^-d}_0$ scattering 
length}

In the low--energy approximation the amplitude $M(K^- p n \to K^- p n
)_{\rm EW}$ is defined in terms of the S--wave scattering lengths of
$\bar{K} N$ scattering.  For the calculation of $M(K^- p n \to K^- p n
)_{\rm EW}$ with the $np$ pair in the ${^3}{\rm S}_1$ state with
isospin zero, $I = 0$, we suggest to use the following effective
Lagrangian
\begin{eqnarray}\label{label4.2}
\hspace{-0.3in}&&{\cal L}_{\rm eff}(x) =\bar{p}(x)(i\gamma^{\mu}
\partial_{\mu} - m_N)p(x) + \bar{n}(x)(i\gamma^{\mu}\partial_{\mu} -
m_N)n(x)\nonumber\\
\hspace{-0.3in}&&+ \,\partial_{\mu}K^{-\dagger}(x)
\partial^{\mu}K^-(x) - m^2_KK^{-\dagger}(x)K^-(x) +
\partial_{\mu}\bar{K}^{0\dagger}(x)\partial^{\mu}\bar{K}^0(x) - m^2_K
\bar{K}^{0\dagger}(x)\bar{K}^0(x) \nonumber\\
\hspace{-0.3in}&&+ \,4\pi \Big(1 +
\frac{m_K}{m_N}\Big)\Big[\frac{1}{2}\,(a^0_0 + a^1_0) K^{-\dagger}(x)
K^-(x) \bar{p}(x) p(x) + \frac{1}{2}\,(a^1_0 - a^0_0)
\bar{K}^{0\dagger}(x) K^-(x) \bar{n}(x)p(x)\Big] \nonumber\\
\hspace{-0.3in}&& + 4\pi \Big(1 + \frac{m_K}{m_N}\Big) \Big[a_1
K^{-\dagger}(x) K^-(x)\bar{n}(x)n(x) +\frac{1}{2}\,(a^1_0 - a^0_0)
K^{-\dagger}(x) \bar{K}^0(x) \bar{p}(x) n(x)\Big],
\end{eqnarray}
where $a^0_0$ and $a^1_0$ are the S--wave scattering lengths of $\bar{K}
N$ scattering with isospin $I = 0$ and $I = 1$, respectively. 

The effective action for the $K^- p n \to K^- p n$ transition in the
one--kaon exchange approximation, described by the effective
Lagrangian (\ref{label4.2}), reads
\begin{eqnarray}\label{label4.3}
&&S^{(K^-pn)}_{\rm eff} = - \int d^4x\,d^4y\, 4\pi^2 \Big(1 +
\frac{m_K}{m_N}\Big)^2 \nonumber\\ &&\times\, [ 4 a^1_0(a^0_0 + a^1_0)
K^{-\dagger}(x) \bar{p}(x) p(x)\,\langle 0|{\rm T}(K^-(x)
K^{-\dagger}(y))|0\rangle\,\bar{n}(y) n(y)\,K^-(y)\nonumber\\&&
\hspace{0.2in}+ (a^0_0 - a^1_0)^2 K^{-\dagger}(x) \bar{p}(x)
n(x)\,\langle 0|{\rm T}(\bar{K}^0(x) \bar{K}^{0\dagger}(y))|0\rangle\,
\bar{n}(y) p(y)\,]\,K^-(y).
\end{eqnarray}
The second term describes the contribution of the first diagram in
Fig.3, which corresponds to the charge--exchange channel. 

Since in the case of isospin symmetry the vacuum expectation values of
the $\bar{K}$-meson fields are equal
\begin{eqnarray}\label{label4.4}
\langle 0|{\rm T}(K^-(x) K^{-\dagger}(y))|0\rangle &=& \langle 0|{\rm
T}(\bar{K}^0(x) \bar{K}^{0\dagger}(y))|0\rangle =\nonumber\\ &=& -\,i
\Delta(x - y) = \int \frac{d^4k}{(2\pi)^4i}\,\frac{\displaystyle
e^{\textstyle\,-ik\cdot (x - y)}}{m^2_K - k^2 - i\,0},
\end{eqnarray}
the r.h.s. of the effective action (\ref{label4.3}) can be transcribed
into the form
\begin{eqnarray}\label{label4.5}
\hspace{-0.3in}S^{(K^-pn)}_{\rm eff} &=& i\int d^4x\,d^4y\,4\pi^2
\Big(1 + \frac{m_K}{m_N}\Big)^2\,\Delta(x - y)\, \Big[4 a^1_0(a^0_0 +
a^1_0) K^{-\dagger}(x) K^-(y)\nonumber\\
\hspace{-0.3in}&&\times\, \bar{p}(x) p(x) \bar{n}(y) n(y) + (a^0_0 -
a^1_0)^2 K^{-\dagger}(x) K^-(y) \bar{p}(x) n(x) \bar{n}(y) p(y)\Big].
\end{eqnarray}
Now we have to take into account that the $np$ pair couples in the
${^3}{\rm S}_1$ state with isospin zero, $I = 0$. This can be carried
out by means of a Fierz transformation. Keeping only the term,
describing the $np$ pair in the ${^3}{\rm S}_1$ state with isospin
zero, we get
\begin{eqnarray}\label{label4.6}
\hspace{-0.5in}&& \bar{p}(x)p(x)\bar{n}(y)n(y) \to
-\,\frac{1}{4}\,\bar{p}(x)\gamma_{\mu}n^c(y) \bar{n^c}(y)\gamma^{\mu}
p(x)+\,\frac{1}{8}\,\bar{p}(x)\sigma_{\mu\nu}n^c(y)
\bar{n^c}(y)\sigma^{\mu\nu} p(x),\nonumber\\
\hspace{-0.5in}&& \bar{p}(x)n(x)\bar{n}(y)p(y) \to +
\,\frac{1}{4}\,\bar{p}(x)\gamma_{\mu}n^c(y) \bar{n^c}(x)\gamma^{\mu}
p(y) - \,\frac{1}{8}\,\bar{p}(x)\sigma_{\mu\nu} n^c(y)
\bar{n^c}(x)\sigma^{\mu\nu} p(y).
\end{eqnarray}
As has been shown in \cite{AI1}--\cite{AI4} the nucleon densities
$\bar{p}(x)\gamma_{\mu}n^c(y)$ and $\bar{n^c}(x)\gamma^{\mu} p(y)$
have the quantum numbers of the deuteron. In the non--relativistic
limit there survives only the ${^3}{\rm S}_1$ component of the $np$
pair, whereas the ${^3}{\rm D}_1$ state is suppressed.

In order to understand the quantum numbers of the components of the
tensor nucleon densities $\bar{p}(x)\sigma_{\mu\nu}n^c(y)$ and
$\bar{n^c}(x)\sigma^{\mu\nu} p(y)$ it is convenient to represent the
product of the tensor nucleon densities as follows
\begin{eqnarray}\label{label4.7}
\hspace{-0.3in}\bar{p}(x)\sigma_{\mu\nu} n^c(y)
\bar{n^c}(y)\sigma^{\mu\nu} p(x) &=& - 2\,\bar{p}(x)
\gamma^0\vec{\gamma}\, n^c(y)\cdot \bar{n^c}(y) \gamma^0 \vec{\gamma}
\,p(x)\nonumber\\
\hspace{-0.3in}&& - 2\,\bar{p}(x) \gamma^0 \vec{\gamma}
\gamma^5 n^c(y)\cdot \bar{n^c}(y)\gamma^0\vec{\gamma}\gamma^5
p(x),\nonumber\\
\hspace{-0.3in}\bar{p}(x)\sigma_{\mu\nu}n^c(y)
\bar{n^c}(x)\sigma^{\mu\nu} p(y) &=& - 2\,
\bar{p}(x)\gamma^0\vec{\gamma}\,n^c(y)\cdot \bar{n^c}(x) \gamma^0
\vec{\gamma} \,p(y) \nonumber\\
\hspace{-0.3in}&& - 2\, \bar{p}(x)\gamma^0\vec{\gamma} \gamma^5 n^c(y)
\cdot \bar{n^c}(x) \gamma^0\vec{\gamma}\gamma^5 p(y).
\end{eqnarray}
One can show that only nucleon densities $\bar{p}(x)
\gamma^0\vec{\gamma} n^c(y)$ and $\bar{n^c}(y) \gamma^0 \vec{\gamma}
p(x)$ have the quantum numbers of the deuteron. However the
contribution of the ${^3}{\rm D}_1$ component enters with the sign
opposite to that of the nucleon densities $\bar{p}(x)\gamma^{\mu}
n^c(y)$ and $\bar{n^c}(x) \gamma^{\mu} p(y)$. Therefore, they coincide
in the non--relativistic limits. Since the nucleon densities
$\bar{p}(x)\gamma^0\vec{\gamma} \gamma^5 n^c(y)$ and $\bar{n^c}(x)
\gamma^0\vec{\gamma}\gamma^5 p(y)$ have no the quantum numbers of the
deuteron, we will drop them from further consideration.

In the low--energy limit, when the masses of nucleons are much greater
their relative 3--momenta, and using (\ref{label4.7}) we reduce the
four--nucleon interaction in the effective action (\ref{label4.6}) to
the form\,\footnote{For the derivation of this effective action we
have taken into account that in the non--relativistic limit $
[\bar{p}(x) \gamma^0\vec{\gamma}\,n^c(y)]\cdot [\bar{n^c}(y) \gamma^0
\vec{\gamma}\,p(x)] \to -\,[\bar{p}(x) \vec{\gamma} \,n^c(y)]\cdot
[\bar{n^c}(y) \vec{\gamma}\,p(x)]$.}
\begin{eqnarray}\label{label4.8}
  S^{(K^-(pn)_{{^3}{\rm S}_1})}_{\rm eff} &=& -\,i\int d^4x\,d^4y\,2\pi^2
  \Big(1 + \frac{m_K}{m_N}\Big)^2\,((a^0_0 - a^1_0)^2 - 4 a^1_0(a^0_0 + a^1_0))\nonumber\\ 
&&\times\, \Delta(x - y)\,K^{-\dagger}(x) K^-(y)
  [\bar{p}(x)\vec{\gamma}\,n^c(x)]\cdot [\bar{n^c}(y)\vec{\gamma}\,
  p(y)].
\end{eqnarray}
Using the effective action (\ref{label4.8}) we obtain the amplitude of
the reaction $K^- (p n)_{{^3}{\rm S}_1} \to K^- (p n)_{{^3}{\rm
S}_1}$, caused by the one--kaon exchanges, with the $np$ pairs coupled
in the ${^3}{\rm S}_1$ states with isospin zero:
\begin{eqnarray}\label{label4.9}
\hspace{-0.3in}&&M(K^-(\vec{0}\,) p(\vec{Q},\sigma_p) n(-
\vec{Q},\sigma_n) \to K^-(\vec{0}\,) p(\vec{K},\sigma_p) n( -
\vec{K},\sigma_n))_{\rm EW} = \nonumber\\
\hspace{-0.3in}&&= -\,2\pi^2 \Big(1 +
\frac{m_K}{m_N}\Big)^2\,((a^0_0 - a^1_0)^2 - 4 a^1_0(a^0_0 + a^1_0))\,\frac{1}{m^2_K + (\vec{K} -
\vec{Q}\,)^2}\,\nonumber\\
\hspace{-0.3in}&&\times\,[\bar{u}(\vec{K},\sigma_p)\vec{\gamma}\,n^c(-
\vec{K},\sigma_n)]\cdot [\bar{u^c}( -\vec{Q},\sigma_n)\vec{\gamma}\,
u(\vec{Q},\sigma_p)].
\end{eqnarray}
At low energies the summation over polarisations of the $np$ pair in
the ${^3}{\rm S}_1$ state gives (see Appendix C)
\begin{eqnarray}\label{label4.10}
  \hspace{-0.5in}&&\frac{1}{3}\sum_{(\sigma_p,
    \sigma_n)}M(K^-(\vec{0}\,) p(\vec{Q},\sigma_p) n(- \vec{Q},\sigma_n)
  \to \pi^-(\vec{0}\,) p(\vec{K},\sigma_p) n( - \vec{K},\sigma_n))_{\rm
    EW} = \nonumber\\
  \hspace{-0.5in}&&= 16\, \pi^2 \,m^2_N\, \Big(1 +
  \frac{m_K}{m_N}\Big)^2\, (4a^1_0(a^0_0 + a^1_0) - (a^0_0 - a^1_0)^2)\,\frac{1}{m^2_K + (\vec{K} - \vec{Q}\,)^2}.
\end{eqnarray}
Substituting (\ref{label4.10}) into (\ref{label2.19}) we get
\begin{eqnarray}\label{label4.11}
&&\tilde{f}^{\,K^-d}_0(0)_{\rm EW} = \frac{1}{4}\,\Big(1 +
\frac{m_K}{m_d}\Big)^{-1}\Big(1 + \frac{m_K}{m_N}\Big)^2\,(4a^1_0(a^0_0 + a^1_0) - (a^0_0 - a^1_0)^2)\nonumber\\ 
&&\times
\int\frac{d^3K}{(2\pi)^3}\frac{d^3Q}{(2\pi)^3}
\frac{m_N}{E_N(\vec{K}\,)} \Phi^*_d(\vec{K} )\frac{4\pi}{m^2_K +
(\vec{K} - \vec{Q}\,)^2}\,\frac{m_N}{E_N(\vec{Q}\,)}\,\Phi_d(\vec{Q}).
\end{eqnarray}
The expression (\ref{label4.11}) can be transcribed into the form
suggested by Ericson and Weise \cite{TE88}
\begin{eqnarray}\label{label4.12}
  \tilde{f}^{\,K^-d}_0(0)_{\rm EW} = \frac{1}{4}\,\Big(1 +
  \frac{m_K}{m_d}\Big)^{-1}\Big(1 + \frac{m_K}{m_N}\Big)^2\,(4a^1_0(a^0_0 + a^1_0) - (a^0_0 - a^1_0)^2)\, 
\Big\langle \frac{1}{r_{12}}\Big\rangle,
\end{eqnarray}
where $r_{12}$ is a distance between two scatterers $n$ and $p$
\cite{TE88}. 

The double scattering contribution to the S--wave amplitude of $K^-d$
scattering has been calculated by Kamalov {\it et al.} \cite{EO01}. In
the notation by Kamalov {\it et al.} the amplitude
$\tilde{f}^{\,K^-d}_0(0)_{\rm EW}$ reads
\begin{eqnarray}\label{labelA.1}
  \tilde{f}^{\,K^-d}_0(0)_{\rm EW} = \Big(1 +
  \frac{m_K}{m_d}\Big)^{-1}\Big(1 + \frac{m_K}{m_N}\Big)^2\,(2 a_p a_n  - a^2_x)\, 
\Big\langle \frac{1}{r_{12}}\Big\rangle,
\end{eqnarray}
where $a_p = (a^0_0 + a^1_0)/2$, $a_n = a^1_0$ and $a_x = (a^1_0 -
a^0_0)/2$\,\footnote{The term proportional to $a^2_x$ is defined by
  the first diagram in Fig.3. It corresponds to the charge--exchanged
  channel, which dominates in the double scattering. In our former
  version nucl--th/0406053v2 the contribution of the double scattering
  contained the factor $(a_p a_n - a^2_x)$ instead of $(2 a_p a_n -
  a^2_x)$. We are grateful to Avraham Gal for calling our attention to
  this discrepancy. The replacement of $a_pa_n$ by $2a_p a_n$ changes
  the contribution of the double scattering by $16\,\%$ only (see
  Eq.(\ref{label4.16})). The change of the total S--wave scattering
  length of $K^-d$ scattering does not go beyond the theoretical
  uncertainty, which is about $18\,\%$.}.

In our approach $\langle 1/r_{12}\rangle$ is defined by
\begin{eqnarray}\label{label4.13}
  \Big\langle \frac{1}{r_{12}}\Big\rangle = \int
  d^3x\,\Psi^*_d(\vec{r}\,)\,\frac{\displaystyle e^{\textstyle\,- m_K
      r}}{r}\,\Psi_d(\vec{r}\,) = 2\gamma_d \,E_1\Big(\frac{m_N}{m_K
    + 2 \gamma_d}\Big) = 0.29\,m_{\pi},
\end{eqnarray}
where $E_1(z)$ is the Exponential Integral \cite{HMF72}, $\gamma_d =
1/r_d = 0.327\,m_{\pi}$ and $\Psi_d(\vec{r}\,)$ is the wave function
of the deuteron in the ground state in the coordinate representation.
We have restricted the spatial region of the integration from below by
the Compton--wavelength of the nucleon \cite{AI2}.

The analogous calculation of the amplitude of $\pi^-d$ scattering
\cite{AI04} in the one--pion exchange approximation reproduces fully
the Ericson--Weise formula \cite{TE88}
\begin{eqnarray}\label{label4.14}
\tilde{f}^{\pi^-d}_0(0)_{\rm EW} &=& 2\,\Big(1 +
\frac{m_{\pi}}{m_d}\Big)^{-1}\Big(1 + \frac{m_{\pi}}{m_N}\Big)^2\,
(b^2_0 - 2 b^2_1)\,\Big\langle \frac{1}{r_{12}}\Big\rangle,
\end{eqnarray}
where $b_0 = (a^{1/2}_0 + 2 a^{3/2}_0)/3$ and $b_0 = (a^{3/2}_0 -
a^{1/2}_0)/3$ are the isoscalar and isovector S--wave scattering
lengths of $\pi N$ scattering, $a^{1/2}_0$ and $a^{3/2}_0$ are the
S--wave scattering lengths of $\pi N$ scattering with isospin $I =
1/2$ and $I = 3/2$ and
\begin{eqnarray}\label{label4.15}
\Big\langle \frac{1}{r_{12}}\Big\rangle = \int
d^3x\,\Psi^*_d(\vec{r}\,)\,\frac{\displaystyle e^{\textstyle\,-
m_{\pi} r}}{r}\,\Psi_d(\vec{r}\,) = 2 \gamma_d
E_1\Big(\frac{m_N}{m_{\pi} + 2\gamma_d}\Big) = 0.69\,m_{\pi},
\end{eqnarray}
The numerical value (\ref{label4.15}) agrees well with the
Ericson--Weise estimate $\langle 1/r_{12}\rangle = 0.64\,m_{\pi}$
\cite{TE88}.

Since the S--wave scattering lengths $a^0_0$ and $a^1_0$ of $\bar{K}N$
scattering are equal to $a^0_0 = (-1.221 \pm 0.072)\,{\rm fm}$ and
$a^1_0 = (0.258 \pm 0.024)\,{\rm fm}$ (see (\ref{label3.14})), the
value of $\tilde{f}^{\,K^-d}_0(0)_{\rm EW}$ amounts to
\begin{eqnarray}\label{label4.16}
\tilde{f}^{\,K^-d}_0(0)_{\rm EW} = (-\, 0.301 \pm 0.021)\,{\rm fm}.
\end{eqnarray}
Thus, the S--wave scattering length of $K^-d$ scattering, defined by
the Ericson--Weise formula, is equal to
\begin{eqnarray}\label{label4.17}
(a^{K^-d}_0)_{\rm EW} = \frac{1 + m_K/m_N}{1 + m_K/m_d}\,(a^{K^-p}_0 +
a^{K^-n}_0) + \tilde{f}^{\,K^-d}_0(0)_{\rm EW} = (-\,0.572 \pm
0.094)\,{\rm fm}.
\end{eqnarray}
The total S--wave scattering length $a^{K^-d}_0$ of $K^-d$ scattering
is defined by
\begin{eqnarray}\label{label4.18}
a^{K^-d}_0 = (a^{K^-d}_0)_{\rm EW} + {\cal
R}e\,\tilde{f}^{\,K^-d}_0(0).
\end{eqnarray}
The S--wave amplitude $\tilde{f}^{\,K^-d}_0(0)$ of the reaction $K^-
(p n)_{{^3}{\rm S}_1} \to K^- (p n)_{{^3}{\rm S}_1}$ is saturated by
the inelastic two--body $K^- (p n)_{{^3}{\rm S}_1}\to NY$ and
three--body $K^- (p n)_{{^3}{\rm S}_1} \to NY \pi$ channels
\begin{eqnarray}\label{label4.19}
\tilde{f}^{\,K^-d}_0(0) &=& \tilde{f}^{\,K^-d}_0(0)_{(\rm two-body)} +
\tilde{f}^{\,K^-d}_0(0)_{(\rm three-body)},
\end{eqnarray}
where we have denoted
\begin{eqnarray}\label{label4.20}
\tilde{f}^{\,K^-d}_0(0)_{(\rm two-body)} &=&
\sum_{NY}\tilde{f}^{\,K^-d}_0(0)_{NY},\nonumber\\
\tilde{f}^{\,K^-d}_0(0)_{(\rm three-body)} &=&
\sum_{NY\pi}\tilde{f}^{\,K^-d}_0(0)_{NY\pi}.
\end{eqnarray}
The contribution of the reactions $K^- (p n)_{{^3}{\rm S}_1} \to N
\Lambda^0 \pi \pi$ can be neglected due to a smallness of the phase
volume.

\section{Inelastic two--body channels $K^- (p n)_{{^3}{\rm S}_1}
 \to NY$. General formulas}
\setcounter{equation}{0}

The part of the width $\Gamma_{1s}$ of the ground state of kaonic
deuterium $A_{Kd}$ is defined by the decays $A_{Kd} \to NY$, where
$NY = n \Lambda^0, n \Sigma^0$ and $p \Sigma^-$. The other possible
two--body decays as $A_{Kd} \to n \Lambda(1405)$ and $A_{Kd} \to N
\Sigma(1385)$ are suppressed by the phase volume relative to the
decays $A_{Kd} \to n \Lambda^0, n \Sigma^0$ and $p \Sigma^-$
\cite{BK59}--\cite{VV70}.

The contribution of the decays $A_{Kd} \to n \Lambda^0, n \Sigma^0$
and $p \Sigma^-$ to the energy level displacement of the ground state
of kaonic deuterium we take into account by computing the amplitude of
the reaction $K^- (p n)_{{^3}{\rm S}_1} \to K^- (p n)_{{^3}{\rm
S}_1}$, defined by the inelastic channels $K^- (p n)_{{^3}{\rm S}_1}
\to N Y \to K^- (p n)_{{^3}{\rm S}_1}$, where $NY = n \Lambda^0, n
\Sigma^0$ and $p \Sigma^-$. At threshold in the reaction $K^- (p
n)_{{^3}{\rm S}_1} \to N Y$ the $NY$ pair can be produced in the
${^3}{\rm P}_1$ and ${^1}{\rm P}_1$ state.

The amplitude of low--energy $K^- (p n)_{{^3}{\rm S}_1} \to K^- (p
n)_{{^3}{\rm S}_1}$ scattering, caused by the contribution of
two--body inelastic channels $K^- (p n)_{{^3}{\rm S}_1} \to N Y \to
K^- (p n)_{{^3}{\rm S}_1}$ with the $NY$ pair coupled in the ${^3}{\rm
P}_1$ and ${^1}{\rm P}_1$ state, we define as \cite{IV1}
\begin{eqnarray}\label{label5.1}
&&\tilde{M}(K^-(\vec{0}\,) p(\vec{Q},\sigma_p) n(- \vec{Q},\sigma_n)
\to K^-(\vec{0}\,) p(\vec{K},\sigma_p) n( - \vec{K},\sigma_n)) =
\nonumber\\ &&= \int \frac{d^3k_1}{(2\pi)^3 2
E_n(\vec{k}_1)}\frac{d^3k_2}{(2\pi)^3 2
E_{\Lambda^0}(\vec{k}_2)}\,\frac{(2\pi)^3\,\delta^{(3)}(\vec{k}_1 +
\vec{k}_2)}{E_n(\vec{k}_1) + E_{\Lambda^0}(\vec{k}_2) - 2 m_N - m_K
-i\,0}\nonumber\\ &&\times \sum_{(\alpha_2,\alpha_1; {^3}{\rm P}_1)}
M(K^-(\vec{0}\,) p(\vec{Q},\sigma_p) n( - \vec{Q},\sigma_n) \to
n(\vec{k}_1,\alpha_1) \Lambda^0(\vec{k}_2,\alpha_2); {^3}{\rm
P}_1)\nonumber\\ &&\times\, M(n(\vec{k}_1,\alpha_1)
\Lambda^0(\vec{k}_2,\alpha_2) \to K^-(\vec{0}\,) p(\vec{K},\sigma_p)
n( - \vec{K},\sigma_n); {^3}{\rm P}_1)\nonumber\\ &&+ \int
\frac{d^3k_1}{(2\pi)^3 2 E_n(\vec{k}_1)}\frac{d^3k_2}{(2\pi)^3 2
E_{\Sigma^0}(\vec{k}_2)}\,\frac{(2\pi)^3\,\delta^{(3)}(\vec{k}_1 +
\vec{k}_2)}{E_n(\vec{k}_1) + E_{\Sigma^0}(\vec{k}_2) - 2 m_N - m_K
-i\,0}\nonumber\\ &&\times \sum_{(\alpha_2,\alpha_1; {^3}{\rm P}_1)}
M(K^-(\vec{0}\,) p(\vec{Q},\sigma_p) n( - \vec{Q},\sigma_n) \to
n(\vec{k}_1,\alpha_1) \Sigma^0(\vec{k}_2,\alpha_2); {^3}{\rm
P}_1)\nonumber\\ &&\times\, M(n(\vec{k}_1,\alpha_1)
\Sigma^0(\vec{k}_2,\alpha_2) \to K^-(\vec{0}\,) p(\vec{K},\sigma_p) n(
- \vec{K},\sigma_n); {^3}{\rm P}_1)\nonumber\\ &&+ \int
\frac{d^3k_1}{(2\pi)^3 2 E_p(\vec{k}_1)}\frac{d^3k_2}{(2\pi)^3 2
E_{\Sigma^-}(\vec{k}_2)}\,\frac{(2\pi)^3\,\delta^{(3)}(\vec{k}_1 +
\vec{k}_2)}{E_p(\vec{k}_1) + E_{\Sigma^-}(\vec{k}_2) - 2 m_N - m_K
-i\,0}\nonumber\\ &&\times \sum_{(\alpha_2,\alpha_1; {^3}{\rm P}_1)}
M(K^-(\vec{0}\,) p(\vec{Q},\sigma_p) n( - \vec{Q},\sigma_n) \to
p(\vec{k}_1,\alpha_1) \Sigma^-(\vec{k}_2,\alpha_2); {^3}{\rm
P}_1)\nonumber\\ &&\times\, M(p(\vec{k}_1,\alpha_1)
\Sigma^-(\vec{k}_2,\alpha_2) \to K^-(\vec{0}\,) p(\vec{K},\sigma_p) n(
- \vec{K},\sigma_n); {^3}{\rm P}_1)\nonumber\\ &&+ \int
\frac{d^3k_1}{(2\pi)^3 2 E_n(\vec{k}_1)}\frac{d^3k_2}{(2\pi)^3 2
E_{\Lambda^0}(\vec{k}_2)}\,\frac{(2\pi)^3\,\delta^{(3)}(\vec{k}_1 +
\vec{k}_2)}{E_n(\vec{k}_1) + E_{\Lambda^0}(\vec{k}_2) - 2 m_N - m_K
-i\,0}\nonumber\\ &&\times \sum_{(\alpha_2,\alpha_1; {^1}{\rm P}_1)}
M(K^-(\vec{0}\,) p(\vec{Q},\sigma_p) n( - \vec{Q},\sigma_n) \to
n(\vec{k}_1,\alpha_1) \Lambda^0(\vec{k}_2,\alpha_2); {^1}{\rm
P}_1)\nonumber\\ &&\times\, M(n(\vec{k}_1,\alpha_1)
\Lambda^0(\vec{k}_2,\alpha_2) \to K^-(\vec{0}\,) p(\vec{K},\sigma_p)
n( - \vec{K},\sigma_n); {^1}{\rm P}_1)\nonumber\\ &&+ \int
\frac{d^3k_1}{(2\pi)^3 2 E_n(\vec{k}_1)}\frac{d^3k_2}{(2\pi)^3 2
E_{\Sigma^0}(\vec{k}_2)}\,\frac{(2\pi)^3\,\delta^{(3)}(\vec{k}_1 +
\vec{k}_2)}{E_n(\vec{k}_1) + E_{\Sigma^0}(\vec{k}_2) - 2 m_N - m_K
-i\,0}\nonumber\\ &&\times \sum_{(\alpha_2,\alpha_1; {^1}{\rm P}_1)}
M(K^-(\vec{0}\,) p(\vec{Q},\sigma_p) n( - \vec{Q},\sigma_n) \to
n(\vec{k}_1,\alpha_1) \Sigma^0(\vec{k}_2,\alpha_2); {^1}{\rm
P}_1)\nonumber\\ &&\times\, M(n(\vec{k}_1,\alpha_1)
\Sigma^0(\vec{k}_2,\alpha_2) \to K^-(\vec{0}\,) p(\vec{K},\sigma_p) n(
- \vec{K},\sigma_n); {^1}{\rm P}_1)\nonumber\\ &&+ \int
\frac{d^3k_1}{(2\pi)^3 2 E_p(\vec{k}_1)}\frac{d^3k_2}{(2\pi)^3 2
E_{\Sigma^-}(\vec{k}_2)}\,\frac{(2\pi)^3\,\delta^{(3)}(\vec{k}_1 +
\vec{k}_2)}{E_p(\vec{k}_1) + E_{\Sigma^-}(\vec{k}_2) - 2 m_N - m_K
-i\,0}\nonumber\\ &&\times \sum_{(\alpha_2,\alpha_1; {^1}{\rm P}_1)}
M(K^-(\vec{0}\,) p(\vec{Q},\sigma_p) n( - \vec{Q},\sigma_n) \to
p(\vec{k}_1,\alpha_1) \Sigma^-(\vec{k}_2,\alpha_2); {^1}{\rm
P}_1)\nonumber\\ &&\times\, M(p(\vec{k}_1,\alpha_1)
\Sigma^-(\vec{k}_2,\alpha_2) \to K^-(\vec{0}\,) p(\vec{K},\sigma_p) n(
- \vec{K},\sigma_n); {^1}{\rm P}_1),
\end{eqnarray}
where we have neglected the contribution of a kinetic energy of a
relative motion of the $np$ pair.

The real and imaginary parts of the S--wave amplitude
$\tilde{f}^{\,K^-d}_0(0)$, caused by the intermediate states
$(NY)_{{^3}{\rm P}_1}$ and $(NY)_{{^1}{\rm P}_1}$, we determine as
\begin{eqnarray}\label{label5.2}
\hspace{-0.5in}&&{\cal R}e\,\tilde{f}^{\,K^-d}_0(0)_{(NY; {^3}{\rm
P}_1)} = \frac{1}{512\pi^4}\,\frac{1}{1 +
m_K/m_d}\,\frac{1}{3}\sum_{(\sigma_p,\sigma_n;{^3}{\rm S}_1)}\sum_{(\alpha_2,\alpha_1; {^3}{\rm P}_1)}
\nonumber\\
\hspace{-0.5in}&&\times \,{\cal P}\int \frac{d^3k}{
E_N(k)E_Y(k)}\frac{1}{E_N(\vec{k}\,) + E_Y(\vec{k}\,) - 2 m_N -
m_K}\nonumber\\
\hspace{-0.5in}&&\times\, \Big|\int\frac{d^3K}{(2\pi)^3}
\frac{\Phi_d(\vec{K})}{E_N(\vec{K}\,)} \, M( K^-(\vec{0}\,)
p(\vec{K},\sigma_p) n( - \vec{K},\sigma_n) \to N(\vec{k},\alpha_1) Y(-
\vec{k},\alpha_2); {^3}{\rm P}_1)\Big|^2,\nonumber\\
\hspace{-0.5in}&&{\cal R}e\,\tilde{f}^{\,K^-d}_0(0)_{(NY; {^1}{\rm
P}_1)} = \frac{1}{512\pi^4}\,\frac{1}{1 +
m_K/m_d}\,\frac{1}{3}\sum_{(\sigma_p,\sigma_n;{^3}{\rm S}_1)}
\sum_{(\alpha_2,\alpha_1; {^1}{\rm P}_1)}\nonumber\\
\hspace{-0.5in}&&\times \,{\cal P}\int \frac{d^3k}{
E_N(k)E_Y(k)}\frac{1}{E_N(\vec{k}\,) +
E_Y(\vec{k}\,) - 2 m_N - m_K}\nonumber\\
\hspace{-0.5in}&&\times\, \Big|\int\frac{d^3K}{(2\pi)^3}
\frac{\Phi_d(\vec{K})}{E_N(\vec{K}\,)} \, M( K^-(\vec{0}\,)
p(\vec{K},\sigma_p) n( - \vec{K},\sigma_n) \to N(\vec{k},\alpha_1) Y(-
\vec{k},\alpha_2); {^1}{\rm P}_1)\Big|^2,
\end{eqnarray}
where ${\cal P}$ means the principle value of the integral, and
\begin{eqnarray}\label{label5.3}
\hspace{-0.5in}&&{\cal I}m\,\tilde{f}^{\,K^-d}_0(0)_{(NY; {^3}{\rm
P}_1)} = \frac{1}{512\pi^3}\,\frac{1}{1 +
m_K/m_d}\,\frac{1}{3}\sum_{(\sigma_p,\sigma_n;{^3}{\rm S}_1)}
\sum_{(\alpha_2,\alpha_1; {^3}{\rm P}_1)}\nonumber\\
\hspace{-0.5in}&&\times \int \frac{d^3k}{
E_N(k)E_Y(k)}\,\delta(E_N(\vec{k}\,) + E_Y(\vec{k}\,) - 2 m_N -
m_K)\nonumber\\
\hspace{-0.5in}&&\times\, \Big|\int\frac{d^3K}{(2\pi)^3}
\frac{\Phi_d(\vec{K})}{E_N(\vec{K}\,)} \,M( K^-(\vec{0}\,)
p(\vec{K},\sigma_p) n( - \vec{K},\sigma_n) \to N(\vec{k},\alpha_1) Y(-
\vec{k},\alpha_2); {^3}{\rm P}_1)\Big|^2,\nonumber\\
\hspace{-0.5in}&&{\cal I}m\,\tilde{f}^{\,K^-d}_0(0)_{(NY; {^1}{\rm
P}_1)} = \frac{1}{512\pi^3}\,\frac{1}{1 +
m_K/m_d}\,\frac{1}{3}\sum_{(\sigma_p,\sigma_n;{^3}{\rm S}_1)}
\sum_{(\alpha_2,\alpha_1; {^1}{\rm P}_1)}\nonumber\\
\hspace{-0.5in}&&\times \int \frac{d^3k}{
E_N(k)E_Y(k)}\,\delta(E_N(\vec{k}\,) + E_Y(\vec{k}\,) - 2 m_N -
m_K)\nonumber\\
\hspace{-0.5in}&&\times\, \Big|\int\frac{d^3K}{(2\pi)^3}
\frac{\Phi_d(\vec{K})}{E_N(\vec{K}\,)} \,M( K^-(\vec{0}\,)
p(\vec{K},\sigma_p) n( - \vec{K},\sigma_n) \to N(\vec{k},\alpha_1) Y(-
\vec{k},\alpha_2); {^1}{\rm P}_1)\Big|^2.
\end{eqnarray}
The amplitudes $ M(K^-(\vec{0}\,) p(\vec{K},\sigma_p) n( -
\vec{K},\sigma_n) \to N(\vec{k},\alpha_1) Y(- \vec{k},\alpha_2)$ we
suggest to compute within the approach developed in \cite{AI01}.

\section{Amplitude of reaction $K^- (p n)_{{^3}{\rm S}_1}
 \to n \Lambda^0 \to K^- (p n)_{{^3}{\rm S}_1}$ and the energy level
 displacement} \setcounter{equation}{0}

The amplitudes of the reactions $K^- (pn)_{{^3}{\rm S}_1} \to n
\Lambda^0 $, where $n \Lambda^0$ pair is coupled in the ${^3}{\rm
P}_1$ and ${^1}{\rm P}_1$ states, we define as
\begin{eqnarray}\label{label6.1}
\hspace{-0.3in}&&M(K^-(\vec{0}\,) p(\vec{K},\sigma_p) n( -
\vec{K},\sigma_n) \to n(\vec{k},\alpha_1) \Lambda^0(-
\vec{k},\alpha_2); {^3}{\rm P}_1) = -\, i\,C^{\,(n\Lambda^0; {^3}{\rm
P}_1)}_{K^-(pn; {^3}{\rm S}_1)}\,\nonumber\\
\hspace{-0.3in}&&\times\, \frac{[\bar{u^c}(- \vec{K},\sigma_n)
\vec{\gamma} u( \vec{K},\sigma_p)]\cdot [\bar{u}(- \vec{k},\alpha_2)
\vec{\gamma} \gamma^5u^c ( \vec{k},\alpha_1)] }{\displaystyle 1 -
\frac{1}{2}\,r^t_{np} a^t_{np} K^2 + i\,a^t_{np} K}\,f^{(n\Lambda^0;
{^3}{\rm P}_1)}_{K^-(pn; {^3}{\rm S}_1)}(k_0),\nonumber\\
\hspace{-0.3in}&&M(K^-(\vec{0}\,) p(\vec{K},\sigma_p) n( -
\vec{K},\sigma_n) \to n(\vec{k},\alpha_1) \Lambda^0(-
\vec{k},\alpha_2); {^1}{\rm P}_1) = -\,i\,C^{\,(n\Lambda^0; {^1}{\rm
P}_1)}_{K^-(pn; {^3}{\rm S}_1)}\,\nonumber\\
\hspace{-0.3in}&&\times\, \frac{[\bar{u^c}(- \vec{K},\sigma_n)
\vec{\gamma} u( \vec{K},\sigma_p)]\cdot [\bar{u}(- \vec{k},\alpha_2)
\gamma^0 \vec{\gamma} \gamma^5u^c ( \vec{k},\alpha_1)] }{\displaystyle
1 - \frac{1}{2}\,r^t_{np} a^t_{np} K^2 + i\,a^t_{np}
K}\,f^{(n\Lambda^0; {^1}{\rm P}_1)}_{K^-(pn; {^3}{\rm S}_1)}(k_0),
\end{eqnarray}
where $a^t_{np} = (5.424 \pm 0.004)\,{\rm fm} = (3.837 \pm
0.003)\,{\rm m^{-1}_{\pi}}$ and $r^t_{np} = (1.759 \pm 0.005)\,{\rm
fm} = (1.244 \pm 0.004)\,{\rm m^{-1}_{\pi}}$ are the spin--triplet
S--wave scattering length and effective range of $np$ scattering in
the ${^3}{\rm S}_1$ state \cite{MN79}; $f^{(n\Lambda^0; X)}_{K^-(pn;
{^3}{\rm S}_1)}(k_0)$ is the amplitude of the final--state
$n\Lambda^0$ interaction near threshold of the reaction $K^-
(pn)_{{^3}{\rm S}_1} \to (n \Lambda^0)_X$ and $C^{\,(n\Lambda^0;
X)}_{K^-(pn; {^3}{\rm S}_1)}$ is the effective coupling constant of
the transition $ K^- (pn)_{{^3}{\rm S}_1} \to (n \Lambda^0)_X$, where
$X = {^3}{\rm P}_1$ or ${^1}{\rm P}_1$.

The spinorial wave functions of the $(np)_{{^3}{\rm S}_1}$,
$(NY)_{{^3}{\rm P}_1}$ and $(NY)_{{^1}{\rm P}_1}$ states are analysed
in Appendix C.

\subsection{Effective coupling constant
$C^{\,n\Lambda^0}_{K^-pn}$}

In the one--meson exchange approximation \cite{AI01} the effective
coupling constant of the transition $n \Lambda^0 \to K^- p n$ is
defined by the Feynman diagrams are depicted in Fig.4.
\begin{figure}
\centering \psfrag{K-}{$K^-$} \psfrag{K0b}{$\bar{K}^0$}
\psfrag{L0}{$\Lambda^0$} \psfrag{p+}{$\pi^+$} \psfrag{p0}{$\pi^0$}
\psfrag{S-}{$\Sigma^-$} \psfrag{S0}{$\Sigma^0$}
\psfrag{Sp}{$\Sigma^+$}\psfrag{p}{$p$} \psfrag{n}{$n$}
\psfrag{e}{$\eta$}
\includegraphics[height=0.90\textheight]{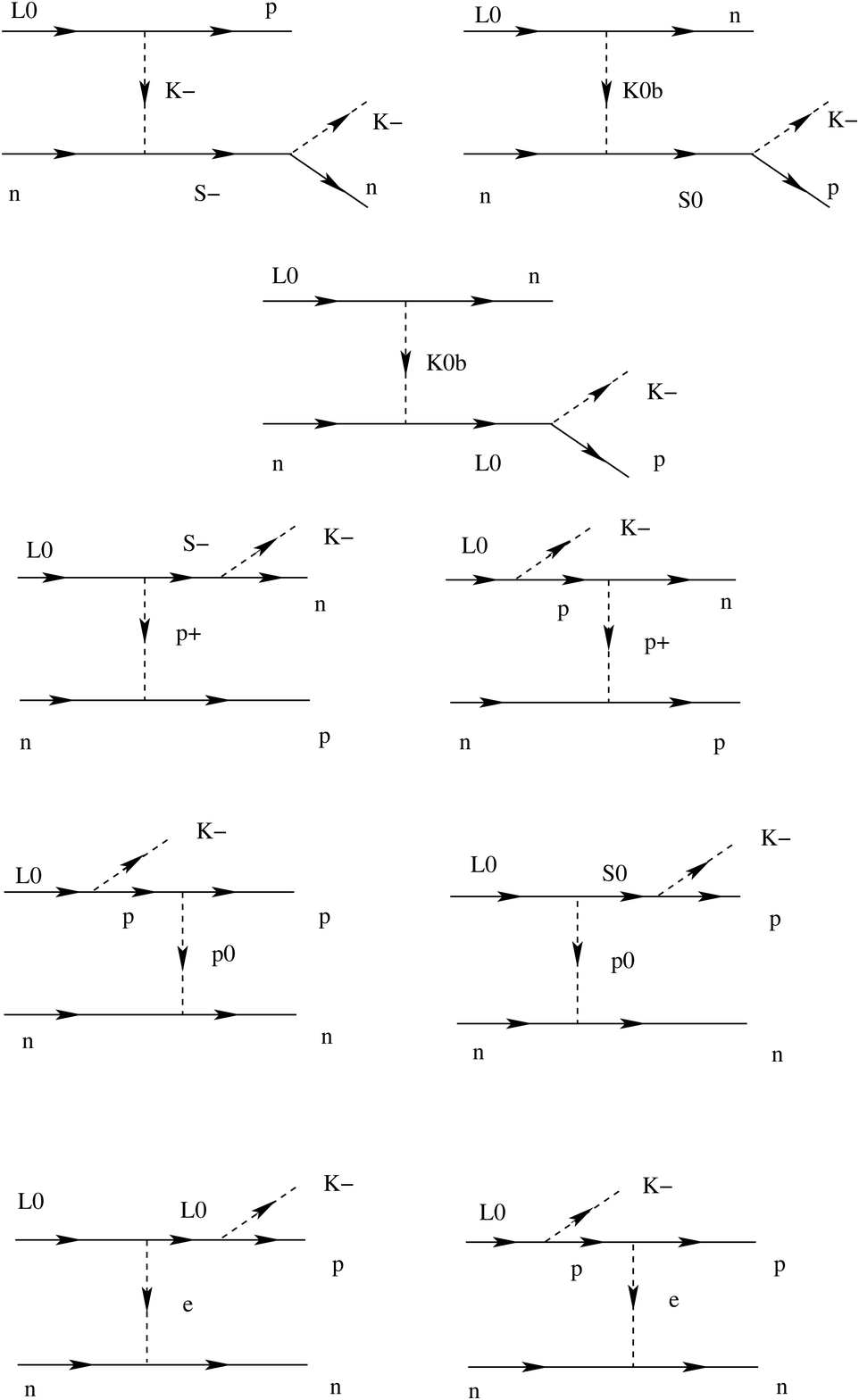}
\caption{Feynman diagrams describing the effective coupling constant
of the transition $n \Lambda^0 \to K^- p n$ in the one--pseudoscalar
meson exchange approximation}
\end{figure}
In the momentum representation the set of Feynman diagrams in Fig.4 is
given by\,\footnote{The coupling constants of the $PBB$, $SBB$ and
$SPP$ interactions, where $P$ and $S$ are nonets of pseudo--scalar and
scalar mesons, the partners under chiral $U(3)\times U(3)$
transformations, are computed in the Appendix D.}
\begin{eqnarray}\label{label6.2}
&&{\cal M}(n(\vec{k},\alpha_1) \Lambda^0(- \vec{k},\alpha_2) \to
K^-(\vec{0}\,) p(\vec{K},\sigma_p) n( - \vec{K},\sigma_n)) =
\nonumber\\ &&= \frac{2}{\sqrt{3}}\,(2\alpha - 1)^2\,(3 -
2\alpha)\,g^3_{\pi NN}\,\times\, [\bar{u}(k_p,\sigma_p)i\gamma^5
u(q_{\Lambda^0},\alpha_{\Lambda^0})]\,\times\,\frac{1}{m^2_K -
(q_{\Lambda^0} - k_p)^2}\nonumber\\
&&\times\,\Big[\bar{u}(k_n,\sigma_n)i\gamma^5 \frac{1}{m_{\Sigma} -
\hat{k}_n - \hat{Q}_K}i\gamma^5 u(q_n,\alpha_n)\Big]\nonumber\\ &&+
\frac{1}{\sqrt{3}}\,(2\alpha - 1)^2\,(3 - 2\alpha)\,g^3_{\pi
NN}\,\times\, [\bar{u}(k_n,\sigma_n)i\gamma^5
u(q_{\Lambda^0},\alpha_{\Lambda^0})]\, \times\,\frac{1}{m^2_K -
(q_{\Lambda^0} - k_n)^2}\nonumber\\
&&\times\,\Big[\bar{u}(k_p,\sigma_p)i\gamma^5 \frac{1}{m_{\Sigma^0} -
\hat{k}_p - \hat{Q}_K}i\gamma^5 u(q_n,\alpha_n)\Big]\nonumber\\
&&-\,\frac{1}{3\sqrt{3}}\,(3 - 2\alpha)^3\,g^3_{\pi NN}\,\times\,
[\bar{u}(k_n,\sigma_n)i\gamma^5 u(q_{\Lambda^0},\alpha_{\Lambda^0})]
\, \times\,\frac{1}{m^2_K - (q_{\Lambda^0} - k_n)^2}\nonumber\\
&&\times\,\Big[\bar{u}(k_p,\sigma_p)i\gamma^5 \frac{1}{m_{\Lambda^0} -
\hat{k}_p - \hat{Q}_K}i\gamma^5 u(q_n,\alpha_n)\Big]\nonumber\\&& +
\frac{4}{\sqrt{3}}\,\alpha\,(2\alpha - 1)\,g^3_{\pi
NN}\,\times\,\Big[\bar{u}(k_n,\sigma_n)i\gamma^5 \frac{1}{m_{\Sigma} -
\hat{k}_n - \hat{Q}_K}i\gamma^5
u(q_{\Lambda^0},\alpha_{\Lambda^0})\Big]\nonumber\\ &&
\times\,\frac{1}{m^2_{\pi} - (k_p -
q_n)^2}\,\times\,[\bar{u}(k_p,\sigma_p)i \gamma^5
u(q_n,\alpha_n)]\nonumber\\ &&- \frac{2}{\sqrt{3}}\,(3 -
2\alpha)\,g^3_{\pi NN}\,\times\,\Big[ \bar{u}(k_n,\sigma_n)i\gamma^5
\frac{1}{m_p - \hat{q}_{\Lambda^0} + \hat{Q}_K}i\gamma^5
u(q_{\Lambda^0},\alpha_{\Lambda^0})\Big]\nonumber\\
&&\times\,\frac{1}{m^2_{\pi} - (k_p -
q_n)^2}\,\times\,[\bar{u}(k_p,\sigma_p)i\gamma^5
u(q_n,\alpha_n)]\nonumber\\ &&+\,\frac{1}{\sqrt{3}}\,(3 - 2\alpha)\,
g^3_{\pi NN}\,\times\,\Big[ \bar{u}(k_p,\sigma_p)i\gamma^5
\frac{1}{m_p - \hat{q}_{\Lambda^0} + \hat{Q}_K}i\gamma^5
u(q_{\Lambda^0},\alpha_{\Lambda^0})\Big]\nonumber\\
&&\times\,\frac{1}{m^2_{\pi} - (k_n -
q_n)^2}\,\times\,[\bar{u}(k_n,\sigma_n)i\gamma^5
u(q_n,\alpha_n)]\nonumber\\ && -
\,\frac{2}{\sqrt{3}}\,\alpha\,(2\alpha - 1)\, g^3_{\pi
NN}\,\times\,\Big[ \bar{u}(k_p,\sigma_p) i \gamma^5
\frac{1}{m_{\Sigma} - \hat{k}_p - \hat{Q}_K} i \gamma^5
u(q_{\Lambda^0},\alpha_{\Lambda^0})\Big]\nonumber\\
&&\times\,\frac{1}{m^2_{\pi} - (k_n -
q_n)^2}\,\times\,[\bar{u}(k_n,\sigma_n) i \gamma^5
u(q_n,\alpha_n)]\nonumber\\ && +\,\frac{2}{3\sqrt{3}}\,\alpha\,(3 -
2\alpha)\,(3 - 4\alpha)\,g^3_{\pi
NN}\,\times\,\Big[\bar{u}(k_p,\sigma_p) i\gamma^5
\frac{1}{m_{\Lambda^0} - \hat{k}_p - \hat{Q}_K} i \gamma^5
u(q_{\Lambda^0},\alpha_{\Lambda^0})\Big]\nonumber\\
&&\times\,\frac{1}{m^2_{\eta} - (k_n -
q_n)^2}\,\times\,[\bar{u}(k_n,\sigma_n) i \gamma^5 u(q_n,
\alpha_n)]\nonumber\\ &&- \,\frac{1}{3\sqrt{3}}\,(3 - 2\alpha)\,(3 -
4\alpha)^2\,g^3_{\pi NN}\,\times\,\Big[ \bar{u}(k_p,\sigma_p)i\gamma^5
\frac{1}{m_p - \hat{q}_{\Lambda^0} + \hat{Q}_K} i\gamma^5
u(q_{\Lambda^0},\alpha_{\Lambda^0})\Big]\nonumber\\
&&\times\,\frac{1}{m^2_{\eta} - (k_n -
q_n)^2}\,\times\,[\bar{u}(k_n,\sigma_n)i\gamma^5 u(q_n,\alpha_n)].
\end{eqnarray}
Following \cite{AI01}, the amplitude of the transition $n\Lambda^0 \to
K^-pn$ (\ref{label6.2}), computed near threshold, we define by the
local effective Lagrangian
\begin{eqnarray}\label{label6.3}
\hspace{-0.3in}&&{\cal L}^{n\Lambda^0 \to K^-pn}_{\rm eff}(x)_P
=\nonumber\\ \hspace{-0.3in}&&= -\,\frac{2}{\sqrt{3}}\,\frac{(3 -
2\alpha)(2\alpha - 1)^2 g^3_{\pi NN}}{m^2_K - (E_{\Lambda^0} - m_N)^2
+ k^2_0}\,\frac{1}{m_{\Sigma} + m_N + m_K}\,[\bar{p}(x)i\gamma^5
\Lambda^0(x)][\bar{n}(x)n(x)]\nonumber\\ \hspace{-0.3in}&&
-\,\frac{1}{\sqrt{3}}\,\frac{(3 - 2\alpha) (2\alpha - 1)^2 g^3_{\pi
NN}}{m^2_K - (E_{\Lambda^0} - m_N)^2 + k^2_0}\,\frac{1}{m_{\Sigma} +
m_N + m_K}\,[\bar{n}(x)i\gamma^5
\Lambda^0(x)][\bar{p}(x)n(x)]\nonumber\\ \hspace{-0.3in}&&
+\,\frac{1}{3\sqrt{3}}\,\frac{(3 - 2\alpha)^3 g^3_{\pi NN}}{m^2_K -
(E_{\Lambda^0} - m_N)^2 + k^2_0}\,\frac{1}{m_{\Lambda^0} + m_N +
m_K}\,[\bar{n}(x)i\gamma^5 \Lambda^0(x)][\bar{p}(x)n(x)]\nonumber\\
\hspace{-0.3in}&& -\,\frac{4}{\sqrt{3}}\,\frac{\alpha (2\alpha - 1)
g^3_{\pi NN}}{m^2_{\pi} - (E_N - m_N)^2 + k^2_0}\,\frac{1}{m_{\Sigma}
+ m_N + m_K}\,[\bar{n}(x) \Lambda^0(x)][\bar{p}(x)i\gamma^5 n(x)]\nonumber\\
\hspace{-0.3in}&& +\,\frac{2}{\sqrt{3}}\,\frac{(3 - 2\alpha) g^3_{\pi
NN}}{m^2_{\pi} - (E_N - m_N)^2 + k^2_0}\,\frac{m_N + m_K -
m_{\Lambda^0}}{m^2_N - (E_{\Lambda^0} - m_K)^2 + k^2_0}\,[\bar{n}(x)
\Lambda^0(x)][\bar{p}(x)i\gamma^5 n(x)]\nonumber\\
\hspace{-0.3in}&& -\,\frac{1}{\sqrt{3}}\,\frac{(3 - 2\alpha) g^3_{\pi
NN}}{m^2_{\pi} - (E_N - m_N)^2 + k^2_0}\,\frac{m_N + m_K -
m_{\Lambda^0}}{m^2_N - (E_{\Lambda^0} - m_K)^2 + k^2_0}\,[\bar{p}(x)
\Lambda^0(x)][\bar{n}(x)i\gamma^5 n(x)]\nonumber\\
\hspace{-0.3in}&& +\,\frac{2}{\sqrt{3}}\,\frac{\alpha (2\alpha - 1)
g^3_{\pi NN}}{m^2_{\pi} - (E_N - m_N)^2 + k^2_0}\,\frac{1}{m_{\Sigma}
+ m_N + m_K}\,[\bar{p}(x) \Lambda^0(x)][\bar{n}(x)i\gamma^5
n(x)]\nonumber\\
\hspace{-0.3in}&& -\,\frac{2}{3\sqrt{3}}\,\frac{\alpha (3 - 4 \alpha)
(3 - 2\alpha) g^3_{\pi NN}}{m^2_{\eta} - (E_N - m_N)^2 +
k^2_0}\,\frac{1}{m_{\Lambda^0} + m_N + m_K}\,[\bar{p}(x)
\Lambda^0(x)][\bar{n}(x)i\gamma^5 n(x)]\nonumber\\
\hspace{-0.3in}&& +\,\frac{1}{3\sqrt{3}}\,\frac{(3
- 2\alpha)(3 - 4\alpha)^2 g^3_{\pi NN}}{m^2_{\eta} - (E_N - m_N)^2 +
k^2_0}\,\frac{m_N + m_K - m_{\Lambda^0}}{m^2_N - (E_{\Lambda^0} - m_K)^2 + k^2_0}\,[\bar{p}(x)
\Lambda^0(x)][\bar{n}(x)i\gamma^5 n(x)],\nonumber\\
\hspace{-0.3in}&&
\end{eqnarray}
where\,\footnote{The index $P$ means that there are no scalar--meson
exchange contributions.} $E_{\Lambda^0} = ((m_K + 2 m_N)^2 +
m^2_{\Lambda^0} - m^2_N)/2(m_K + 2m_N) = 1262\,{\rm MeV}$, $E_N =
((m_K + 2 m_N)^2 + m^2_N - m^2_{\Lambda^0})/2(m_K + 2m_N) = 1108\,{\rm
MeV}$ and $k_0 = ((2m_N + m_K)^2 - (m_{\Lambda^0} + m_N)^2)^{1/2}
((2m_N + m_K)^2 - (m_{\Lambda^0} - m_N)^2)^{1/2}/2(2m_N + m_K) =
592\,{\rm MeV}$ are the energies and the relative momentum of the
$\Lambda^0$--hyperon and the neutron at threshold. Then, for numerical
calculations we set $g_{\pi NN} = 13.21$ \cite{PSI2} and $\alpha =
0.635$ \cite{MS89} (see also \cite{AI01}).

Now we have to take into account that the $np$ pair couples in the
${^3}{\rm S}_1$ state with isospin zero. This can be carried out by
means of Fierz transformation:
\begin{eqnarray}\label{label6.4}
\hspace{-0.3in}&&[\bar{p}(x)\gamma^5 \Lambda^0(x)] [\bar{n}(x)n(x)]
\to -\,\frac{1}{4}\,[\bar{p}(x)\gamma_{\mu} n^c(x)][\bar{n^c}(x)
\gamma^{\mu}\gamma^5 \Lambda^0(x)]\nonumber\\
\hspace{-0.3in}&&\hspace{2in}
+\,\frac{1}{8}\,[\bar{p}(x)\sigma_{\mu\nu} n^c(x)][\bar{n^c}(x)
\sigma^{\mu\nu}\gamma^5 \Lambda^0(x)],\nonumber\\
\hspace{-0.3in}&&[\bar{n}(x)\gamma^5 \Lambda^0(x)] [\bar{p}(x)n(x)]
\to +\,\frac{1}{4}\,[\bar{p}(x)\gamma_{\mu} n^c(x)][\bar{n^c}(x)
\gamma^{\mu}\gamma^5 \Lambda^0(x)], \nonumber\\
\hspace{-0.3in}&&\hspace{2in}-\,\frac{1}{8}\,[\bar{p}(x)
\sigma_{\mu\nu} n^c(x)] [\bar{n^c}(x) \sigma^{\mu\nu}\gamma^5
\Lambda^0(x)] \nonumber\\
\hspace{-0.3in}&&[\bar{p}(x)\Lambda^0(x)][\bar{n}(x)\gamma^5 n(x)] \to
+\,\frac{1}{4}\,[\bar{p}(x)\gamma_{\mu} n^c(x)][\bar{n^c}(x)
\gamma^{\mu}\gamma^5 \Lambda^0(x)] \nonumber\\
\hspace{-0.3in}&&\hspace{2in}+\,\frac{1}{8}\,
[\bar{p}(x)\sigma_{\mu\nu} n^c(x)] [\bar{n^c}(x)
\sigma^{\mu\nu}\gamma^5 \Lambda^0(x)],\nonumber\\
\hspace{-0.3in}&&[\bar{n}(x)\Lambda^0(x)] [\bar{p}(x)\gamma^5 n(x)]
\to -\,\frac{1}{4}\,[\bar{p}(x)\gamma_{\mu} n^c(x)][\bar{n^c}(x)
\gamma^{\mu}\gamma^5 \Lambda^0(x)]\nonumber\\
\hspace{-0.3in}&&\hspace{2in} - \,\frac{1}{8}\,[\bar{p}(x)
\sigma_{\mu\nu} n^c(x)][\bar{n^c}(x) \sigma^{\mu\nu}\gamma^5
\Lambda^0(x)].
\end{eqnarray}
In the non--relativistic limit the r.h.s. of these products can be
reduced to the form
\begin{eqnarray}\label{label6.5}
\hspace{-0.3in}&&[\bar{p}(x)\gamma^5 \Lambda^0(x)] [\bar{n}(x)n(x)]
\to +\,\frac{1}{4}\,[\bar{p}(x)\vec{\gamma}\, n^c(x)] \cdot
[\bar{n^c}(x) \vec{\gamma}\,\gamma^5 \Lambda^0(x)]\nonumber\\
\hspace{-0.3in}&&\hspace{2in}
-\,\frac{1}{4}\,[\bar{p}(x)\vec{\gamma}\,n^c(x)] \cdot [\bar{n^c}(x)
\gamma^0\vec{\gamma}\,\gamma^5 \Lambda^0(x)],\nonumber\\
\hspace{-0.3in}&&[\bar{n}(x)\gamma^5 \Lambda^0(x)] [\bar{p}(x)n(x)]
\to -\,\frac{1}{4}\,[\bar{p}(x)\vec{\gamma}\, n^c(x)]\cdot
[\bar{n^c}(x) \vec{\gamma}\,\gamma^5 \Lambda^0(x)], \nonumber\\
\hspace{-0.3in}&&\hspace{2in}+\,\frac{1}{4}\,
[\bar{p}(x)\vec{\gamma}\,n^c(x)] \cdot [\bar{n^c}(x)
\gamma^0\vec{\gamma}\,\gamma^5 \Lambda^0(x)] \nonumber\\
\hspace{-0.3in}&&[\bar{p}(x)\Lambda^0(x)][\bar{n}(x)\gamma^5 n(x)] \to
-\,\frac{1}{4}\, [\bar{p}(x)\vec{\gamma}\, n^c(x)]\cdot [\bar{n^c}(x)
\vec{\gamma}\,\gamma^5 \Lambda^0(x)]\nonumber\\
\hspace{-0.3in}&&\hspace{2in}-\,\frac{1}{4}\,[\bar{p}(x)
\vec{\gamma}\,n^c(x)] \cdot [\bar{n^c}(x)
\gamma^0\vec{\gamma}\,\gamma^5 \Lambda^0(x)],\nonumber\\
\hspace{-0.3in}&&[\bar{n}(x)\Lambda^0(x)] [\bar{p}(x)\gamma^5 n(x)]
\to +\,\frac{1}{4}\, [\bar{p}(x)\vec{\gamma}\, n^c(x)]\cdot
[\bar{n^c}(x) \vec{\gamma}\,\gamma^5 \Lambda^0(x)]\nonumber\\
\hspace{-0.3in}&&\hspace{2in} + \,\frac{1}{4}\,[\bar{p}(x)
\vec{\gamma}\,n^c(x)] \cdot [\bar{n^c}(x)
\gamma^0\vec{\gamma}\,\gamma^5 \Lambda^0(x)].
\end{eqnarray}
For the derivation of this Lagrangian we have taken into account that
in the non--relativistic limit the four--baryon product reduces as
follows \[[\bar{p}(x)\gamma^0\vec{\gamma}\,n^c(x)] \cdot [\bar{n^c}(x)
\gamma^0\vec{\gamma}\gamma^5 \Lambda^0(x)] \to
[\bar{p}(x)\vec{\gamma}\,n^c(x)] \cdot [\bar{n^c}(x)
\gamma^0\vec{\gamma}\gamma^5 \Lambda^0(x)].\] The effective
low--energy four--baryon interactions $[\bar{p}(x)\vec{\gamma}\,
n^c(x)]\cdot [\bar{n^c}(x) \vec{\gamma}\,\gamma^5 \Lambda^0(x)]$ and
$[\bar{p}(x) \vec{\gamma}\,n^c(x)] \cdot [\bar{n^c}(x)
\gamma^0\vec{\gamma}\,\gamma^5 \Lambda^0(x)]$ describe the transitions
$(n\Lambda^0)_{{^3}{\rm P}_1} \to K^- (pn)_{{^3}{\rm S}_1}$ and
$(n\Lambda^0)_{{^1}{\rm P}_1} \to K^- (pn)_{{^3}{\rm S}_1}$, where the
$n\Lambda^0$ pair couples in the ${^3}{\rm P}_1$ and ${^1}{\rm P}_1$
state, respectively.

\subsection{Reaction $(n\Lambda^0)_{{^3}{\rm P}_1} \to K^-
(pn)_{{^3}{\rm S}_1}$}

The amplitude of the reaction $(n\Lambda^0)_{{^3}{\rm P}_1}\to K^-
(pn)_{{^3}{\rm S}_1}$, where $n \Lambda^0$ pair couples to the $np$
pair, which is in the ${^3}{\rm S}_1$ state, is defined by
\begin{eqnarray}\label{label6.6}
\hspace{-0.3in}&&M(n(\vec{k},\alpha_1) \Lambda^0(- \vec{k},\alpha_2)
\to K^-(\vec{0}\,) p(\vec{K},\sigma_p) n( - \vec{K},\sigma_n);
{^3}{\rm P}_1) = i\,C^{\,(n\Lambda^0; {^3}{\rm
P}_1)}_{K^-(pn; {^3}{\rm S}_1)}\,\nonumber\\
\hspace{-0.3in}&&\times\, \frac{[\bar{u}(\vec{K},\sigma_p)
\vec{\gamma} u^c( - \vec{K},\sigma_n)]\cdot
[\bar{u^c}(\vec{k},\alpha_1) \vec{\gamma} \gamma^5u(-
\vec{k},\alpha_2)] }{\displaystyle 1 - \frac{1}{2}\,r^t_{np} a^t_{np}
K^2 - i\,a^t_{np} K}\,f^{(n\Lambda^0; {^3}{\rm P}_1)}_{K^-(pn; {^3}{\rm S}_1)}(k_0),
\end{eqnarray}
where $f^{(n\Lambda^0; {^3}{\rm P}_1)}_{K^-(pn; {^3}{\rm S}_1)}(k_0)$
is the amplitude, describing the $n\Lambda^0$ rescattering in the
${^3}{\rm P}_1$ state near threshold of the $K^-(p n)_{{^3}{\rm S}_1}$
system production and $C^{\,(n\Lambda^0; {^3}{\rm P}_1)}_{K^-(pn;
{^3}{\rm S}_1)}$ is the effective coupling constant of the
transition $(n\Lambda^0)_{{^3}{\rm P}_1} \to K^- (pn)_{{^3}{\rm
S}_1}$.

Using (\ref{label6.5}) and (\ref{label6.3}) we obtain the effective
Lagrangian of the transition $(n\Lambda^0)_{{^3}{\rm P}_1} \to K^-
(pn)_{{^3}{\rm S}_1}$ near threshold:
\begin{eqnarray}\label{label6.7}
{\cal L}^{(n\Lambda^0)_{{^3}{\rm P}_1}\to K^-(pn_; {^3}{\rm
S}_1)}_{\rm eff}(x)_P = i\,C^{\,(n\Lambda^0; {^3}{\rm P}_1)}_{K^-(pn;
{^3}{\rm S}_1)}\, K^{-\dagger}(x)\, [\bar{p}(x)\vec{\gamma}\, n^c(x)]
\cdot [\bar{n^c}(x) \vec{\gamma}\, \gamma^5 \Lambda^0(x)],
\end{eqnarray}
where we have denoted
\begin{eqnarray}\label{label6.8}
\hspace{-0.3in}C^{\,(n\Lambda^0; {^3}{\rm P}_1)}_{K^-(pn; {^3}{\rm
S}_1)} &=&-\,\frac{1}{2\sqrt{3}}\, \frac{(3
- 2\alpha)(2\alpha - 1)^2 g^3_{\pi NN}}{m^2_K - (E_{\Lambda^0} -
m_N)^2 + k^2_0}\,\frac{1}{m_{\Sigma} + m_N + m_K}\nonumber\\
\hspace{-0.3in}&& +\,\frac{1}{4\sqrt{3}}\,\frac{(3 - 2\alpha)(2\alpha
- 1)^2 g^3_{\pi NN}}{m^2_K - (E_{\Lambda^0} - m_N)^2 +
k^2_0}\,\frac{1}{m_{\Sigma} + m_N + m_K}\nonumber\\
\hspace{-0.3in}&& -\,\frac{1}{12\sqrt{3}}\,\frac{(3 - 2\alpha)^3
g^3_{\pi NN}}{m^2_K - (E_{\Lambda^0} - m_N)^2 +
k^2_0}\,\frac{1}{m_{\Lambda^0} + m_N + m_K}\nonumber\\
\hspace{-0.3in}&& -\,\frac{1}{\sqrt{3}}\,\frac{\alpha (2\alpha - 1)
g^3_{\pi NN}}{m^2_{\pi} - (E_N - m_N)^2 + k^2_0}\,\frac{1}{m_{\Sigma}
+ m_N + m_K}\nonumber\\
\hspace{-0.3in}&& +\,\frac{1}{2\sqrt{3}}\,\frac{(3 - 2\alpha) g^3_{\pi
NN}}{m^2_{\pi} - (E_N - m_N)^2 + k^2_0}\,\frac{m_N + m_K -
m_{\Lambda^0}}{m^2_N - (E_{\Lambda^0} - m_K)^2 + k^2_0}\nonumber\\
\hspace{-0.3in}&& +\,\frac{1}{4\sqrt{3}}\,\frac{(3 - 2\alpha) g^3_{\pi
NN}}{m^2_{\pi} - (E_N - m_N)^2 + k^2_0}\,\frac{m_N + m_K -
m_{\Lambda^0}}{m^2_N - (E_{\Lambda^0} - m_K)^2 + k^2_0}\nonumber\\
\hspace{-0.3in}&& -\,\frac{1}{2\sqrt{3}}\,\frac{\alpha (2\alpha - 1)
g^3_{\pi NN}}{m^2_{\pi} - (E_N - m_N)^2 + k^2_0}\,\frac{1}{m_{\Sigma}
+ m_N + m_K}\nonumber\\
\hspace{-0.3in}&& +\,\frac{1}{6\sqrt{3}}\,\frac{\alpha (3 - 4 \alpha)
(3 - 2\alpha) g^3_{\pi NN}}{m^2_{\eta} - (E_N - m_N)^2 +
k^2_0}\,\frac{1}{m_{\Lambda^0} + m_N + m_K}\nonumber\\
\hspace{-0.3in}&& -\,\frac{1}{12\sqrt{3}}\,\frac{(3
- 2\alpha)(3 - 4\alpha)^2 g^3_{\pi NN}}{m^2_{\eta} - (E_N - m_N)^2 +
k^2_0}\,\frac{m_N + m_K - m_{\Lambda^0}}{m^2_N - (E_{\Lambda^0} - m_K)^2 + k^2_0} = \nonumber\\
\hspace{-0.3in}&&= 1.7\times 10^{-6}\,{\rm MeV}^{-3}.
\end{eqnarray}
The Lagrangian (\ref{label6.7}) describes the interaction of the
$n\Lambda^0$ pair in the ${^3}{\rm P}_1$ state with the $np$ pair in
the ${^3}{\rm S}_1$ state through the emission of the $K^-$--meson.

Using the results obtained in \cite{IV3} one can show that the
contribution of the resonances $\Lambda(1405)$ and $\Sigma(1750)$ to
the effective coupling constant of the transition $n\Lambda^0 \to K^-
p n$ is negligible small.

In Fig.5 we have depicted Feynman diagrams describing the
contributions of the scalar mesons 
\begin{figure}
\centering \psfrag{K-}{$K^-$} 
\psfrag{L0}{$\Lambda^0$} 
\psfrag{p+}{$\pi^+$} \psfrag{p0}{$\pi^0$}
\psfrag{p}{$p$} \psfrag{n}{$n$}
\psfrag{s}{$\sigma$}
\psfrag{e}{$\eta$}
\psfrag{k-}{$\kappa^-$}
\psfrag{k0}{$\bar{\kappa}^0$}
\includegraphics[height=0.30\textheight]{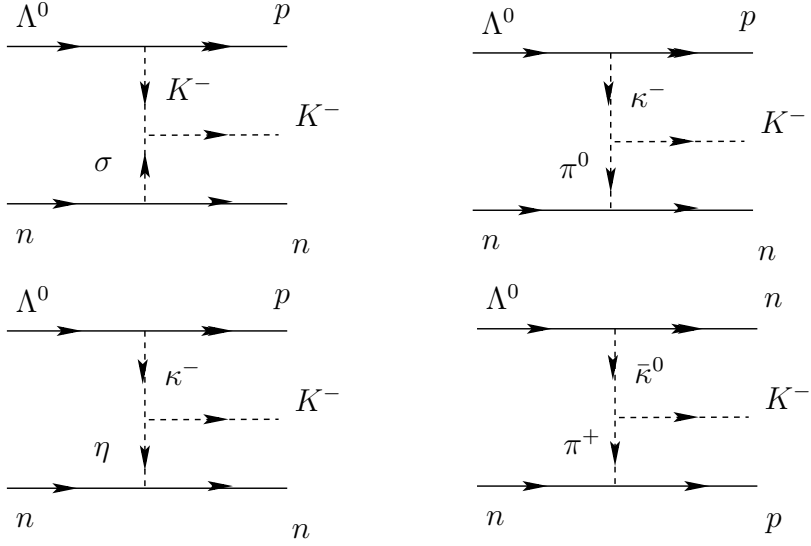}
\caption{Feynman diagrams describing the contribution of scalar mesons
to the effective coupling constant of the transition $n \Lambda^0 \to
K^- p n$ in the one--pseudoscalar meson exchange approximation}
\end{figure}
The effective Lagrangian of the transition $n\Lambda^0 \to K^- p n$,
caused by the scalar--meson exchange, is equal to
\begin{eqnarray}\label{label6.9}
{\cal L}^{\,n\Lambda^0}_{K^-pn}(x)_S &=&
\frac{1}{2\sqrt{3}}\,\frac{1}{g_A F_{\pi}}\,\frac{(3 -
2\alpha)\,g^2_{\pi NN}}{m^2_K - (E_{\Lambda^0} - m_N)^2 +
k^2_0}\,[\bar{p}(x)i\gamma^5 \Lambda^0(x)][\bar{n}(x)n(x)]\nonumber\\
&+& \frac{1}{2\sqrt{3}}\,\frac{1}{g_A F_{\pi}}\,\frac{(3 -
2\alpha)\,g^2_{\pi NN}}{m^2_{\pi} - (E_N - m_N)^2 +
k^2_0}\,[\bar{p}(x)\Lambda^0(x)][\bar{n}(x)i\gamma^5 n(x)]\nonumber\\
&+& \frac{1}{6\sqrt{3}}\,\frac{1}{g_A F_{\pi}}\,\frac{(3 -
2\alpha)\,(3 - 4\alpha)\,g^2_{\pi NN}}{m^2_{\eta} - (E_N -
m_N)^2 + k^2_0}\,[\bar{p}(x)\Lambda^0(x)][\bar{n}(x)i\gamma^5
n(x)]\nonumber\\ &-& \frac{1}{\sqrt{3}}\,\frac{1}{g_A
F_{\pi}}\,\frac{(3 - 2\alpha)\,g^2_{\pi NN}}{m^2_{\pi} -
(E_N - m_N)^2 +
k^2_0}\,[\bar{n}(x)\Lambda^0(x)][\bar{p}(x)i\gamma^5 n(x)].
\end{eqnarray}
The effective coupling constant of the $(n\Lambda^0)_{{^3}{\rm P}_1}
\to K^-(pn)_{{^3}{\rm S}_1}$ transition, induced by the scalar--meson
exchanges, reads
\begin{eqnarray}\label{label6.10}
\hspace{-0.3in}\delta C^{\,(n\Lambda^0; {^3}{\rm P}_1)}_{K^-(pn;
{^3}{\rm S}_1)} &=& \frac{1}{8\sqrt{3}}\,\frac{1}{g_A
F_{\pi}}\,\frac{(3 - 2\alpha)\,g^2_{\pi NN}}{m^2_K - (E_{\Lambda^0} -
m_N)^2 + k^2_0}\nonumber\\ \hspace{-0.3in}&-&
\frac{1}{8\sqrt{3}}\,\frac{F_K}{g_A F^2_{\pi}}\,\frac{(3 -
2\alpha)\,g^2_{\pi NN}}{m^2_{\pi} - (E_N - m_N)^2 +
k^2_0}\nonumber\\ \hspace{-0.3in}&-&
\frac{1}{24\sqrt{3}}\,\frac{F_K}{g_A F^2_{\pi}}\,\frac{(3 -
2\alpha)\,(3 - 4\alpha)\,g^2_{\pi NN}}{m^2_{\eta} - (E_N -
m_N)^2 + k^2_0}\nonumber\\ \hspace{-0.3in}&-&
\frac{1}{4\sqrt{3}}\,\frac{F_K}{g_A F^2_{\pi}}\,\frac{(3 -
2\alpha)\,g^2_{\pi NN}}{m^2_{\pi} - (E_N - m_N)^2 + k^2_0}
=\nonumber\\ &=& -\,1.3\times 10^{-6}\,{\rm MeV}^{-3}.
\end{eqnarray}
Thus, the total effective coupling constant $C^{\,(n\Lambda^0; {^3}{\rm
P}_1)}_{K^-(pn; {^3}{\rm S}_1)}$ of the $(n\Lambda^0)_{{^3}{\rm P}_1}
\to {K^-(pn)_{{^3}{\rm S}_1}}$ transition, is equal to
\begin{eqnarray}\label{label6.11}
C^{\,(n\Lambda^0; {^3}{\rm P}_1)}_{K^-(pn; {^3}{\rm S}_1)} = 4\times
10^{-7}\,{\rm MeV}^{-3}.
\end{eqnarray}
The contribution of the scalar mesons we have computed in the infinite
mass limit. This corresponds to the non--linear realization of chiral
symmetry used within ChPT by Gasser and Leutwyler \cite{JG99}.

Now we proceed to computing the amplitude of the $n\Lambda^0$
rescattering in the ${^3}{\rm P}_1$ state.

\subsection{Amplitude of $n\Lambda^0$
rescattering in the ${^3}{\rm P}_1$ state}

The amplitude $f^{(n\Lambda^0; {^3}{\rm P}_1)}_{K^-(pn)_{{^3}{\rm
S}_1}}(k_0)$, describing the rescattering of the $n\Lambda^0$ pair in
the ${^3}{\rm P}_1$ state near threshold of the $K^-(pn)_{{^3}{\rm
S}_1}$ system production, is defined by the Feynman diagrams depicted
in Fig.6\,\footnote{Within the dispersion relation approach the
final--state interaction of the baryon--baryon pairs (or
baryon--baryon rescattering in the initial state) has been elaborated
by Anisovich {\it et al.} \cite{VA93}.}.
\begin{figure}
\centering \psfrag{K-}{$K^-$} 
\psfrag{L0}{$\Lambda^0$}
 \psfrag{p+}{$\pi^+$}
\psfrag{p0}{$\pi^0$} \psfrag{S1}{$\Sigma^-(1750)$}
\psfrag{S10}{$\Sigma^0(1750)$} \psfrag{Sp}{$\Sigma^+$}
\psfrag{p}{$p$}
\psfrag{n}{$n$} \psfrag{e}{$\eta$}
\includegraphics[height=0.25\textheight]{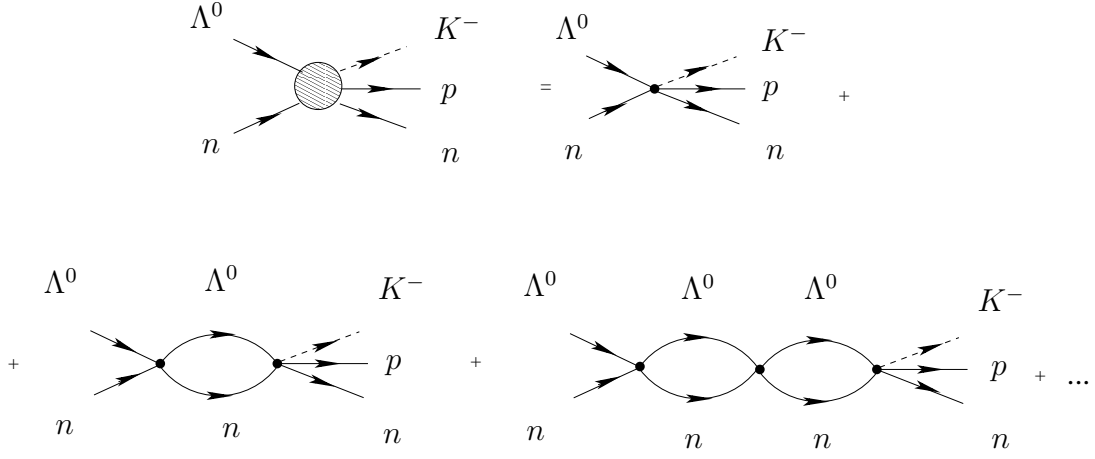}
\caption{Feynman diagrams describing the $(n\Lambda^0)_{{^3}{\rm
P}_1}$ rescattering in the initial state of the reaction
$(n\Lambda^0)_{{^3}{\rm P}_1} \to K^- (np)_{{^3}{\rm S}_1}$.}
\end{figure}
The result of the
calculation of these diagrams reads (see Appendix E and \cite{AI01})
\begin{eqnarray}\label{label6.12}
\hspace{-0.3in}\Big|f^{(n\Lambda^0; {^3}{\rm P}_1)}_{K^-(pn; {^3}{\rm
S}_1)}(k_0)\Big| = \Big|\Big\{1 - \frac{C_{n\Lambda^0}({^3}{\rm
P}_1)}{12\pi^2}\,\frac{k^3_0}{E(k_0)}\,\Big[{\ell n}\Big(\frac{E(k_0)
+ k_0}{E(k_0) - k_0} \Big) - i\,\pi\Big]\Big\}^{-1}\Big| \simeq 1,
\end{eqnarray}
where $E(k_0) = \sqrt{k^2_0 + m^2_B}$\,\footnote{For simplicity we use
the equal masses of baryons for the calculation of the rescattering of
the $n\Lambda^0$ pair, where $m_B = \sqrt{(2 m_N + m_K)^2 - 4 k^2_0}/2
= 1030\,{\rm MeV}$.} and the effective coupling constant
$C_{n\Lambda^0}({^3}{\rm P}_1)$ is equal to \cite{AI01}
\begin{eqnarray}\label{label6.13}
C_{n\Lambda^0}({^3}{\rm P}_1) &=& (3 - 2 \alpha)^2\,\frac{g^2_{\pi
NN}}{12 k^2_0}\,{\ell n}\Big(1 + \frac{4 k^2_0}{~m^2_K}\Big) - \alpha
(3 - 4\alpha)\,\frac{g^2_{\pi NN}}{6 k^2_0}\,{\ell n}\Big(1 + \frac{4
k^2_0}{~m^2_{\eta}}\Big)=\nonumber\\ &=& 2.0\times  10^{-4}\,{\rm
MeV}^{-2}.
\end{eqnarray}
The rescattering of the $n\Lambda^0$ pair in the ${^3}{\rm P}_1$ state
is defined by the interaction, computed in the one--meson exchange
approximation (see \cite{AI01}):
\begin{eqnarray}\label{label6.14}
{\cal L}^{(n\Lambda^0)_{{^3}{\rm P}_1} \to (n\Lambda^0)_{{^3}{\rm
P}_1}}_{\rm eff}(x) = -\,\frac{1}{4}\,C_{n\Lambda^0}({^3}{\rm
P}_1)\,[\bar{\Lambda}^0(x) \vec{\gamma}\,\gamma^5 n^c(x)]\cdot
[\bar{n^c}(x)\vec{\gamma}\,\gamma^5 \Lambda^0(x)].
\end{eqnarray}
For the derivation of the effective Lagrangian (\ref{label6.14}) we
have used the relations
\begin{eqnarray}\label{label6.15}
\hspace{-0.3in}[\bar{\Lambda}^0(x)\gamma^5
\Lambda^0(x)][\bar{n}(x)\gamma^5 n(x)] &=& +
\frac{1}{4}\,[\bar{\Lambda}^0(x)\vec{\gamma}\,\gamma^5 n^c(x)] \cdot
[\bar{n^c}(x)\vec{\gamma}\,\gamma^5 \Lambda^0(x)]\nonumber\\
\hspace{-0.3in}&&+
\frac{1}{4}\,[\bar{\Lambda}^0(x)\gamma^0\vec{\gamma}\,\gamma^5 n^c(x)]
\cdot [\bar{n^c}(x)\gamma^0\vec{\gamma}\,\gamma^5 \Lambda^0(x)] +
\ldots,\nonumber\\
\hspace{-0.3in}[\bar{\Lambda}^0(x)\gamma^5 n(x)][\bar{n}(x)\gamma^5
\Lambda^0(x)] &=& +
\frac{1}{4}\,[\bar{\Lambda}^0(x)\vec{\gamma}\,\gamma^5 n^c(x)]\cdot
[\bar{n^c}(x) \vec{\gamma}\,\gamma^5 \Lambda^0(x)]\nonumber\\
\hspace{-0.3in}&&+
\frac{1}{4}\,[\bar{\Lambda}^0(x)\gamma^0\vec{\gamma}\,\gamma^5 n^c(x)]
\cdot [\bar{n^c}(x)\gamma^0\vec{\gamma}\,\gamma^5 \Lambda^0(x)] +
\ldots
\end{eqnarray}
caused by Fierz transformation.

\subsection{S--wave amplitude $\tilde{f}^{\,K^-d}_0(0)_{(n\Lambda^0;
{^3}{\rm P}_1)}$ of $K^-d$ scattering}

For the calculation of the S--wave amplitude
$\tilde{f}^{\,K^-d}_0(0)_{(n\Lambda^0; {^3}{\rm P}_1)}$ of $K^-d$
scattering near threshold, caused by the exchange of the $n\Lambda^0$
pair in the ${^3}{\rm P}_1$ state, we have to square the spinorial
wave functions of the coupled baryons and to sum over polarisations,
taking into account that $np$ and $n\Lambda^0$ pairs are in the
${^3}{\rm S}_1$ and ${^3}{\rm P}_1$ state, respectively. This yields
(see Appendix C)
\begin{eqnarray}\label{label6.16}
\hspace{-0.3in}&&\frac{1}{3}\sum_{(\sigma_p,\sigma_n; {^3}{\rm S}_1))}
\sum_{(\alpha_2,\alpha_1; {^3}{\rm P}_1)}|[\bar{u}(\vec{K},\sigma_p)
\vec{\gamma}\, u^c( - \vec{K},\sigma_n)]\cdot
[\bar{u^c}(\vec{k},\alpha_1) \vec{\gamma}\gamma^5u(-
\vec{k},\alpha_2)]|^2 =
\frac{128}{3}\,m^2_N\,\vec{k}^{\;2}.\nonumber\\ 
\hspace{-0.3in}&&
\end{eqnarray}
First, let us compute the imaginary part of the amplitude
$\tilde{f}^{\,K^-d}_0(0)_{(n\Lambda^0; {^3}{\rm P}_1)}$. From
(\ref{label5.3}) and taking into account the result of the summation
over polarisations of baryons (\ref{label6.16}) we get
\begin{eqnarray}\label{label6.17}
\hspace{-0.3in}{\cal I}m\,\tilde{f}^{\,K^-d}_0(0)_{(n\Lambda^0;
{^3}{\rm P}_1)} &=& \frac{1}{3\pi^2}\,\frac{1}{1 +
m_K/m_d}\,\frac{k^3_0}{2m_N + m_K}\, [C^{\,(n\Lambda^0; {^3}{\rm
P}_1)}_{K^-(pn; {^3}{\rm S}_1)}]^2\, \Big|f^{(n\Lambda^0; {^3}{\rm
P}_1)}_{K^-(pn; {^3}{\rm S}_1)}(k_0) \Big|^2\nonumber\\
\hspace{-0.3in}&&\times\,\Bigg|\int\frac{d^3K}{(2\pi)^3}
\frac{\Phi_d(\vec{K})}{\displaystyle 1 -
\frac{1}{2}r^t_{np}a^t_{np}K^2 - i\,a^t_{np} K}\Bigg|^2,
\end{eqnarray}
where we have integrated over $\vec{k}$.  The integral over $\vec{K}$
can be calculated analytically and result reads
\begin{eqnarray}\label{label6.18}
&&\int\frac{d^3K}{(2\pi)^3} \frac{\Phi_d(\vec{K})}{\displaystyle 1 -
\frac{1}{2}r^t_{np}a^t_{np}K^2 - i\,a^t_{np} K} =\nonumber\\ &&=
\sqrt{\frac{2\gamma_d}{\pi^3}}\, \Big[\frac{\pi}{2}\,\frac{a b
\gamma_d}{(a + \gamma_d)(b + \gamma_d)} + i\,\frac{ab}{b -
a}\,\Big(\frac{b^2}{b^2 - \gamma^2_d}\,{\ell n}\frac{b}{\gamma_d} -
\frac{a^2}{a^2 - \gamma^2_d}\,{\ell n}\frac{a}{\gamma_d}\Big)\Big] =
\nonumber\\ &&= (0.030 + i\,0.061)\,m^{3/2}_{\pi},
\end{eqnarray}
where we have denoted
\begin{eqnarray}\label{label6.19}
a = \frac{1}{r^t_{np}}\,\Bigg(1 - \sqrt{1 - \frac{2
r^t_{np}}{a^t_{np}}}\Bigg) = 0.327\,m_{\pi}\,,\, b =
\frac{1}{r^t_{np}}\,\Bigg(1 + \sqrt{1 - \frac{2
r^t_{np}}{a^t_{np}}}\Bigg) = 1.280\,m_{\pi}.
\end{eqnarray}
Thus, at threshold the imaginary part of the S--wave amplitude of
$K^-d$ scattering, caused by the two--body inelastic channel $K^-(p
n)_{{^3}{\rm S}_1} \to (n \Lambda^0)_{{^3}{\rm P}_1} \to K^-(p
n)_{{^3}{\rm S}_1} $, is equal to
\begin{eqnarray}\label{label6.20}
{\cal I}m\,\tilde{f}^{\,K^-d}_0(0)_{(n\Lambda^0; {^3}{\rm P}_1)} &=&
4.6\times 10^{-3}\, \frac{1}{3\pi^2}\,\frac{m^3_{\pi}}{1 +
m_K/m_d}\,\frac{k^3_0}{2m_N + m_K}\, \nonumber\\
&&\times\,[C^{\,(n\Lambda^0; {^3}{\rm P}_1)}_{K^-(pn; {^3}{\rm
S}_1)}]^2\, |f^{(n\Lambda^0; {^3}{\rm P}_1)}_{K^-(pn; {^3}{\rm
S}_1)}(k_0)|^2 = 0.9 \times 10^{-3}\,{\rm fm}.
\end{eqnarray}
The real part of $\tilde{f}^{\,K^-d}_0(0)_{(n\Lambda^0; {^3}{\rm
P}_1)}$ is defined by
\begin{eqnarray}\label{label6.21}
\hspace{-0.5in}&&{\cal R}e\,\tilde{f}^{\,K^-d}_0(0)_{(n\Lambda^0;
{^3}{\rm P}_1)} = \frac{1}{256\pi^3}\,\frac{1}{1 +
m_K/m_d}\,\frac{1}{3}\sum_{(\sigma_p,\sigma_n;{^3}{\rm S}_1)}
\sum_{(\alpha_1,\alpha_2; {^3}{\rm P}_1)}\nonumber\\
\hspace{-0.5in}&&\times \,\frac{1}{2\pi}{\cal P}\int \frac{d^3k}{
E_N(k)E_{\Lambda^0}(k)}\frac{1}{E_N(\vec{k}\,) +
E_{\Lambda^0}(\vec{k}\,) - 2 m_N - m_K}\nonumber\\
\hspace{-0.5in}&&\times\, \Big|\int\frac{d^3K}{(2\pi)^3}
\frac{\Phi_d(\vec{K})}{E_N(\vec{K}\,)} \, M(n(\vec{k},\alpha_1)
\Lambda^0(- \vec{k},\alpha_2) \to K^-(\vec{0}\,) p(\vec{K},\sigma_p)
n( - \vec{K},\sigma_n); {^3}{\rm P}_1)\Big|^2.
\end{eqnarray}
The real part of the integral over $\vec{k}$ is divergent. For the
regularization of the divergent integral we introduce the cut--off
$\Lambda$. Subtracting the divergent part and keeping the finite part
dependent on the cut--off $\Lambda$ the result of the integration over
$\vec{k}$ reads\,\footnote{We assume also that after the subtraction
of the divergent part the integrand is a peaked function around the
point $E_N(\vec{k}\,) + E_{\Lambda^0}(\vec{k}\,) - 2 m_N - m_K = 0$,
i.e. around $|\vec{k}\,| = k_0$. That is valid for the imaginary part
of the S--wave amplitude $\tilde{f}^{\,K^-d}_0(0)_{(n\Lambda^0;
{^3}{\rm P}_1)}$.  Due to this assumption we can take away the squared
amplitude of the reaction $K^-(pn)_{{^3}{\rm S}_1} \to
(n\Lambda^0)_{{^3}{\rm P}_1}$ at $|\vec{k}\,| = k_0$.}
\begin{eqnarray}\label{label6.22}
\hspace{-0.3in}&&\frac{1}{2\pi}\,{\cal P}\int \frac{d^3k}{
E^2_N(\vec{k}\,)}\frac{1}{E_N(\vec{k}\,) + E_{\Lambda^0}(\vec{k}\,) -
2 m_N - m_K -i\,0} = \frac{\pi m_B}{2 m_B +
m_K}\nonumber\\
\hspace{-0.3in}&&\times\,\left\{\frac{2}{\pi}\,
\arctan\Big(\frac{\Lambda}{m_B}\Big) - \frac{2}{\pi}\,\frac{k_0}{ m_B}
\,{\ell n}\left[\displaystyle \frac{\displaystyle \frac{\Lambda}{m_B +
\sqrt{\Lambda^2 + m^2_N}} + \frac{k_0}{m_B + \sqrt{k^2_0 +
m^2_B}}}{\displaystyle \frac{\Lambda}{m_B + \sqrt{\Lambda^2 + m^2_B}}
- \frac{k_0}{m_B + \sqrt{k^2_0 + m^2_B}}}\right]\right\} = \nonumber\\
\hspace{-0.3in}&& = \frac{\pi}{2}\,\frac{1}{1 + m_K/2
m_B}\,F\Big(\frac{\Lambda}{m_B},\frac{k_0}{m_B}\Big),
\end{eqnarray}
where $m_B = \sqrt{(2 m_N + m_K)^2 - 4k^2_0}/2 = 1030\,{\rm MeV}$. For
numerical analysis we set $\Lambda = m_N$\,\footnote{We assume that
the cut--off $\Lambda = m_N$ is an universal cut--off for the analysis
of low--energy interactions of the deuteron near threshold. We
relegate the readers to Section 4, the analysis of the Ericson--Weise
formula for the S--wave scattering length of $K^-d$ scattering, and
\cite{AI2,AI04}.}.

Hence, at threshold the part part of the S--wave amplitude of $K^-d$
scattering, caused by the two--body inelastic channel $K^- (p
n)_{{^3}{\rm S}_1} \to (n\Lambda^0)_{{^3}{\rm P}_1} \to K^- (p
n)_{{^3}{\rm S}_1}$, reads
\begin{eqnarray}\label{label6.23}
\hspace{-0.5in}&&{\cal R}e\,\tilde{f}^{\,K^-d}_0(0)_{(n\Lambda^0;
{^3}{\rm P}_1)} = 4.6\times 10^{-3}\,
\frac{1}{12\pi^2}\,\frac{m^3_{\pi}}{1 + m_K/m_d}\,\frac{k^2_0}{1 +
m_K/2m_B}\,[C^{\,(n\Lambda^0)_{{^3}{\rm P}_1}}_{K^-(pn; {^3}{\rm
S}_1)}]^2\nonumber\\
\hspace{-0.5in}&&\times\, |f^{(n\Lambda^0)_{{^3}{\rm P}_1}}_{K^-(pn;
{^3}{\rm
S}_1)}(k_0)|^2\,F\Big(\frac{\Lambda}{m_B},\frac{k_0}{m_B}\Big) =
\,-\,0.1\times 10^{-3}\,{\rm fm}.
\end{eqnarray}
The S--wave amplitude $\tilde{f}^{\,K^-d}_0(0)_{(n\Lambda^0; {^3}{\rm
P}_1)}$ of $K^-d$ scattering, caused by the two--body inelastic
channel $K^- (p n)_{{^3}{\rm S}_1} \to (n\Lambda^0)_{{^3}{\rm P}_1}
\to K^- (p n)_{{^3}{\rm S}_1}$, is resulted in
\begin{eqnarray}\label{label6.24}
\tilde{f}^{\,K^-d}_0(0)_{(n\Lambda^0; {^3}{\rm P}_1)} = (-\,0.1 + i\,0.9)
\times 10^{-3}\,{\rm fm}.
\end{eqnarray}
Now we proceed to computing the S--wave amplitude
$\tilde{f}^{\,K^-d}_0(0)_{(n\Lambda^0; {^1}{\rm P}_1)}$ of $K^-d$
scattering near threshold, saturated by the intermediate
$(n\Lambda^0)_{{^1}{\rm P}_1}$ state.

\subsection{\bf Reaction $(n\Lambda^0)_{{^1}{\rm P}_1} \to K^-
(pn)_{{^3}{\rm S}_1}$}

The amplitude of the reaction $(n \Lambda^0)_{{^1}{\rm P}_1} \to K^-
(pn)_{{^3}{\rm S}_1}$ is defined by
\begin{eqnarray}\label{label6.25}
\hspace{-0.3in}&&M(n(\vec{k},\alpha_1) \Lambda^0(- \vec{k},\alpha_2)
\to K^-(\vec{0}\,) p(\vec{K},\sigma_p) n( - \vec{K},\sigma_n);
{^1}{\rm P}_1) = i\,C^{\,(n\Lambda^0; {^1}{\rm P}_1)}_{K^- (pn;
{^3}{\rm S}_1)}\,\nonumber\\
\hspace{-0.3in}&&\times\, \frac{[\bar{u}(\vec{K},\sigma_p)
\vec{\gamma} u^c( - \vec{K},\sigma_n)]\cdot
[\bar{u^c}(\vec{k},\alpha_1) \vec{\gamma} \gamma^5u(-
\vec{k},\alpha_2)] }{\displaystyle 1 - \frac{1}{2}\,r^t_{np} a^t_{np}
K^2 + i\,a^t_{np} K}\,f^{(n\Lambda^0; {^1}{\rm P}_1)}_{K^- (pn;
{^3}{\rm S}_1)}(k_0),
\end{eqnarray}
where $f^{\,(n\Lambda^0; {^1}{\rm P}_1)}_{K^- (pn; {^3}{\rm
S}_1)}(k_0)$ is the amplitude, describing the $n\Lambda^0$
rescattering near threshold of the $K^-(p n)_{{^3}{\rm S}_1}$ system
production, and $C^{\,(n\Lambda^0; {^1}{\rm P}_1)}_{K^- (pn; {^3}{\rm
S}_1)}$ is the effective coupling constant of the transition
$(n \Lambda^0)_{{^1}{\rm P}_1} \to K^- (pn)_{{^3}{\rm S}_1}$.

The effective Lagrangian of the transition $(n\Lambda^0)_{{^1}{\rm
P}_1} \to K^- (pn)_{{^3}{\rm S}_1}$ at threshold can be defined by
\begin{eqnarray}\label{label6.26}
{\cal L}^{\,(n\Lambda^0; {^1}{\rm P}_1) \to K^-(pn; {^3}{\rm
S}_1)}_{\rm eff}(x) = i\,C^{\,(n\Lambda^0; {^1}{\rm P}_1)}_{K^- (pn;
{^3}{\rm S}_1)}\, K^{-\dagger}(x)\, [\bar{p}(x)\vec{\gamma}\, n^c(x)]
\cdot [\bar{n^c}(x) \gamma^0\vec{\gamma}\, \gamma^5 \Lambda^0(x)].
\end{eqnarray}
The effective coupling constant $C^{\,(n\Lambda^0; {^1}{\rm P}_1)}_{K^-
(pn; {^3}{\rm S}_1)}$ is equal to 
\begin{eqnarray}\label{label6.27}
\hspace{-0.3in}C^{\,(n\Lambda^0; {^1}{\rm P}_1)}_{K^- (pn;
{^3}{\rm S}_1)}
&=& \,\frac{1}{2\sqrt{3}}\, \frac{(3
- 2\alpha)(2\alpha - 1)^2 g^3_{\pi NN}}{m^2_K - (E_{\Lambda^0} -
m_N)^2 + k^2_0}\,\frac{1}{m_{\Sigma} + m_N + m_K}\nonumber\\
\hspace{-0.3in}&& -\,\frac{1}{4\sqrt{3}}\,\frac{(3 - 2\alpha)(2\alpha
- 1)^2 g^3_{\pi NN}}{m^2_K - (E_{\Lambda^0} - m_N)^2 +
k^2_0}\,\frac{1}{m_{\Sigma} + m_N + m_K}\nonumber\\
\hspace{-0.3in}&& +\,\frac{1}{12\sqrt{3}}\,\frac{(3 - 2\alpha)^3
g^3_{\pi NN}}{m^2_K - (E_{\Lambda^0} - m_N)^2 +
k^2_0}\,\frac{1}{m_{\Lambda^0} + m_N + m_K}\nonumber\\
\hspace{-0.3in}&& -\,\frac{1}{\sqrt{3}}\,\frac{\alpha (2\alpha - 1)
g^3_{\pi NN}}{m^2_{\pi} - (E_N - m_N)^2 + k^2_0}\,\frac{1}{m_{\Sigma}
+ m_N + m_K}\nonumber\\
\hspace{-0.3in}&& +\,\frac{1}{2\sqrt{3}}\,\frac{(3 - 2\alpha) g^3_{\pi
NN}}{m^2_{\pi} - (E_N - m_N)^2 + k^2_0}\,\frac{m_N + m_K -
m_{\Lambda^0}}{m^2_N - (E_{\Lambda^0} - m_K)^2 + k^2_0}\nonumber\\
\hspace{-0.3in}&& +\,\frac{1}{4\sqrt{3}}\,\frac{(3 - 2\alpha) g^3_{\pi
NN}}{m^2_{\pi} - (E_N - m_N)^2 + k^2_0}\,\frac{m_N + m_K -
m_{\Lambda^0}}{m^2_N - (E_{\Lambda^0} - m_K)^2 + k^2_0}\nonumber\\
\hspace{-0.3in}&& -\,\frac{1}{2\sqrt{3}}\,\frac{\alpha (2\alpha - 1)
g^3_{\pi NN}}{m^2_{\pi} - (E_N - m_N)^2 + k^2_0}\,\frac{1}{m_{\Sigma}
+ m_N + m_K}\nonumber\\
\hspace{-0.3in}&& +\,\frac{1}{6\sqrt{3}}\,\frac{\alpha (3 - 4 \alpha)
(3 - 2\alpha) g^3_{\pi NN}}{m^2_{\eta} - (E_N - m_N)^2 +
k^2_0}\,\frac{1}{m_{\Lambda^0} + m_N + m_K}\nonumber\\
\hspace{-0.3in}&& -\,\frac{1}{12\sqrt{3}}\,\frac{(3 - 2\alpha)(3 -
4\alpha)^2 g^3_{\pi NN}}{m^2_{\eta} - (E_N - m_N)^2 +
k^2_0}\,\frac{m_N + m_K - m_{\Lambda^0}}{m^2_N - (E_{\Lambda^0} -
m_K)^2 + k^2_0}\nonumber\\ &&-\,\frac{1}{8\sqrt{3}}\,\frac{1}{g_A
F_{\pi}}\,\frac{(3 - 2\alpha)\,g^2_{\pi NN}}{m^2_K - (E_{\Lambda^0} -
m_N)^2 + k^2_0}\nonumber\\ \hspace{-0.3in}&&-
\frac{1}{8\sqrt{3}}\,\frac{1}{g_A F_{\pi}}\,\frac{(3 -
2\alpha)\,g^2_{\pi NN}}{m^2_{\pi} - (E_N - m_N)^2 + k^2_0}\nonumber\\
\hspace{-0.3in}&&- \frac{1}{24\sqrt{3}}\,\frac{1}{g_A
F_{\pi}}\,\frac{(3 - 2\alpha)\,(3 - 4\alpha)\,g^2_{\pi NN}}{m^2_{\eta}
- (E_N - m_N)^2 + k^2_0}\nonumber\\ \hspace{-0.3in}&&-
\frac{1}{4\sqrt{3}}\,\frac{1}{g_A F_{\pi}}\,\frac{(3 -
2\alpha)\,g^2_{\pi NN}}{m^2_{\pi} - (E_N - m_N)^2 + k^2_0}
=\nonumber\\
\hspace{-0.3in}&& = 6\times 10^{-7}\,{\rm MeV}^{-3}.
\end{eqnarray}
The effective coupling constant (\ref{label6.27}) contains the
contribution of the scalar--meson exchanges computed in the infinite
mass limit.

The Lagrangian (\ref{label6.26}) describes the interaction of the
$n\Lambda^0$ pair in the ${^1}{\rm P}_1$ state with the $np$ pair in
the ${^3}{\rm S}_1$ through the emission of the $K^-$--meson.

\subsection{ Amplitude of $(n\Lambda^0)_{{^1}{\rm P}_1}$
rescattering}

The amplitude $f^{\,(n\Lambda^0; {^1}{\rm P}_1)}_{K^-pn}(k_0)$,
describing the rescattering of the $n\Lambda^0$ pair in the ${^1}{\rm
P}_1$ state near threshold of the $K^- (p n)_{{^3}{\rm S}_1}$ system
production, is defined by the Feynman diagrams analogous to those
depicted in Fig.6.  The result of the calculation of these diagrams
reads (see Appendix E and \cite{AI01})
\begin{eqnarray}\label{label6.28}
\Big|f^{\,(n\Lambda^0; {^1}{\rm P}_1)}_{K^-(pn; {^3}{\rm
S}_1)}(k_0)\Big| = \Big|\Big\{1 - \frac{C_{n\Lambda^0}({^1}{\rm
P}_1)}{24\pi^2}\,\frac{k^3_0}{E(k_0)}\,\Big[{\ell n}\Big(\frac{E(k_0)
+ k_0}{E(k_0) - k_0} \Big) - i\,\pi\Big]\Big\}^{-1}\Big| \simeq 1,
\end{eqnarray}
where $E(k_0) = \sqrt{k^2_0 + m^2_B}$\,\footnote{For simplicity we use
the equal masses of baryons for the calculation of the rescattering of
the $n\Lambda^0$ pair, where $m_B = \sqrt{(2 m_N + m_K)^2 - 4 k^2_0}/2
= 1030\,{\rm MeV}$.} and the effective coupling constant
$C_{(n\Lambda^0)}({^1}{\rm P}_1)$ is equal to
\cite{AI01}
\begin{eqnarray}\label{label6.29}
C_{n\Lambda^0}({^1}{\rm P}_1) &=& (3 - 2 \alpha)^2\,\frac{g^2_{\pi
NN}}{12 k^2_0}\,{\ell n}\Big(1 + \frac{4 k^2_0}{~m^2_K}\Big) - \alpha
(3 - 4\alpha)\,\frac{g^2_{\pi NN}}{6 k^2_0}\,{\ell n}\Big(1 + \frac{4
k^2_0}{~m^2_{\eta}}\Big)=\nonumber\\ &=& 2.0\times 10^{-4}\,{\rm
MeV}^{-2}.
\end{eqnarray}
The rescattering of the $n\Lambda^0$ pair in the ${^1}{\rm P}_1$ state
is defined by the interaction, computed in the one--meson exchange
approximation (see \cite{AI01}):
\begin{eqnarray}\label{label6.30}
{\cal L}^{(n\Lambda^0; {^1}{\rm P}_1) \to (n\Lambda^0; {^1}{\rm
P}_1)}_{\rm eff}(x) = -\,\frac{1}{4}\,C_{n\Lambda^0}({^1}{\rm
P}_1))\,[\bar{\Lambda}^0(x) \gamma^0\vec{\gamma}\,\gamma^5
n^c(x)]\cdot [\bar{n^c}(x)\gamma^0\vec{\gamma}\,\gamma^5
\Lambda^0(x)].
\end{eqnarray}
Now we can proceed to computing the S--wave amplitude
$\tilde{f}^{\,K^-d}_0(0)_{(n\Lambda^0; {^1}{\rm P}_1)}$ of $K^-d$
scattering near threshold, saturated by the intermediate
$(n\Lambda^0)_{{^1}{\rm P}_1}$ state.

\subsection{S--wave amplitude
$\tilde{f}^{\,K^-d}_0(0)_{(n\Lambda^0; {^1}{\rm P}_1)}$ of $K^-d$
scattering}

For the calculation of the S--wave amplitude
$\tilde{f}^{\,K^-d}_0(0)_{(n\Lambda^0; {^1}{\rm P}_1)}$ of $K^-d$
scattering near threshold, caused by the inelastic channel
$K^-(pn)_{{^3}{\rm S}_1} \to (n\Lambda^0)_{{^1}{\rm P}_1} \to
K^-(pn)_{{^3}{\rm S}_1}$, we have to square the amplitude
(\ref{label6.25}) and to sum over polarisations of baryons, taking
into account that $np$ and $n\Lambda^0$ pairs are in the ${^3}{\rm
S}_1$ and ${^1}{\rm P}_1$ state, respectively. This yields (see
Appendix C)
\begin{eqnarray}\label{label6.31}
\hspace{-0.3in}\frac{1}{3}\sum_{(\sigma_p,\sigma_n)}
\sum_{(\alpha_2,\alpha_1; {^1}{\rm P}_1)}|[\bar{u}(\vec{K},\sigma_p)
\vec{\gamma}\, u^c( - \vec{K},\sigma_n)]\cdot
[\bar{u^c}(\vec{k},\alpha_1) \gamma^0\vec{\gamma}\gamma^5u(-
\vec{k},\alpha_2)]|^2 = \frac{64}{3}\,m^2_N\,\vec{k}^{\;2}.
\end{eqnarray}
At threshold the imaginary part of the S--wave amplitude of $K^-d$
scattering, caused by the two--body inelastic channel $K^- (p
n)_{{^3}{\rm S}_1} \to (n \Lambda^0)_{{^1}{\rm P}_1} \to K^- (p
n)_{{^3}{\rm S}_1}$, is equal to
\begin{eqnarray}\label{label6.32}
{\cal I}m\,\tilde{f}^{\,K^-d}_0(0)_{(n\Lambda^0; {^1}{\rm P}_1)} &=&
4.6\times 10^{-3}\, \frac{1}{6\pi^2}\,\frac{m^3_{\pi}}{1 +
m_K/m_d}\,\frac{k^3_0}{2m_N + m_K}\, \nonumber\\
&&\times\,[C^{\,(n\Lambda^0; {^1}{\rm P}_1)}_{K^-(pn; {^3}{\rm
S}_1)}]^2\, |f^{(n\Lambda^0; {^1}{\rm P}_1)}_{K^-(pn; {^3}{\rm
S}_1)}(k_0)|^2 = 1.0\times 10^{-3}\,{\rm fm}.
\end{eqnarray}
The real part of $\tilde{f}^{\,K^-d}_0(0)_{(n\Lambda^0; {^1}{\rm
P}_1)}$ reads
\begin{eqnarray}\label{label6.33}
\hspace{-0.5in}&&{\cal
R}e\,\tilde{f}^{\,K^-d}_0(0)_{(n\Lambda^0; {^1}{\rm P}_1)} =
4.6\times 10^{-3}\, \frac{1}{24\pi^2}\,\frac{m^3_{\pi}}{1 +
m_K/m_d}\,\frac{k^2_0}{1 + m_K/2m_B}\,[C^{\,(n\Lambda^0; {^1}{\rm
P}_1)}_{K^-(pn; {^3}{\rm S}_1)}]^2
\nonumber\\
\hspace{-0.5in}&&\times\,|f^{(n\Lambda^0; {^1}{\rm P}_1)}_{K^-(pn;
{^3}{\rm
S}_1)}(k_0)|^2\,F\Big(\frac{\Lambda}{m_B},\frac{k_0}{m_B}\Big) =
\,-\,0.1\times 10^{-3}\,{\rm fm}.
\end{eqnarray}
Thus, for the S--wave amplitude
$\tilde{f}^{\,K^-d}_0(0)_{(n\Lambda^0)_{{^1}{\rm P}_1}}$ we get
\begin{eqnarray}\label{label6.34}
\tilde{f}^{\,K^-d}_0(0)_{(n\Lambda^0; {^1}{\rm P}_1)} = (-\,0.1 +
i\,1.0) \times 10^{-3}\,{\rm fm}.
\end{eqnarray}
Now we can estimate the contribution of the two--body inelastic
channel $K^- (pn)_{{^3}{\rm S}_1} \to n \Lambda^0 \to K^-
(pn)_{{^3}{\rm S}_1}$ to the S--wave amplitude
$f^{K^-d}_0(0)_{n\Lambda^0}$ of $K^-d$ scattering near threshold and
the energy level displacement of the ground state of kaonic deuterium.

\subsection{S--wave amplitude $f^{K^-d}_0(0)_{n\Lambda^0}$ 
and the energy level displacement}

The S--wave amplitude of $K^-d$ scattering saturated at threshold by
the intermediate $n\Lambda^0$ state in the ${^3}{\rm P}_1$ and
${^1}{\rm P}_1$ is equal to the sum of the contributions
(\ref{label6.24}) and (\ref{label6.34})
\begin{eqnarray}\label{label6.35}
\tilde{f}^{\,K^-d}_0(0)_{n\Lambda^0} = (-\,0.2 + i\,1.9) \times
10^{-3}\,{\rm fm}.
\end{eqnarray}
The contribution of the decay $A_{Kd} \to n \Lambda^0$ to the energy
level displacement of the ground state of kaonic deuterium amounts to
\begin{eqnarray}\label{label6.36}
-\,\epsilon^{(n\Lambda^0)}_{1s} +
i\,\frac{\Gamma^{(n\Lambda^0)}_{1s}}{2} =
602\,\tilde{f}^{\,K^-d}_0(0)_{n\Lambda^0} = (-\,0.1 + i\,1.1)\,{\rm eV}.
\end{eqnarray}
Hence, the partial width of the decay $A_{Kd} \to n \Lambda^0$ is
equal to $\Gamma^{(n\Lambda^0)}_{1s} = 2.2\,{\rm eV}$. 

According to \cite{VV70}, the experimental rate of the production of
the $n\Lambda^0$ pair at threshold of the reaction $K^-d \to
n\Lambda^0$ is equal to $R(K^-d \to n \Lambda^0) = (0.387 \pm
0.041)\,\%$.

Using our estimate of the partial width, $\Gamma^{(n\Lambda^0)}_{1s} =
2.2\,{\rm eV}$, and the experimental rate, $R(K^-d \to n \Lambda^0) =
(0.387 \pm 0.041)\,\%$, we can estimate the expected value of the total
width of the energy level of the ground state of kaonic deuterium
\begin{eqnarray}\label{label6.37}
\Gamma_{1s} = \frac{\Gamma^{(n\Lambda^0)}_{1s} }{(0.387 \pm
0.041)\times 10^{-2}} = (570 \pm 130)\,{\rm eV}.
\end{eqnarray}
We have taken into account the theoretical accuracy of the energy
level displacement, which is about 20$\%$:
$-\,\epsilon^{(n\Lambda^0)}_{1s} + i\, \Gamma^{(n\Lambda^0)}_{1s}/2 =
(- \,0.10 \pm 0.02) + i\,(1.10 \pm 0.22)\,{\rm eV}$.

Our expected value of the total width of the ground state of kaonic
deuterium is by a factor of 2 smaller compared with the value of the
total width predicted by Barrett and Deloff \cite{RB99}.

\section{Amplitude of reaction $K^- (p n)_{{^3}{\rm S}_1}
 \to n \Sigma^0 \to K^- (p n)_{{^3}{\rm S}_1}$ and the energy level
 displacement} \setcounter{equation}{0}

The amplitudes of the reactions $K^- (pn)_{{^3}{\rm S}_1} \to (n
\Sigma^0)_X $, where $n \Sigma^0$ pair couples in the $X = {^3}{\rm
P}_1$ and and ${^1}{\rm P}_1$ state, we define as
\begin{eqnarray}\label{label7.1}
\hspace{-0.3in}&&M(K^-(\vec{0}\,) p(\vec{K},\sigma_p) n( -
\vec{K},\sigma_n) \to n(\vec{k},\alpha_1) \Sigma^0(-
\vec{k},\alpha_2); {^3}{\rm P}_1) = -\,i\,C^{\,(n\Sigma^0; {^3}{\rm
P}_1)}_{K^-(pn; {^3}{\rm S}_1)}\,\nonumber\\
\hspace{-0.3in}&&\times\, \frac{[\bar{u^c}(- \vec{K},\sigma_n)
\vec{\gamma} u( \vec{K},\sigma_p)]\cdot [\bar{u}(- \vec{k},\alpha_2)
\vec{\gamma} \gamma^5u^c ( \vec{k},\alpha_1)] }{\displaystyle 1 -
\frac{1}{2}\,r^t_{np} a^t_{np} K^2 + i\,a^t_{np} K}\,f^{(n\Sigma^0;
{^3}{\rm P}_1)}_{K^-(pn; {^3}{\rm S}_1)}(k_0),\nonumber\\
\hspace{-0.3in}&&M(K^-(\vec{0}\,) p(\vec{K},\sigma_p) n( -
\vec{K},\sigma_n) \to n(\vec{k},\alpha_1) \Sigma^0(-
\vec{k},\alpha_2); {^1}{\rm P}_1) = -\,i\,C^{\,(n\Sigma^0; {^1}{\rm
P}_1)}_{K^-(pn; {^3}{\rm S}_1)}\,\nonumber\\
\hspace{-0.3in}&&\times\, \frac{[\bar{u^c}(- \vec{K},\sigma_n)
\vec{\gamma} u( \vec{K},\sigma_p)]\cdot [\bar{u}(- \vec{k},\alpha_2)
\gamma^0 \vec{\gamma} \gamma^5u^c ( \vec{k},\alpha_1)] }{\displaystyle
1 - \frac{1}{2}\,r^t_{np} a^t_{np} K^2 + i\,a^t_{np}
K}\,f^{(n\Sigma^0; {^1}{\rm P}_1)}_{K^-(pn; {^3}{\rm S}_1)}(k_0),
\end{eqnarray}
where $f^{(n\Sigma^0; X)}_{K^-(pn; {^3}{\rm S}_1)}(k_0)$ is the
amplitude of the final--state $n\Sigma^0$ interaction near threshold
of the reaction $K^- (pn)_{{^3}{\rm S}_1} \to (n \Sigma^0)_X$ and
$C^{\,(n\Sigma^0; X)}_{K^-(pn; {^3}{\rm S}_1)}$ is the effective
coupling constant of the transition $ K^- (pn)_{{^3}{\rm S}_1} \to (n
\Sigma^0)_X$, where $X = {^3}{\rm P}_1$ or ${^1}{\rm P}_1$.

\subsection{Effective coupling constant $C^{\,n\Sigma^0}_{K^-pn}$}

In the one--meson exchange approximation \cite{AI01} the effective
coupling constant $C^{\,n\Sigma^0}_{K^-pn}$ of the transition $n
\Sigma^0 \to K^- p n$ is defined by the Feynman diagrams depicted in
Fig.7. 
\begin{figure}
\centering \psfrag{K-}{$K^-$} \psfrag{K0b}{$\bar{K}^0$}
\psfrag{L0}{$\Lambda^0$} \psfrag{p+}{$\pi^+$} \psfrag{p0}{$\pi^0$}
\psfrag{S-}{$\Sigma^-$} \psfrag{S0}{$\Sigma^0$}
\psfrag{Sp}{$\Sigma^+$}\psfrag{p}{$p$} \psfrag{n}{$n$}
\psfrag{e}{$\eta$}
\includegraphics[height=0.90\textheight]{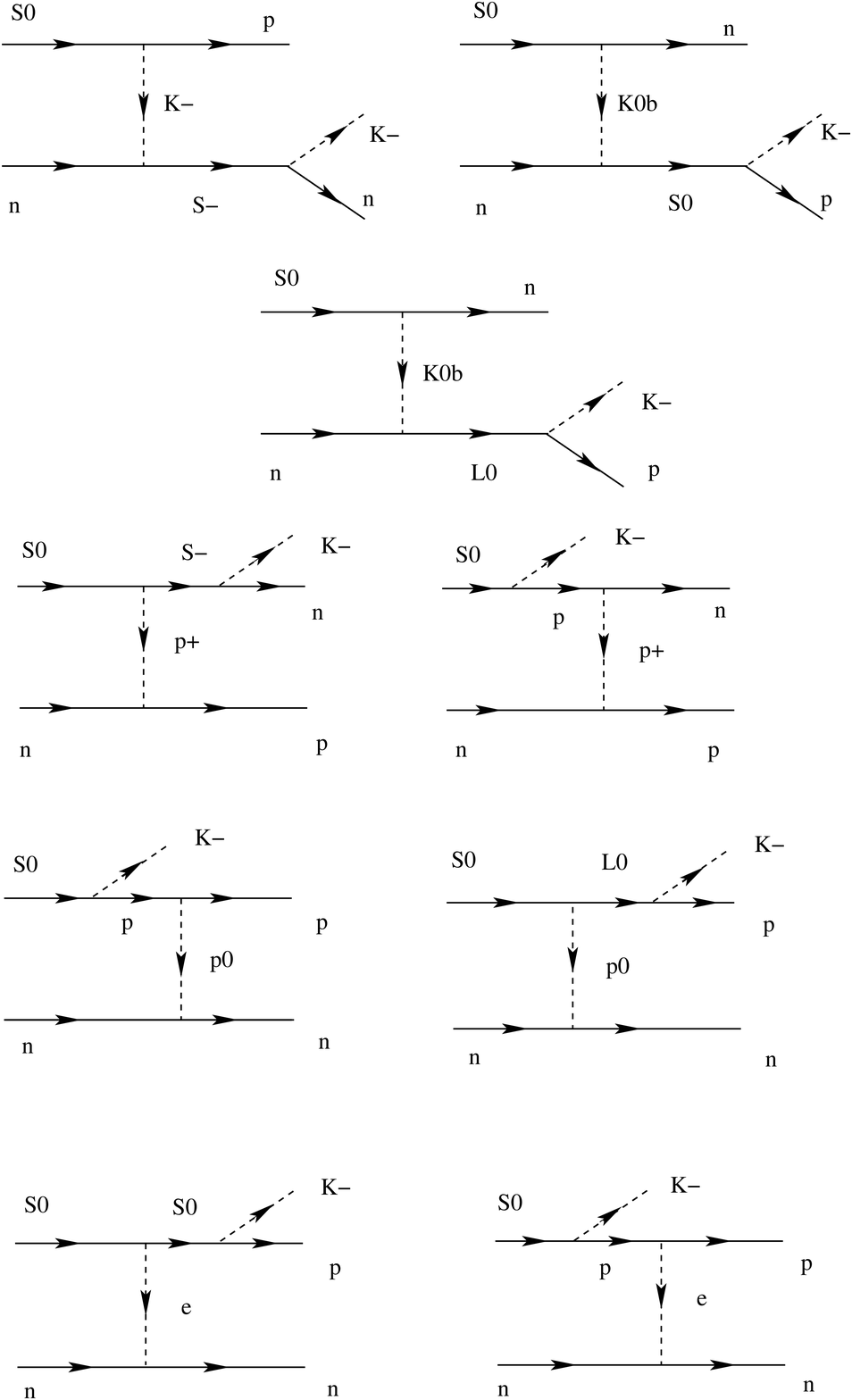}
\caption{Feynman diagrams describing the effective coupling constant
of the transition $n \Sigma^0 \to K^- p n$ in the one--pseudoscalar
meson exchange approximation}
\end{figure}
The amplitude of the transition $n \Sigma^0 \to K^- p n$,
determined by the Feynman diagrams in Fig.7, is equal to 
\begin{eqnarray}\label{label7.2}
&&{\cal M}( n(\vec{k},\alpha_1) \Sigma^0(- \vec{k},\alpha_2) \to
K^-(\vec{0}\,) p(\vec{K},\sigma_p) n( - \vec{K},\sigma_n))
=\nonumber\\ &&= -\,2\,(2\alpha - 1)^3\,g^3_{\pi NN}\times
\Big[\bar{u}(k_n,\sigma_n)i\gamma^5\frac{1}{m_{\Sigma^-} - \hat{k}_p -
\hat{Q}_K}i\gamma^5 u(q_n,\alpha_n)\Big]\nonumber\\
&&\times\,\frac{1}{m^2_K - (q_{\Sigma^0} - k_p)^2}\times
[\bar{u}(k_p,\sigma_p)i\gamma^5
u(q_{\Sigma^0},\alpha_{\Sigma^0})]\nonumber\\ &&+ \,(2\alpha -
1)^3\,g^3_{\pi NN}\times
\Big[\bar{u}(k_p,\sigma_p)i\gamma^5\frac{1}{m_{\Sigma^0} - \hat{k}_p -
\hat{Q}_K}i\gamma^5 u(q_n,\alpha_n)\Big]\nonumber\\
&&\times\,\frac{1}{m^2_K - (q_{\Sigma^0} - k_n)^2}\times
[\bar{u}(k_n,\sigma_n)i\gamma^5
u(q_{\Sigma^0},\alpha_{\Sigma^0})]\nonumber\\ &&-
\,\frac{1}{3}\,(2\alpha - 1)\,(3 - 2 \alpha)^2\,g^3_{\pi NN}\times
\Big[\bar{u}(k_p,\sigma_p)i\gamma^5\frac{1}{m_{\Lambda^0} - \hat{k}_p
- \hat{Q}_K}i\gamma^5 u(q_n,\alpha_n)\Big]\nonumber\\
&&\times\,\frac{1}{m^2_K - (q_{\Sigma^0} - k_n)^2}\times
[\bar{u}(k_n,\sigma_n)i\gamma^5
u(q_{\Sigma^0},\alpha_{\Sigma^0})]\nonumber\\ &&+ \,4\,(1 -
\alpha)\,(2\alpha - 1)\,g^3_{\pi NN}\times
\Big[\bar{u}(k_n,\sigma_n)i\gamma^5\frac{1}{m_{\Sigma^-} - \hat{k}_n -
\hat{Q}_K}i\gamma^5 u(q_{\Sigma^0},\alpha_{\Sigma^0})\Big]\nonumber\\
&&\times\,\frac{1}{m^2_{\pi} - (q_n - k_p)^2}\times
[\bar{u}(k_p,\sigma_p)i\gamma^5 u(q_n,\alpha_n)]\nonumber\\ &&+
\,2\,(2\alpha - 1)\,g^3_{\pi NN}\times
\Big[\bar{u}(k_n,\sigma_n)i\gamma^5\frac{1}{m_p - \hat{q}_{\Sigma^0} +
\hat{Q}_K}i\gamma^5 u(q_{\Sigma^0},\alpha_{\Sigma^0})\Big]\nonumber\\
&&\times\,\frac{1}{m^2_{\pi} - (q_n - k_p)^2}\times
[\bar{u}(k_p,\sigma_p)i\gamma^5 u(q_n,\alpha_n)]\nonumber\\ &&-
\,(2\alpha - 1)\,g^3_{\pi NN}\times
\Big[\bar{u}(k_p,\sigma_p)i\gamma^5\frac{1}{m_p - \hat{q}_{\Sigma^0} +
\hat{Q}_K}i\gamma^5 u(q_{\Sigma^0},\alpha_{\Sigma^0})\Big]\nonumber\\
&&\times\,\frac{1}{m^2_{\pi} - (q_n - k_n)^2}\times
[\bar{u}(k_n,\sigma_n)i\gamma^5 u(q_n,\alpha_n)]\nonumber\\ &&+
\,\frac{2}{3}\,\alpha\,(3 - 2\alpha)\,g^3_{\pi NN}\times
\Big[\bar{u}(k_p,\sigma_p)i\gamma^5\frac{1}{m_p - \hat{q}_{\Sigma^0} +
\hat{Q}_K}i\gamma^5 u(q_{\Sigma^0},\alpha_{\Sigma^0})\Big]\nonumber\\
&&\times\,\frac{1}{m^2_{\pi} - (q_n - k_n)^2}\times
[\bar{u}(k_n,\sigma_n)i\gamma^5 u(q_n,\alpha_n)]\nonumber\\ &&+
\,\frac{2}{3}\,\alpha\,(2\alpha - 1)\,(3 - 2\alpha)\,g^3_{\pi
NN}\times \Big[\bar{u}(k_p,\sigma_p)i\gamma^5\frac{1}{m_{\Sigma^0} -
\hat{k}_p - \hat{Q}_K}i\gamma^5
u(q_{\Sigma^0},\alpha_{\Sigma^0})\Big]\nonumber\\
&&\times\,\frac{1}{m^2_{\eta} - (q_n - k_n)^2}\times
[\bar{u}(k_n,\sigma_n)i\gamma^5 u(q_n,\alpha_n)]\nonumber\\ &&+
\,\frac{1}{3}\,(3 - 4\alpha)^2\,g^3_{\pi NN}\times
\Big[\bar{u}(k_p,\sigma_p)i\gamma^5\frac{1}{m_p - \hat{q}_{\Sigma^0} +
\hat{Q}_K}i\gamma^5 u(q_{\Sigma^0},\alpha_{\Sigma^0})\Big]\nonumber\\
&&\times\,\frac{1}{m^2_{\eta} - (q_n - k_n)^2}\times
[\bar{u}(k_n,\sigma_n)i\gamma^5 u(q_n,\alpha_n)].
\end{eqnarray}
The Feynman diagrams, describing the transition $n\Sigma^0 \to K^- p
n$ through the scalar--meson exchanges, are depicted in Fig.8.
\begin{figure}
\centering \psfrag{K-}{$K^-$} 
\psfrag{S0}{$\Sigma^0$} 
\psfrag{p+}{$\pi^+$} \psfrag{p0}{$\pi^0$}
\psfrag{p}{$p$} \psfrag{n}{$n$}
\psfrag{s}{$\sigma$}
\psfrag{e}{$\eta$}
\psfrag{k-}{$\kappa^-$}
\psfrag{k0}{$\bar{\kappa}^0$}
\includegraphics[height=0.30\textheight]{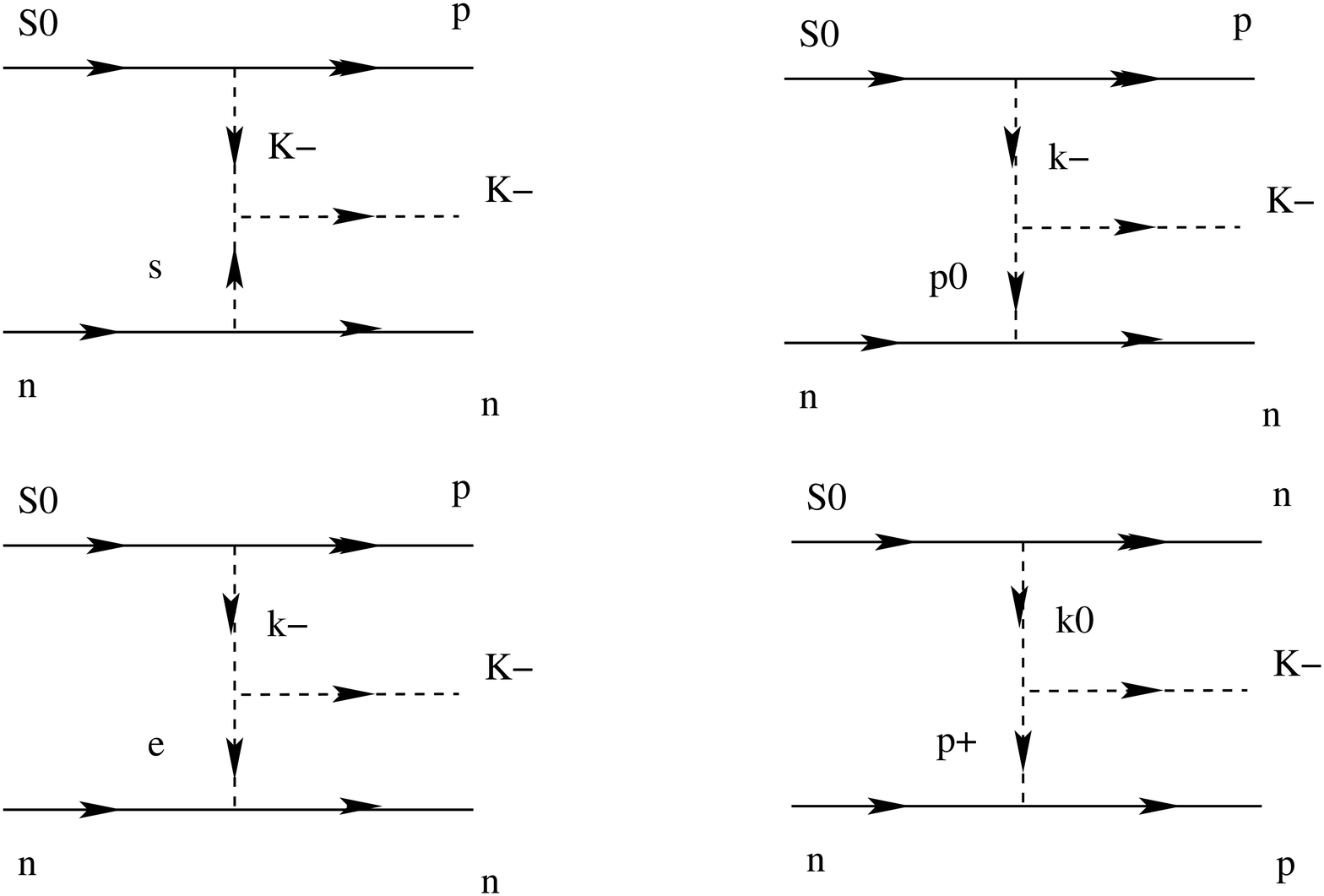}
\caption{Feynman diagrams describing the contribution of scalar mesons
to the effective coupling constant of the transition $n \Sigma^0 \to
K^- p n$ in the one--pseudoscalar meson exchange approximation}
\end{figure}

Taking the amplitude of the transition $n\Sigma^0 \to K^- p n$ at
threshold, we can represent (\ref{label7.2}) in the form of the
effective local Lagrangian
\begin{eqnarray}\label{label7.3}
\hspace{-0.3in}&&{\cal L}^{n\Sigma^0 \to K^-pn}_{\rm eff}(x)_P =
\frac{2\,(2\alpha - 1)^3\,g^3_{\pi NN}}{m^2_K - (E_{\Sigma^0} - m_N)^2
+ k^2_0}\,\frac{1}{m_{\Sigma} + m_N + m_K}\,
[\bar{p}(x)i\gamma^5\Sigma^0(x)][\bar{n}(x)n(x)]\nonumber\\
\hspace{-0.3in}&&- \,\frac{(2\alpha - 1)^3\,g^3_{\pi NN}}{m^2_K -
(E_{\Sigma^0} - m_N)^2 + k^2_0}\,\frac{1}{m_{\Sigma} + m_N + m_K}\,
[\bar{n}(x)i\gamma^5\Sigma^0(x)][\bar{p}(x)n(x)]\nonumber\\
\hspace{-0.3in}&&+\,\frac{1}{3}\,\frac{(2\alpha - 1)\,(3 -
2\alpha)^2\,g^3_{\pi NN}}{m^2_K - (E_{\Sigma^0} - m_N)^2 +
k^2_0}\,\frac{1}{m_{\Lambda^0} + m_N + m_K}\,
[\bar{n}(x)i\gamma^5\Sigma^0(x)][\bar{p}(x)n(x)]\nonumber\\
\hspace{-0.3in}&&- \,\frac{4\,(1 - \alpha)\,(2\alpha - 1)\,g^3_{\pi
NN}}{m^2_{\pi} - (E_N - m_N)^2 + k^2_0}\,\frac{1}{m_{\Sigma} + m_N +
m_K}\,[\bar{n}(x)\Sigma^0(x)][\bar{p}(x)i\gamma^5 n(x)]\nonumber\\
\hspace{-0.3in}&&- \,\frac{2\,(2\alpha - 1)\,g^3_{\pi NN}}{m^2_{\pi} -
(E_N - m_N)^2 + k^2_0}\,\frac{m_N + m_K - m_{\Sigma}}{m^2_N -
(E_{\Sigma^0} - m_K)^2 +
k^2_0}\,[\bar{n}(x)\Sigma^0(x)][\bar{p}(x)i\gamma^5 n(x)]\nonumber\\
\hspace{-0.3in}&&+\,\frac{(2\alpha - 1)\,g^3_{\pi NN}}{m^2_{\pi} - (E_N - m_N)^2 +
k^2_0}\,\frac{m_N + m_K - m_{\Sigma}}{m^2_N - (E_{\Sigma^0} - m_K)^2 +
k^2_0}\,[\bar{p}(x)\Sigma^0(x)][\bar{n}(x)i\gamma^5 n(x)]\nonumber\\
\hspace{-0.3in}&& -\,\frac{2}{3}\,\frac{\alpha\,(3 -
2\alpha)\,g^3_{\pi NN}}{m^2_{\pi} - (E_N - m_N)^2 + k^2_0}\,\frac{m_N
+ m_K - m_{\Sigma}}{m^2_N - (E_{\Sigma^0} - m_K)^2 +
k^2_0}\,[\bar{p}(x)\Sigma^0(x)][\bar{n}(x)i\gamma^5 n(x)]\nonumber\\
\hspace{-0.3in}&&- \,\frac{2}{3}\,\frac{\alpha\,(2\alpha - 1)\,(3 -
2\alpha)\,g^3_{\pi NN}}{m^2_{\eta} - (E_N - m_N)^2 +
k^2_0}\,\frac{1}{m_{\Sigma} + m_N + m_K}\,[\bar{p}(x)\Sigma^0(x)]
[\bar{n}(x)i\gamma^5 n(x)]\nonumber\\ \hspace{-0.3in}&&-
\,\frac{1}{3}\,\frac{(3 - 4\alpha)^2\,g^3_{\pi NN}}{m^2_{\eta} - (E_N
- m_N)^2 + k^2_0}\,\frac{m_N + m_K - m_{\Sigma}}{m^2_N - (E_{\Sigma^0}
- m_K)^2 + k^2_0}\,[\bar{p}(x)\Sigma^0(x)][\bar{n}(x)i\gamma^5 n(x)].
\end{eqnarray}
The effective Lagrangian of the transition $n\Lambda^0 \to K^- p n$,
caused by the scalar--meson exchanges, is equal to
\begin{eqnarray}\label{label7.4}
{\cal L}^{\,n\Sigma^0}_{K^-pn}(x)_S &=& -\,\frac{1}{2}\,\frac{1}{g_A
F_{\pi}}\,\frac{(2\alpha - 1)\,g^2_{\pi NN}}{m^2_K - (E_{\Sigma^0} -
m_N)^2 + k^2_0}\,[\bar{p}(x)i\gamma^5
\Sigma^0(x)][\bar{n}(x)n(x)]\nonumber\\ &&-
\frac{1}{2}\,\frac{F_K}{g_A F^2_{\pi}}\,\frac{(2\alpha - 1)\,g^2_{\pi
NN}}{m^2_{\pi} - (E_N - m_N)^2 +
k^2_0}\,[\bar{p}(x)\Sigma^0(x)][\bar{n}(x)i\gamma^5 n(x)]\nonumber\\
&&- \frac{1}{6}\,\frac{F_K}{g_A F^2_{\pi}}\,\frac{(2\alpha - 1)\,(3 -
4\alpha)\,g^2_{\pi NN}}{m^2_{\eta} - (E_N - m_N)^2 +
k^2_0}\,[\bar{p}(x)\Sigma^0(x)][\bar{n}(x)i\gamma^5 n(x)]\nonumber\\
&&-\,\frac{F_K}{g_A F^2_{\pi}}\,\frac{(2\alpha - 1)\,g^2_{\pi
NN}}{m^2_{\pi} - (E_N - m_N)^2 +
k^2_0}\,[\bar{n}(x)\Sigma^0(x)][\bar{p}(x)i\gamma^5 n(x)].
\end{eqnarray}
Making the Fierz transformation (see (\ref{label6.4})) we reduce the
four--baryon operators to the form
\begin{eqnarray}\label{label7.5}
\hspace{-0.3in}&&[\bar{p}(x)\gamma^5 \Sigma^0(x)] [\bar{n}(x)n(x)]
\to +\,\frac{1}{4}\,[\bar{p}(x)\vec{\gamma}\, n^c(x)] \cdot
[\bar{n^c}(x) \vec{\gamma}\,\gamma^5 \Sigma^0(x)]\nonumber\\
\hspace{-0.3in}&&\hspace{2in}
-\,\frac{1}{4}\,[\bar{p}(x)\vec{\gamma}\,n^c(x)] \cdot [\bar{n^c}(x)
\gamma^0\vec{\gamma}\,\gamma^5 \Sigma^0(x)],\nonumber\\
\hspace{-0.3in}&&[\bar{n}(x)\gamma^5 \Sigma^0(x)] [\bar{p}(x)n(x)]
\to -\,\frac{1}{4}\,[\bar{p}(x)\vec{\gamma}\, n^c(x)]\cdot
[\bar{n^c}(x) \vec{\gamma}\,\gamma^5 \Sigma^0(x)], \nonumber\\
\hspace{-0.3in}&&\hspace{2in}+\,\frac{1}{4}\,
[\bar{p}(x)\vec{\gamma}\,n^c(x)] \cdot [\bar{n^c}(x)
\gamma^0\vec{\gamma}\,\gamma^5 \Sigma^0(x)] \nonumber\\
\hspace{-0.3in}&&[\bar{p}(x) \Sigma^0(x)][\bar{n}(x)\gamma^5 n(x)] \to
-\,\frac{1}{4}\, [\bar{p}(x)\vec{\gamma}\, n^c(x)]\cdot [\bar{n^c}(x)
\vec{\gamma}\,\gamma^5 \Sigma^0(x)]\nonumber\\
\hspace{-0.3in}&&\hspace{2in}-\,\frac{1}{4}\,[\bar{p}(x)
\vec{\gamma}\,n^c(x)] \cdot [\bar{n^c}(x)
\gamma^0\vec{\gamma}\,\gamma^5 \Sigma^0(x)],\nonumber\\
\hspace{-0.3in}&&[\bar{n}(x) \Sigma^0(x)] [\bar{p}(x)\gamma^5 n(x)]
\to +\,\frac{1}{4}\, [\bar{p}(x)\vec{\gamma}\, n^c(x)]\cdot
[\bar{n^c}(x) \vec{\gamma}\,\gamma^5 \Sigma^0(x)]\nonumber\\
\hspace{-0.3in}&&\hspace{2in} + \,\frac{1}{4}\,[\bar{p}(x)
\vec{\gamma}\,n^c(x)] \cdot [\bar{n^c}(x)
\gamma^0\vec{\gamma}\,\gamma^5 \Sigma^0(x)].
\end{eqnarray}
These relations make a projection of the four--baryon states onto the
${^3}{\rm S}_1 \otimes {^3}{\rm P}_1$ and ${^3}{\rm S}_1 \otimes
{^1}{\rm P}_1$ states. Taking into account the relations
(\ref{label7.5}) we can extract from the effective Lagrangian
(\ref{label7.3}) the effective interactions, responsible for the
transitions $(n\Sigma^0)_{{^3}{\rm P}_1} \to K^-(pn)_{{^3}{\rm S}_1}$
and $(n\Sigma^0)_{{^1}{\rm P}_1} \to K^-(pn)_{{^3}{\rm S}_1}$ with the
$n\Sigma^0$ pair in the $ {^3}{\rm P}_1$ and ${^1}{\rm P}_1$ state.

\subsection{Reaction $(n\Sigma^0)_{{^3}{\rm P}_1} \to 
K^- (pn)_{{^3}{\rm S}_1}$}

The amplitude of the reaction $(n\Sigma^0)_{{^3}{\rm P}_1}\to K^-
(pn)_{{^3}{\rm S}_1}$, where the $n\Sigma^0$ pair in the $ {^3}{\rm
P}_1$ state couples to the $np$ pair in the ${^3}{\rm S}_1$ state, is
defined as
\begin{eqnarray}\label{label7.6}
\hspace{-0.3in}&&M(n(\vec{k},\alpha_1) \Sigma^0(- \vec{k},\alpha_2)
\to K^-(\vec{0}\,) p(\vec{K},\sigma_p) n( - \vec{K},\sigma_n);
{^3}{\rm P}_1) = i\,C^{\,(n\Sigma^0; {^3}{\rm P}_1)}_{K^-(pn; {^3}{\rm
S}_1)}\,\nonumber\\ \hspace{-0.3in}&&\times\,
\frac{[\bar{u}(\vec{K},\sigma_p) \vec{\gamma} u^c( -
\vec{K},\sigma_n)]\cdot [\bar{u^c}(\vec{k},\alpha_1) \vec{\gamma}
\gamma^5u(- \vec{k},\alpha_2)] }{\displaystyle 1 -
\frac{1}{2}\,r^t_{np} a^t_{np} K^2 - i\,a^t_{np} K}\,f^{(n \Sigma^0;
{^3}{\rm P}_1)}_{K^-(pn; {^3}{\rm
S}_1)}(k_0),
\end{eqnarray}
where $f^{(n\Sigma^0; {^3}{\rm P}_1)}_{K^-(pn; {^3}{\rm S}_1)}(k_0)$
is the amplitude, describing the $n\Sigma^0$ rescattering in the
${^3}{\rm P}_1$ state near threshold of the $K^-(p n)_{{^3}{\rm S}_1}$
system production, and $C^{\,(n\Sigma^0; {^3}{\rm P}_1)}_{K^-(pn;
{^3}{\rm S}_1)}$ is the effective coupling constant of the transition
$(n\Sigma^0)_{{^3}{\rm P}_1} \to K^- (pn)_{{^3}{\rm S}_1}$.

Using (\ref{label7.3}) and (\ref{label7.4}) we obtain the effective
Lagrangian of the transition $(n\Lambda^0)_{{^3}{\rm P}_1} \to K^-
(pn)_{{^3}{\rm S}_1}$ near threshold:
\begin{eqnarray}\label{label7.7}
{\cal L}^{(n\Sigma^0; {^3}{\rm P}_1)\to K^-(pn; {^3}{\rm S}_1)}_{\rm
eff}(x) = i\,C^{\,(n\Sigma^0; {^3}{\rm P}_1)}_{K^-(pn; {^3}{\rm
S}_1)}\, K^{-\dagger}(x)\, [\bar{p}(x)\vec{\gamma}\, n^c(x)] \cdot
[\bar{n^c}(x) \vec{\gamma}\, \gamma^5 \Sigma^0(x)].
\end{eqnarray}
The effective coupling constant $C^{\,(n\Sigma^0; {^3}{\rm
P}_1)}_{K^-(pn; {^3}{\rm S}_1)}$ is equal to 
\begin{eqnarray}\label{label7.8}
\hspace{-0.3in}C^{\,(n\Sigma^0; {^3}{\rm P}_1)}_{K^-(pn; {^3}{\rm
S}_1)} &=& \frac{1}{2}\,\frac{(2\alpha - 1)^3\,g^3_{\pi NN}}{m^2_K -
(E_{\Sigma^0} - m_N)^2 + k^2_0}\,\frac{1}{m_{\Sigma} + m_N +
m_K}\nonumber\\
\hspace{-0.3in}&+&\frac{1}{4} \,\frac{(2\alpha - 1)^3\,g^3_{\pi
NN}}{m^2_K - (E_{\Sigma^0} - m_N)^2 + k^2_0}\,\frac{1}{m_{\Sigma} +
m_N + m_K}\nonumber\\
\hspace{-0.3in}&-&\,\frac{1}{12}\,\frac{(2\alpha - 1)\,(3 -
2\alpha)^2\,g^3_{\pi NN}}{m^2_K - (E_{\Sigma^0} - m_N)^2 +
k^2_0}\,\frac{1}{m_{\Lambda^0} + m_N + m_K}\nonumber\\
\hspace{-0.3in}&-& \,\frac{(1 - \alpha)\,(2\alpha - 1)\,g^3_{\pi
NN}}{m^2_{\pi} - (E_N - m_N)^2 + k^2_0}\,\frac{1}{m_{\Sigma} + m_N +
m_K}\nonumber\\
\hspace{-0.3in}&-& \,\frac{1}{2}\,\frac{(2\alpha - 1)\,g^3_{\pi
NN}}{m^2_{\pi} - (E_N - m_N)^2 + k^2_0}\,\frac{m_N + m_K -
m_{\Sigma}}{m^2_N - (E_{\Sigma^0} - m_K)^2 + k^2_0}\nonumber\\
\hspace{-0.3in}&-&\,\frac{1}{4}\,\frac{(2\alpha - 1)\,g^3_{\pi
NN}}{m^2_{\pi} - (E_N - m_N)^2 + k^2_0}\,\frac{m_N + m_K -
m_{\Sigma}}{m^2_N - (E_{\Sigma^0} - m_K)^2 + k^2_0}\nonumber\\
\hspace{-0.3in}&+&\,\frac{1}{6}\,\frac{\alpha\,(3 -
2\alpha)\,g^3_{\pi NN}}{m^2_{\pi} - (E_N - m_N)^2 + k^2_0}\,\frac{m_N
+ m_K - m_{\Sigma}}{m^2_N - (E_{\Sigma^0} - m_K)^2 + k^2_0}\nonumber\\
\hspace{-0.3in}&+& \,\frac{1}{6}\,\frac{\alpha\,(2\alpha - 1)\,(3 -
2\alpha)\,g^3_{\pi NN}}{m^2_{\eta} - (E_N - m_N)^2 +
k^2_0}\,\frac{1}{m_{\Sigma} + m_N + m_K}\nonumber\\ \hspace{-0.3in}&+&
\,\frac{1}{12}\,\frac{(3 - 4\alpha)^2\,g^3_{\pi NN}}{m^2_{\eta} - (E_N
- m_N)^2 + k^2_0}\,\frac{m_N + m_K - m_{\Sigma}}{m^2_N - (E_{\Sigma^0}
- m_K)^2 + k^2_0}\nonumber\\ &-&\,\frac{1}{8}\,\frac{1}{g_A
F_{\pi}}\,\frac{(2\alpha - 1)\,g^2_{\pi NN}}{m^2_K - (E_{\Sigma^0} -
m_N)^2 + k^2_0}\nonumber\\ &+&\, \frac{1}{8}\,\frac{1}{g_A
F_{\pi}}\,\frac{(2\alpha - 1)\,g^2_{\pi NN}}{m^2_{\pi} -
(E_N - m_N)^2 + k^2_0}\nonumber\\ &+&\,
\frac{1}{24}\,\frac{1}{g_A F_{\pi}}\,\frac{(2\alpha - 1)\,(3 -
4\alpha)\,g^2_{\pi NN}}{m^2_{\eta} - (E_N - m_N)^2 +
k^2_0}\nonumber\\ &-&\,\frac{1}{4}\,\frac{1}{g_A
F_{\pi}}\,\frac{(2\alpha - 1)\,g^2_{\pi NN}}{m^2_{\pi} -
(E_N - m_N)^2 + k^2_0} = -\,7\times 10^{-7}\,{\rm MeV}^{-3},
\end{eqnarray}
where $E_{\Sigma^0} = 1302\,{\rm MeV}$, $E_N = 1072\,{\rm MeV}$ and
$k_0 = 516\,{\rm MeV}$.

The amplitude $f^{\,(n\Sigma^0; {^3}{\rm P}_1)}_{K^- (pn; {^3}{\rm
S}_1)}(k_0)$, describing the rescattering of the $n\Sigma^0$ pair in
the ${^3}{\rm P}_1$ state near threshold of the $K^- (p n)_{{^3}{\rm
S}_1}$ system production, is defined by the Feynman diagrams depicted
in Fig.6 with a replacement $\Lambda^0 \to \Sigma^0$ and reads (see
(\ref{label6.12}))
\begin{eqnarray}\label{label7.9}
\Big|f^{\,(n\Sigma^0; {^3}{\rm P}_1)}_{K^-(pn; {^3}{\rm
S}_1)}(k_0)\Big| = \Big|\Big\{1 - \frac{C_{n\Sigma^0}({^3}{\rm
P}_1)}{12\pi^2}\,\frac{k^3_0}{E(k_0)}\,\Big[{\ell
n}\Big(\frac{E(k_0) + k_0}{E(k_0) - k_0} \Big) -
i\,\pi\Big]\Big\}^{-1}\Big| \simeq 1,
\end{eqnarray}
where $E(k_0) = \sqrt{k^2_0 + m^2_B}$\,\footnote{For simplicity we use
the equal masses of baryons for the calculation of the rescattering of
the $n\Sigma^0$ pair, where $m_B = \sqrt{(2 m_N + m_K)^2 - 4 k^2_0}/2
= 1070\,{\rm MeV}$.} and the effective coupling constant
$C_{n\Sigma^0}({^3}{\rm P}_1)$ is equal to 
\begin{eqnarray}\label{label7.10}
C_{n\Sigma^0}({^3}{\rm P}_1) &=& (2\alpha - 1)^2\,\frac{g^2_{\pi
NN}}{4 k^2_0}\,{\ell n}\Big(1 + \frac{4 k^2_0}{~m^2_K}\Big) + \alpha
(3 -4\alpha)\,\frac{g^2_{\pi NN}}{6 k^2_0}\,{\ell n}\Big(1 + \frac{4
k^2_0}{~m^2_{\eta}}\Big) = \nonumber\\ &=& 0.7\times 10^{-4}\,{\rm
MeV}^{-2}.
\end{eqnarray}
The rescattering of the $n\Sigma^0$ pair in the ${^3}{\rm P}_1$ state
is defined by the interaction, computed in the one--meson exchange
approximation (see \cite{AI01}):
\begin{eqnarray}\label{label7.11}
{\cal L}^{(n\Sigma^0; {^3}{\rm P}_1) \to (n\Sigma^0; {^3}{\rm
P}_1)}_{\rm eff}(x) = -\,\frac{1}{4}\,C_{n\Sigma^0}({^3}{\rm
P}_1))\,[\bar{\Sigma}^0(x) \vec{\gamma}\,\gamma^5 n^c(x)]\cdot
[\bar{n^c}(x)\vec{\gamma}\,\gamma^5 \Sigma^0(x)].
\end{eqnarray}
In our approach the effective Lagrangian (\ref{label7.11}) describes
also the final--state $(n\Sigma^0)_{{^3}{\rm P}_1}$ interaction near
threshold of the reaction $K^- (pn)_{{^3}{\rm S}_1} \to
(n\Sigma^0)_{{^3}{\rm P}_1}$.

\subsection{S--wave amplitude $\tilde{f}^{\,K^-d}_0(0)_{(n\Sigma^0; 
{^3}{\rm P}_1)}$ of $K^-d$ scattering}

The amplitude $\tilde{f}^{\,K^-d}_0(0)_{(n\Sigma^0; {^3}{\rm P}_1)}$
can be computed similar to $\tilde{f}^{\,K^-d}_0(0)_{(n\Lambda^0;
{^3}{\rm P}_1)}$. The summation over polarisations of the coupled
baryons is defined by (\ref{label6.16}). This gives the imaginary part
of the amplitude $\tilde{f}^{\,K^-d}_0(0)_{(n\Sigma^0; {^3}{\rm
P}_1)}$ equal to
\begin{eqnarray}\label{label7.12}
{\cal I}m\,\tilde{f}^{\,K^-d}_0(0)_{(n\Sigma^0; {^3}{\rm P}_1)} &=&
4.6\times 10^{-3}\, \frac{1}{3\pi^2}\,\frac{m^3_{\pi}}{1 +
m_K/m_d}\,\frac{k^3_0}{2m_N + m_K}\, \nonumber\\
&&\times\,[C^{\,(n\Sigma^0; {^3}{\rm P}_1)}_{K^-pn}]^2\,
|f^{(n\Sigma^0; {^3}{\rm P}_1)}_{K^-(pn; {^3}{\rm S}_1)}(k_0)|^2 = 1.9
\times 10^{-3}\,{\rm fm}.
\end{eqnarray}
The real part of $\tilde{f}^{\,K^-d}_0(0)_{(n\Sigma^0; {^3}{\rm
P}_1)}$ reads
\begin{eqnarray}\label{label7.13}
\hspace{-0.5in}&&{\cal R}e\,\tilde{f}^{\,K^-d}_0(0)_{(n\Sigma^0;
{^3}{\rm P}_1)} = 4.6\times 10^{-3}\,
\frac{1}{12\pi^2}\,\frac{m^3_{\pi}}{1 + m_K/m_d}\,\frac{k^2_0}{1 +
m_K/2m_B}\,[C^{\,(n\Sigma^0; {^3}{\rm P}_1)}_{K^-(pn; {^3}{\rm
S}_1)}]^2\nonumber\\
\hspace{-0.5in}&&\times\, |f^{(n\Sigma^0; {^3}{\rm P}_1)}_{K^-(pn;
{^3}{\rm
S}_1)}(k_0)|^2\,F\Big(\frac{\Lambda}{m_B},\frac{k_0}{m_B}\Big) =
\,0.05\times 10^{-3}\,{\rm fm},
\end{eqnarray}
where $\Lambda = m_N$ and $m_B = \sqrt{(2 m_N + m_K)^2 -4k^2_0}/2 =
1070\,{\rm MeV}$.

The S--wave amplitude $\tilde{f}^{\,K^-d}_0(0)_{(n\Sigma^0; {^3}{\rm
P}_1)}$ of $K^-d$ scattering near threshold, caused by the two--body
inelastic channel $K^-(pn)_{{^3}{\rm S}_1} \to (n \Sigma^0)_{{^3}{\rm
P}_1}\to K^-(pn)_{{^3}{\rm S}_1} $, is given by
\begin{eqnarray}\label{label7.14}
\tilde{f}^{\,K^-d}_0(0)_{(n \Sigma^0; {^3}{\rm P}_1)} = (0.05 +
i\,1.90) \times 10^{-3}\,{\rm fm}.
\end{eqnarray}
Now we proceed to computing the S--wave amplitude
$\tilde{f}^{\,K^-d}_0(0)_{(n\Sigma^0; {^1}{\rm P}_1)}$ of $K^-d$
scattering near threshold, saturated by the intermediate
$(n\Sigma^0)_{{^1}{\rm P}_1}$ state.

\subsection{\bf Reaction $(n\Sigma^0)_{{^1}{\rm P}_1} \to K^-
(pn)_{{^3}{\rm S}_1}$}

The amplitude of the reaction $(n \Sigma^0)_{{^1}{\rm P}_1} \to K^-
(pn)_{{^3}{\rm S}_1}$ is defined by
\begin{eqnarray}\label{label7.15}
\hspace{-0.3in}&&M(n(\vec{k},\alpha_1) \Sigma^0(- \vec{k},\alpha_2)
\to K^-(\vec{0}\,) p(\vec{K},\sigma_p) n( - \vec{K},\sigma_n);
{^1}{\rm P}_1) = i\,C^{\,(n\Sigma^0; {^1}{\rm P}_1)}_{K^- (pn;
{^3}{\rm S}_1)}\,\nonumber\\
\hspace{-0.3in}&&\times\, \frac{[\bar{u}(\vec{K},\sigma_p)
\vec{\gamma} u^c( - \vec{K},\sigma_n)]\cdot
[\bar{u^c}(\vec{k},\alpha_1) \vec{\gamma} \gamma^5u(-
\vec{k},\alpha_2)] }{\displaystyle 1 - \frac{1}{2}\,r^t_{np} a^t_{np}
K^2 + i\,a^t_{np} K}\,f^{(n\Sigma^0; {^1}{\rm P}_1)}_{K^- (pn;
{^3}{\rm S}_1)}(k_0).
\end{eqnarray}
The effective Lagrangian of the transition $(n\Sigma^0)_{{^1}{\rm
P}_1} \to K^- (pn)_{{^3}{\rm S}_1}$ at threshold can be defined by
\begin{eqnarray}\label{label7.16}
{\cal L}^{\,(n\Sigma^0; {^1}{\rm P}_1) \to K^-(pn; {^3}{\rm
S}_1)}_{\rm eff}(x) = i\,C^{\,(n\Sigma^0; {^1}{\rm P}_1)}_{K^- (pn;
{^3}{\rm S}_1)}\, K^{-\dagger}(x)\, [\bar{p}(x)\vec{\gamma}\, n^c(x)]
\cdot [\bar{n^c}(x) \gamma^0\vec{\gamma}\, \gamma^5 \Sigma^0(x)].
\end{eqnarray}
Using the effective Lagrangians (\ref{label7.3}) and (\ref{label7.4}),
and the prescription for the projection of the four--baryon operators
onto the ${^3}{\rm S}_1 \otimes {^3}{\rm P}_1$ and ${^3}{\rm S}_1
\otimes {^1}{\rm P}_1$ states (\ref{label7.5}) we obtain the effective
coupling constant $C^{\,(n\Sigma^0; {^1}{\rm P}_1)}_{K^- (pn; {^3}{\rm
S}_1)}$ equal to
\begin{eqnarray}\label{label7.17}
C^{\,(n\Lambda^0; {^1}{\rm P}_1)}_{K^- (pn; {^3}{\rm S}_1)} =
-\,2\times 10^{-7}\,{\rm MeV}^{-3}.
\end{eqnarray}
The Lagrangian (\ref{label7.16}) describes the interaction of the
$n\Sigma^0$ pair in the ${^1}{\rm P}_1$ state with the $np$ pair in
the ${^1}{\rm S}_1$ through the emission of the $K^-$--meson.

\subsection{ Amplitude of $(n\Sigma^0)_{{^1}{\rm P}_1}$
rescattering}

The amplitude $f^{\,(n\Sigma^0; {^1}{\rm P}_1)}_{K^-pn}(k_0)$,
describing the rescattering of the $n\Sigma^0$ pair in the ${^1}{\rm
P}_1$ state near threshold of the $K^-p n$ system production, is
defined by the Feynman diagrams analogous to those depicted in Fig.6.
The result of the calculation reads (see Appendix E)
\begin{eqnarray}\label{label7.18}
\Big|f^{\,(n\Sigma^0; {^1}{\rm P}_1)}_{K^-(pn; {^3}{\rm
S}_1)}(k_0)\Big| = \Big|\Big\{1 - \frac{C_{n\Sigma^0}({^1}{\rm
P}_1)}{24\pi^2}\,\frac{k^3_0}{E(k_0)}\,\Big[{\ell n}\Big(\frac{E(k_0)
+ k_0}{E(k_0) - k_0} \Big) - i\,\pi\Big]\Big\}^{-1}\Big| \simeq 1,
\end{eqnarray}
where $E(k_0) = \sqrt{k^2_0 + m^2_B}$\,\footnote{For simplicity we use
the equal masses of baryons for the calculation of the rescattering of
the $n\Lambda^0$ pair, where $m_B = \sqrt{(2 m_N + m_K)^2 - 4 k^2_0}/2
= 1070\,{\rm MeV}$.} and the effective coupling constant
$C_{(n\Lambda^0)}({^1}{\rm P}_1)$ is equal to 
\begin{eqnarray}\label{label7.19}
C_{n\Sigma^0}({^3}{\rm P}_1) = C_{(n\Lambda^0)}({^1}{\rm P}_1) =
0.7\times 10^{-4}\,{\rm MeV}^{-2}.
\end{eqnarray}
The rescattering of the $n\Sigma^0$ pair in the ${^3}{\rm P}_1$ state
is defined by the interaction
\begin{eqnarray}\label{label7.20}
{\cal L}^{(n\Sigma^0; {^3}{\rm P}_1) \to (n\Sigma^0; {^3}{\rm
P}_1)}_{\rm eff}(x) = -\,\frac{1}{4}\,C_{n\Sigma^0}({^3}{\rm
P}_1))\,[\bar{\Sigma}^0(x) \vec{\gamma}\,\gamma^5 n^c(x)]\cdot
[\bar{n^c}(x)\vec{\gamma}\,\gamma^5 \Sigma^0(x)].
\end{eqnarray}
The amplitude $\tilde{f}^{\,K^-d}_0(0)_{(n\Sigma^0; {^1}{\rm P}_1)}$,
caused by the intermediate $(n\Sigma^0)_{{^1}{\rm P}_1}$ state we
define as follows.

\subsection{S--wave amplitude
$\tilde{f}^{\,K^-d}_0(0)_{(n\Sigma^0; {^1}{\rm P}_1)}$ of $K^-d$
scattering}

The result of the summation over polarisations of interacting baryons
is given by (\ref{label6.31}). Hence, at threshold the imaginary part
of the S--wave amplitude of $K^-d$ scattering, caused by the
intermediate $(n\Sigma^0)_{{^3}{\rm P}_1}$ state, is equal to
\begin{eqnarray}\label{label7.21}
{\cal I}m\,\tilde{f}^{\,K^-d}_0(0)_{(n\Sigma^0; {^1}{\rm P}_1)} &=&
4.6\times 10^{-3}\, \frac{1}{6\pi^2}\,\frac{m^3_{\pi}}{1 +
m_K/m_d}\,\frac{k^3_0}{2m_N + m_K}\, \nonumber\\
&&\times\,[C^{\,(n\Sigma^0; {^1}{\rm P}_1)}_{K^-(pn; {^3}{\rm
S}_1)}]^2\, |f^{(n\Sigma^0; {^1}{\rm P}_1)}_{K^-(pn; {^3}{\rm
S}_1)}(k_0)|^2 = 0.1\times 10^{-3}\,{\rm fm}.
\end{eqnarray}
The real part of $\tilde{f}^{\,K^-d}_0(0)_{(n\Lambda^0; {^1}{\rm
P}_1)}$ reads
\begin{eqnarray}\label{label7.22}
\hspace{-0.5in}&&{\cal
R}e\,\tilde{f}^{\,K^-d}_0(0)_{(n\Lambda^0; {^1}{\rm P}_1)} =
4.6\times 10^{-3}\, \frac{1}{24\pi^2}\,\frac{m^3_{\pi}}{1 +
m_K/m_d}\,\frac{k^2_0}{1 + m_K/2m_B}\,[C^{\,(n\Lambda^0; {^1}{\rm
P}_1)}_{K^-(pn; {^3}{\rm S}_1)}]^2
\nonumber\\
\hspace{-0.5in}&&\times\,|f^{(n\Lambda^0; {^1}{\rm P}_1)}_{K^-(pn;
{^3}{\rm
S}_1)}(k_0)|^2\,F\Big(\frac{\Lambda}{m_B},\frac{k_0}{m_B}\Big) =
\,2\times 10^{-6}\,{\rm fm}.
\end{eqnarray}
Thus, the S--wave amplitude $\tilde{f}^{\,K^-d}_0(0)_{(n\Sigma^0;
{^1}{\rm P}_1)}$ of $K^-d$ scattering, caused by the two--body
inelastic channel $K^- (pn)_{{^3}{\rm S}_1} \to (n\Sigma^0)_{{^1}{\rm
P}_1} \to K^- (pn)_{{^3}{\rm S}_1}$, is equal to
\begin{eqnarray}\label{label7.23}
\tilde{f}^{\,K^-d}_0(0)_{(n\Sigma^0; {^1}{\rm P}_1)} = (0.02 +
i\,1.00)\times 10^{-4}\,{\rm fm}.
\end{eqnarray}
Now we can estimate the contribution of the two--body inelastic
channel $K^- (pn)_{{^3}{\rm S}_1} \to n \Sigma^0 \to K^-
(pn)_{{^3}{\rm S}_1}$ to the S--wave amplitude $\tilde{f}^{K^-d}_0(0)$
of $K^-d$ scattering near threshold and the energy level displacement
of the ground state of kaonic deuterium.

\subsection{S--wave amplitude $f^{K^-d}_0(0)_{n\Sigma^0}$ 
and the energy level displacement}

The S--wave amplitude of $K^-d$ scattering at threshold, saturated by
the inelastic channel $K^- (pn)_{{^3}{\rm S}_1} \to n\Sigma^0 \to K^-
(pn)_{{^3}{\rm S}_1}$ with the $n\Sigma^0$ pair in the ${^3}{\rm P}_1$
and ${^1}{\rm P}_1$ state, is equal to the sum of the contributions
(\ref{label7.14}) and (\ref{label7.23})
\begin{eqnarray}\label{label7.24}
\tilde{f}^{\,K^-d}_0(0)_{n \Sigma^0} = (0.05 + i\,2.00)\times
10^{-3}\,{\rm fm}.
\end{eqnarray}
The contribution of the decay $A_{Kd} \to n \Lambda^0$ to the energy
level displacement of the ground state of kaonic deuterium amounts to
\begin{eqnarray}\label{label7.25}
-\,\epsilon^{(n\Sigma^0)}_{1s} +
i\,\frac{\Gamma^{(n\Sigma^0)}_{1s}}{2} =
602\,\tilde{f}^{\,K^-d}_0(0)_{n\Sigma^0} = (0.03 + i\,1.2)\,{\rm eV}.
\end{eqnarray}
Hence, the partial width of the decay $A_{Kd} \to n \Lambda^0$ is
equal to $\Gamma^{(n\Sigma^0)}_{1s} = 2.4\,{\rm eV}$.

According to \cite{VV70}, the experimental rate of the production of
the $n\Sigma^0$ pair at threshold of the reaction $K^-d \to n\Sigma^0$
is equal to $R(K^-d \to n \Sigma^0) = (0.337 \pm 0.070)\,\%$.

Using our estimate of the partial width, $\Gamma^{(n\Sigma^0)}_{1s} =
(2.40\pm 0.48)\,{\rm eV}$, where $\pm 0.48\,{\rm eV}$ is a
theoretical accuracy of the result, and the experimental rate,
$R(K^-d \to n \Sigma^0) = (0.337 \pm 0.070)\,\%$, we can estimate the
expected value of the total width of the energy level of the ground
state of kaonic deuterium
\begin{eqnarray}\label{label7.26}
\Gamma_{1s} = \frac{\Gamma^{(n\Sigma^0)}_{1s} }{(0.337 \pm
0.070)\times 10^{-2}} = (700 \pm 200)\,{\rm eV}.
\end{eqnarray}
This value is compared well with our estimate $\Gamma_{1s} = (570 \pm
130)\,{\rm eV}$, which we have made in Section 6 using the theoretical
value of the partial width of the decay $A_{Kd} \to n\Lambda^0$ and
the experimental rate of the $ n \Lambda^0$ production in the reaction
$K^-d \to n \Lambda^0$.

\section{Amplitude of reaction $K^- (p n)_{{^3}{\rm S}_1}
 \to p \Sigma^- \to K^- (p n)_{{^3}{\rm S}_1}$ and the energy level
 displacement} 
\setcounter{equation}{0}

The amplitudes of the reactions $K^- (pn)_{{^3}{\rm S}_1} \to (p
\Sigma^-)_X $, where the $p \Sigma^-$ pair couples in the $X =
{^3}{\rm P}_1$ and ${^1}{\rm P}_1$ state, we define as
\begin{eqnarray}\label{label8.1}
\hspace{-0.3in}&&M(K^-(\vec{0}\,) p(\vec{K},\sigma_p) n( -
\vec{K},\sigma_n) \to p(\vec{k},\alpha_1) \Sigma^-(-
\vec{k},\alpha_2); {^3}{\rm P}_1) = -\,i\,C^{\,(p\Sigma^-; {^3}{\rm
P}_1)}_{K^-(pn; {^3}{\rm S}_1)}\,\nonumber\\
\hspace{-0.3in}&&\times\, \frac{[\bar{u^c}(- \vec{K},\sigma_n)
\vec{\gamma} u( \vec{K},\sigma_p)]\cdot [\bar{u}(- \vec{k},\alpha_2)
\vec{\gamma} \gamma^5u^c ( \vec{k},\alpha_1)] }{\displaystyle 1 -
\frac{1}{2}\,r^t_{np} a^t_{np} K^2 + i\,a^t_{np} K}\,f^{(p\Sigma^-;
{^3}{\rm P}_1)}_{K^-(pn; {^3}{\rm S}_1)}(k_0),\nonumber\\
\hspace{-0.3in}&&M(K^-(\vec{0}\,) p(\vec{K},\sigma_p) n( -
\vec{K},\sigma_n) \to p(\vec{k},\alpha_1) \Sigma^-(-
\vec{k},\alpha_2); {^1}{\rm P}_1) = -\,i\,C^{\,(p\Sigma^-; {^1}{\rm
P}_1)}_{K^-(pn; {^3}{\rm S}_1)}\,\nonumber\\
\hspace{-0.3in}&&\times\, \frac{[\bar{u^c}(- \vec{K},\sigma_n)
\vec{\gamma} u( \vec{K},\sigma_p)]\cdot [\bar{u}(- \vec{k},\alpha_2)
\gamma^0 \vec{\gamma} \gamma^5u^c ( \vec{k},\alpha_1)] }{\displaystyle
1 - \frac{1}{2}\,r^t_{np} a^t_{np} K^2 + i\,a^t_{np}
K}\,f^{(p\Sigma^-; {^1}{\rm P}_1)}_{K^-(pn; {^3}{\rm S}_1)}(k_0),
\end{eqnarray}
where $f^{(p\Sigma^-; X)}_{K^-(pn; {^3}{\rm S}_1)}(k_0)$ is the
amplitude of the final--state $p\Sigma^-$ interaction near threshold
of the reaction $K^- (pn)_{{^3}{\rm S}_1} \to (p \Sigma^-)_X$ and
$C^{\,(p\Sigma^0; X)}_{K^-(pn; {^3}{\rm S}_1)}$ is the effective
coupling constant of the transition $ K^- (pn)_{{^3}{\rm S}_1} \to (p
\Sigma^-)_X$, where $X = {^3}{\rm P}_1$ or ${^1}{\rm P}_1$.

\subsection{Effective coupling constant
$C^{\,p\Sigma^-}_{K^-pn}$} 

The transition $p \Sigma^- \to K^- p n$, induced by the
one--pseudoscalar meson exchange, is defined by the set of Feynman
diagrams depicted in Fig.9. The Feynman diagrams for the transition $p
\Sigma^- \to K^- p n$, determined in the one--pseudoscalar meson
approximation with scalar--meson exchanges, we adduce in Fig.10.
\begin{figure}
\centering \psfrag{K-}{$K^-$} \psfrag{K0b}{$\bar{K}^0$}
\psfrag{L0}{$\Lambda^0$} \psfrag{p+}{$\pi^+$}\psfrag{p-}{$\pi^-$} 
\psfrag{p0}{$\pi^0$}
\psfrag{S-}{$\Sigma^-$} \psfrag{S0}{$\Sigma^0$}
\psfrag{Sp}{$\Sigma^+$}\psfrag{p}{$p$} \psfrag{n}{$n$}
\psfrag{e}{$\eta$}
\includegraphics[height=0.90\textheight]{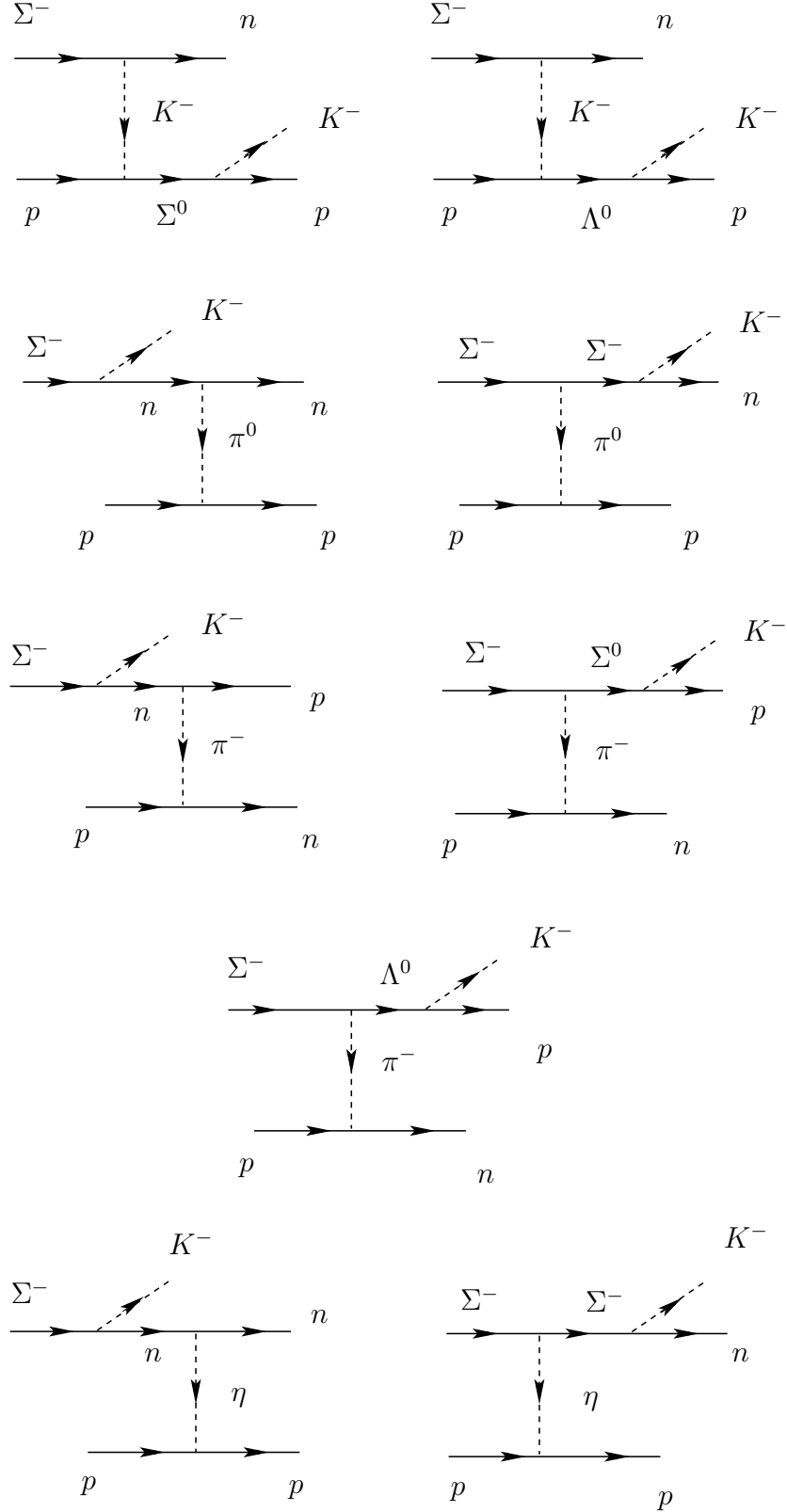}
\caption{Feynman diagrams describing the effective coupling constant
$C^{\,p\Sigma^-}_{K^-pn}$ of the transition $p\Sigma^- \to K^- p n$ in
the one--pseudoscalar meson exchange approximation.}
\end{figure}
\begin{figure}
\centering \psfrag{K-}{$K^-$}
 \psfrag{S-}{$\Sigma^-$}
\psfrag{L0}{$\Lambda^0$} 
\psfrag{p+}{$\pi^+$} \psfrag{p0}{$\pi^0$}
\psfrag{p}{$p$} \psfrag{n}{$n$}
\psfrag{s}{$\sigma$}
\psfrag{e}{$\eta$}
\psfrag{k-}{$\kappa^-$}
\psfrag{k0}{$\bar{\kappa}^0$}
\includegraphics[height=0.30\textheight]{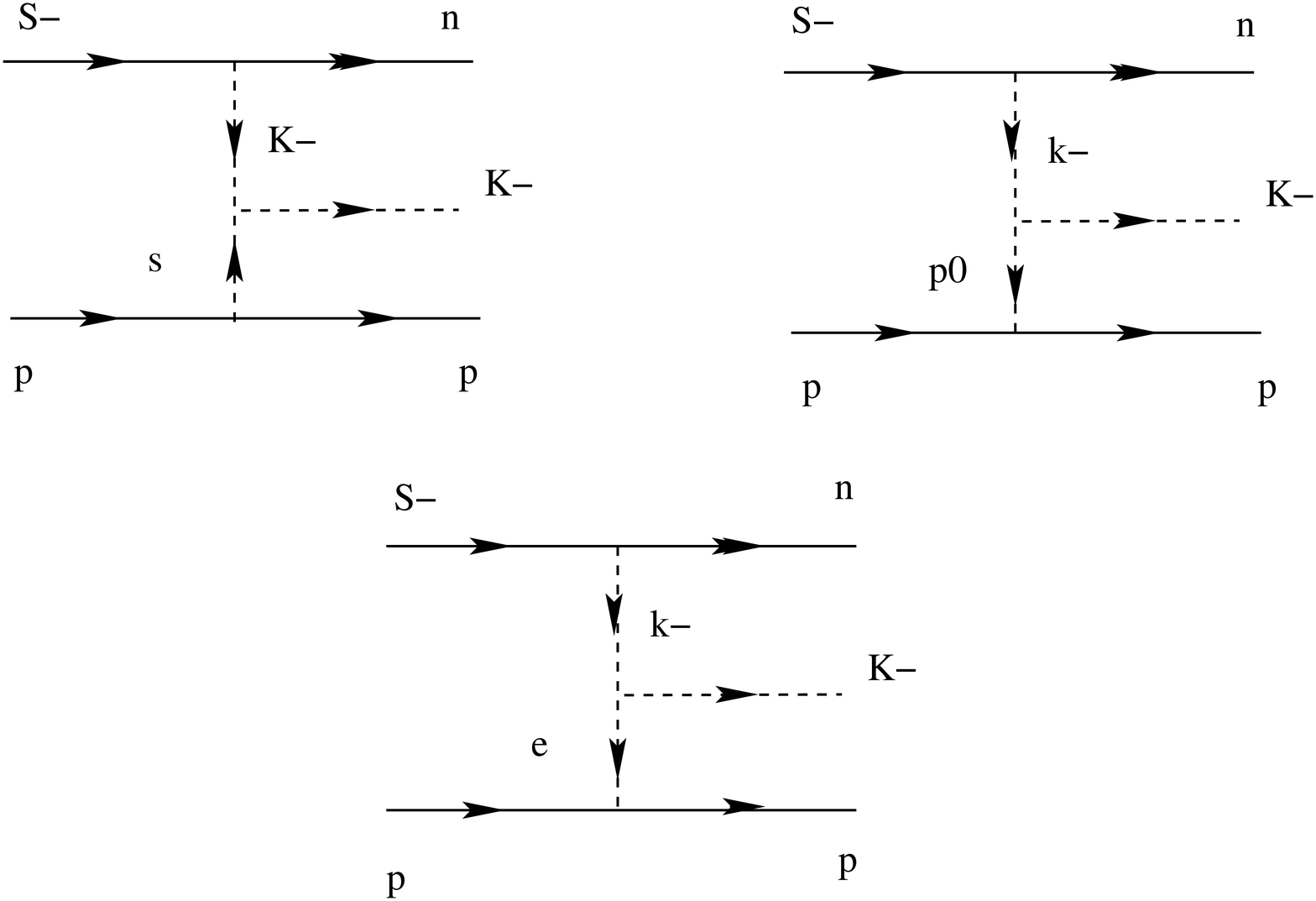}
\caption{Feynman diagrams describing the contribution of scalar mesons
to the effective coupling constant of the transition $p \Sigma^- \to
K^- p n$ in the one--pseudoscalar meson exchange approximation.}
\end{figure}

In the momentum representation the amplitude of the transition $p
\Sigma^- \to K^- p n$, determined by the Feynman diagrams in Fig.9,
reads
\begin{eqnarray}\label{label8.2}
&&{\cal M}( p(\vec{k},\alpha_1) \Sigma^-(- \vec{k},\alpha_2) \to
K^-(\vec{0}\,) p(\vec{K},\sigma_p) n( - \vec{K},\sigma_n))_P
=\nonumber\\ &&= \sqrt{2}\,(2\alpha - 1)^3\,g^3_{\pi NN}\times
\Big[\bar{u}(k_p,\sigma_p)i\gamma^5\frac{1}{m_{\Sigma^0} - \hat{k}_p -
\hat{Q}_K}i\gamma^5 u(q_p,\alpha_p)\Big]\nonumber\\
&&\times\,\frac{1}{m^2_K - (q_{\Sigma^-} - k_n)^2}\times
[\bar{u}(k_n,\sigma_n)i\gamma^5
u(q_{\Sigma^-},\alpha_{\Sigma^-})]\nonumber\\ &&-
\,\frac{\sqrt{2}}{3}\,(2\alpha - 1)\,(3- 2\alpha)^2\,g^3_{\pi
NN}\times \Big[\bar{u}(k_p,\sigma_p)i\gamma^5\frac{1}{m_{\Lambda^0} -
\hat{k}_p - \hat{Q}_K}i\gamma^5 u(q_p,\alpha_p)\Big]\nonumber\\
&&\times\,\frac{1}{m^2_K - (q_{\Sigma^-} - k_n)^2}\times
[\bar{u}(k_n,\sigma_n)i\gamma^5
u(q_{\Sigma^-},\alpha_{\Sigma^-})]\nonumber\\ &&- \,\sqrt{2}\,(2\alpha
- 1)\,g^3_{\pi NN}\times
\Big[\bar{u}(k_n,\sigma_n)i\gamma^5\frac{1}{m_n - \hat{q}_{\Sigma^-} +
\hat{Q}_K}i\gamma^5 u(q_{\Sigma^-},\alpha_{\Sigma^-})\Big]\nonumber\\
&&\times\,\frac{1}{m^2_{\pi} - (q_p - k_p)^2}\times
[\bar{u}(k_p,\sigma_p)i\gamma^5 u(q_p,\alpha_p)]\nonumber\\ &&-
\,2\sqrt{2}\,(1 - \alpha)\,(2\alpha - 1)\,g^3_{\pi NN}\times
\Big[\bar{u}(k_n,\sigma_n)i\gamma^5\frac{1}{m_{\Sigma^-} - \hat{k}_n -
\hat{Q}_K}i\gamma^5 u(q_{\Sigma^-},\alpha_{\Sigma^-})\Big]\nonumber\\
&&\times\,\frac{1}{m^2_{\pi} - (q_p - k_p)^2}\times
[\bar{u}(k_p,\sigma_p)i\gamma^5 u(q_p,\alpha_p)]\nonumber\\ &&+
\,2\sqrt{2}\,(2\alpha - 1)\,g^3_{\pi NN}\times
\Big[\bar{u}(k_p,\sigma_p)i\gamma^5\frac{1}{m_n - \hat{q}_{\Sigma^-} +
\hat{Q}_K}i\gamma^5 u(q_{\Sigma^-},\alpha_{\Sigma^-})\Big]\nonumber\\
&&\times\,\frac{1}{m^2_{\pi} - (q_p - k_p)^2}\times
[\bar{u}(k_n,\sigma_n)i\gamma^5 u(q_p,\alpha_p)]\nonumber\\ &&+
\,2\sqrt{2}\,(1 - \alpha)\,(2\alpha - 1)\,g^3_{\pi NN}\times
\Big[\bar{u}(k_p,\sigma_p)i\gamma^5\frac{1}{m_{\Sigma^0} - \hat{k}_p -
\hat{Q}_K}i\gamma^5 u(q_{\Sigma^-},\alpha_{\Sigma^-})\Big]\nonumber\\
&&\times\,\frac{1}{m^2_{\pi} - (q_p - k_p)^2}\times
[\bar{u}(k_n,\sigma_n)i\gamma^5 u(q_p,\alpha_p)]\nonumber\\ &&-
\,\frac{2\sqrt{2}}{3}\,\alpha\,(3 - 2\alpha)\,g^3_{\pi NN}\times
\Big[\bar{u}(k_p,\sigma_p)i\gamma^5\frac{1}{m_{\Lambda^0} - \hat{k}_p
- \hat{Q}_K}i\gamma^5 u(q_{\Sigma^-},\alpha_{\Sigma^-})\Big]
\nonumber\\ &&\times\,\frac{1}{m^2_{\pi} - (q_p - k_p)^2}\times
[\bar{u}(k_n,\sigma_n)i\gamma^5 u(q_p,\alpha_p)]\nonumber\\ &&+
\,\frac{\sqrt{2}}{3}\,(2 \alpha - 1)\,g^3_{\pi NN}\times
\Big[\bar{u}(k_n,\sigma_n)i\gamma^5\frac{1}{m_n - \hat{q}_{\Sigma^-} +
\hat{Q}_K}i\gamma^5 u(q_{\Sigma^-},\alpha_{\Sigma^-})\Big] \nonumber\\
&&\times\,\frac{1}{m^2_{\eta} - (q_p - k_p)^2}\times
[\bar{u}(k_p,\sigma_p)i\gamma^5 u(q_p,\alpha_p)]\nonumber\\ &&+
\,\frac{2\sqrt{2}}{3}\,\alpha\,(2 \alpha - 1)\,(3 - 4\alpha)\,g^3_{\pi
NN}\times \Big[\bar{u}(k_n,\sigma_n)i\gamma^5\frac{1}{m_{\Sigma^-} -
\hat{k}_n - \hat{Q}_K}i\gamma^5
u(q_{\Sigma^-},\alpha_{\Sigma^-})\Big] \nonumber\\
&&\times\,\frac{1}{m^2_{\eta} - (q_p - k_p)^2}\times
[\bar{u}(k_p,\sigma_p)i\gamma^5 u(q_p,\alpha_p)].
\end{eqnarray}
Near threshold of the transition $p \Sigma^- \to K^- p n$ the set of
diagrams in Fig.9 we represent in the form of the local effective
Lagrangian
\begin{eqnarray}\label{label8.3}
\hspace{-0.3in}&&{\cal L}^{\,p\Sigma^- \to K^- pn}_{\rm eff}(x)_P
=\nonumber\\
\hspace{-0.3in}&&= -\,\sqrt{2}\,\frac{(2\alpha - 1)^3\,g^3_{\pi
NN}}{m^2_K - (E_{\Sigma^-} - m_N)^2 + k^2_0}\,\frac{1}{m_{\Sigma} +
m_N +
m_K}\,[\bar{n}(x)i\gamma^5\Sigma^-(x)][\bar{p}(x)p(x)]\nonumber\\
\hspace{-0.3in}&&+ \,\frac{\sqrt{2}}{3}\,\frac{(2\alpha - 1)\,(3-
2\alpha)^2\,g^3_{\pi NN}}{m^2_K - (E_{\Sigma^-} - m_N)^2 +
k^2_0}\,\frac{1}{m_{\Lambda^0} + m_N +
m_K}\,[\bar{n}(x)i\gamma^5\Sigma^-(x)][\bar{p}(x)p(x)]\nonumber\\ &&+
\,\sqrt{2}\,\frac{(2\alpha - 1)\,g^3_{\pi NN}}{m^2_{\pi} - (E_N -
m_N)^2 + k^2_0}\,\frac{m_N + m_K - m_{\Sigma}}{m^2_N - (E_{\Sigma^-} -
m_K)^2 + k^2_0}\,[\bar{n}(x)\Sigma^-(x)][\bar{p}(x)i\gamma^5
p(x)]\nonumber\\ &&+ \,2\sqrt{2}\,\frac{(1 - \alpha)\,(2\alpha -
1)\,g^3_{\pi NN}}{m^2_{\pi} - (E_N - m_N)^2 +
k^2_0}\,\frac{1}{m_{\Sigma} + m_N +
m_K}\,[\bar{n}(x)\Sigma^-(x)][\bar{p}(x)i\gamma^5 p(x)]\nonumber\\ &&-
\,2\sqrt{2}\,\frac{(2\alpha - 1)\,g^3_{\pi NN}}{m^2_{\pi} - (E_N -
m_N)^2 + k^2_0}\,\frac{m_N + m_K - m_{\Sigma}}{m^2_N - (E_{\Sigma^-} -
m_K)^2 + k^2_0}\,[\bar{p}(x)\Sigma^-(x)][\bar{n}(x)i\gamma^5
p(x)]\nonumber\\ &&-\,2\sqrt{2}\,\frac{(1 - \alpha)\,(2\alpha -
1)\,g^3_{\pi NN}}{m^2_{\pi} - (E_N - m_N)^2 +
k^2_0}\,\frac{1}{m_{\Sigma} + m_N + m_K}\,[\bar{p}(x)\Sigma^-(x)]
[\bar{n}(x)i\gamma^5 p(x)]\nonumber\\ &&+
\,\frac{2\sqrt{2}}{3}\,\frac{\alpha\,(3 - 2\alpha)\,g^3_{\pi
NN}}{m^2_{\pi} - (E_N - m_N)^2 + k^2_0}\,\frac{1}{m_{\Lambda^0} + m_N
+ m_K}\,[\bar{p}(x)\Sigma^-(x)] [\bar{n}(x)i\gamma^5 p(x)]\nonumber\\
&&- \,\frac{\sqrt{2}}{3}\,\frac{(2 \alpha - 1)\,g^3_{\pi
NN}}{m^2_{\eta} - (E_N - m_N)^2 + k^2_0}\,\frac{m_N + m_K -
m_{\Sigma}}{m^2_N - (E_{\Sigma^-} - m_K)^2 + k^2_0}\,
[\bar{n}(x)\Sigma^-(x)] [\bar{p}(x)i\gamma^5 p(x)]\nonumber\\ &&-
\,\frac{2\sqrt{2}}{3}\,\frac{\alpha\,(2 \alpha - 1)\,(3 -
4\alpha)\,g^3_{\pi NN}}{m^2_{\eta} - (E_N - m_N)^2 +
k^2_0}\,\frac{1}{m_{\Sigma} + m_N + m_K}\,[\bar{n}(x)\Sigma^-(x)]
[\bar{p}(x)i\gamma^5 p(x)],
\end{eqnarray}
where $E_{\Sigma^-} = 1302\,{\rm MeV}$, $E_N = 1072\,{\rm MeV}$ and
$k_0 = 516\,{\rm MeV}$.

The effective Lagrangian of the transition $p \Sigma^- \to K^- p n$,
defined by the Feynman diagrams in Fig.10 with the scalar--meson
exchanges, is equal to
\begin{eqnarray}\label{label8.4}
&&{\cal L}^{\,p\Sigma^- \to K^- pn}_{\rm eff}(x)_S =\nonumber\\ &&= -
\,\frac{1}{\sqrt{2}}\,\frac{1}{g_A F_{\pi}}\,\frac{(2\alpha -
1)\,g^2_{\pi NN}}{m^2_K - (E_{\Sigma^-} - m_N)^2 +
k^2_0}\,[\bar{n}(x)i\gamma^5 \Sigma^-(x)][\bar{p}(x) p(x)]\nonumber\\
&&~~+ \,\frac{1}{\sqrt{2}}\,\frac{1}{g_A F_{\pi}}\,\frac{(2\alpha -
1)\,g^2_{\pi NN}}{m^2_{\pi} - (E_N - m_N)^2 +
k^2_0}\,[\bar{n}(x)\Sigma^-(x)][\bar{p}(x)i\gamma^5 p(x)]\nonumber\\
&&~~- \,\frac{1}{3\sqrt{2}}\,\frac{1}{g_A F_{\pi}}\,\frac{(2\alpha -
1)\,(3 - 4\alpha)\,g^2_{\pi NN}}{m^2_{\eta} - (E_N - m_N)^2 +
k^2_0}\,[\bar{n}(x)\Sigma^-(x)][\bar{p}(x)i\gamma^5 p(x)].
\end{eqnarray}
Recall that the contribution of the scalar mesons is computed in the
infinite mass limit corresponding to the non--linear realization of
chiral symmetry. The effective Lagrangians defining the transitions
$(p \Sigma^-)_{{^3}{\rm P}_1} \to K^- (pn)_{{^3}{\rm S}_1}$ and $(p
\Sigma^-)_{{^1}{\rm P}_1} \to K^- (pn)_{{^3}{\rm S}_1}$ can be derived
from (\ref{label8.3}) and (\ref{label8.4}) by means of the Fierz
transformation (see (\ref{label6.4}) and (\ref{label6.5}))
\begin{eqnarray}\label{label8.5}
\hspace{-0.3in}&&[\bar{n}(x)\gamma^5 \Sigma^-(x)] [\bar{p}(x)p(x)] \to
+\,\frac{1}{4}\,[\bar{n}(x)\vec{\gamma}\, p^c(x)] \cdot [\bar{p^c}(x)
\vec{\gamma}\,\gamma^5 \Sigma^-(x)]\nonumber\\
\hspace{-0.3in}&&\hspace{2in}
-\,\frac{1}{4}\,[\bar{n}(x)\vec{\gamma}\,p^c(x)] \cdot [\bar{p^c}(x)
\gamma^0\vec{\gamma}\,\gamma^5 \Sigma^-(x)],\nonumber\\
\hspace{-0.3in}&&[\bar{n}(x)\Sigma^-(x)][\bar{p}(x)\gamma^5 p(x)] \to
-\,\frac{1}{4}\, [\bar{n}(x)\vec{\gamma}\, p^c(x)]\cdot [\bar{p^c}(x)
\vec{\gamma}\,\gamma^5 \Sigma^-(x)]\nonumber\\
\hspace{-0.3in}&&\hspace{2in}-\,\frac{1}{4}\,[\bar{n}(x)
\vec{\gamma}\,p^c(x)] \cdot [\bar{p^c}(x)
\gamma^0\vec{\gamma}\,\gamma^5 \Sigma^-(x)],\nonumber\\
\hspace{-0.3in}&&[\bar{p}(x)\Sigma^-(x)] [\bar{n}(x)\gamma^5 p(x)] \to
+\,\frac{1}{4}\, [\bar{n}(x)\vec{\gamma}\, p^c(x)]\cdot [\bar{p^c}(x)
\vec{\gamma}\,\gamma^5 \Sigma^-(x)]\nonumber\\
\hspace{-0.3in}&&\hspace{2in} + \,\frac{1}{4}\,[\bar{n}(x)
\vec{\gamma}\,p^c(x)] \cdot [\bar{p^c}(x)
\gamma^0\vec{\gamma}\,\gamma^5 \Sigma^-(x)].
\end{eqnarray}
Substituting (\ref{label8.5}) into (\ref{label8.3}) and (\ref{label8.4})
we obtain the effective Lagrangians responsible for the transitions
$(p \Sigma^-)_{{^3}{\rm P}_1} \to K^- (pn)_{{^3}{\rm S}_1}$ and $(p
\Sigma^-)_{{^1}{\rm P}_1} \to K^- (pn)_{{^3}{\rm S}_1}$.

\subsection{Reaction $(p \Sigma^-)_{{^3}{\rm P}_1} \to K^-
(pn)_{{^3}{\rm S}_1}$}

The amplitude of the reaction $(p \Sigma^-)_{{^3}{\rm P}_1}\to K^-
(pn)_{{^3}{\rm S}_1}$ is defined by
\begin{eqnarray}\label{label8.6}
\hspace{-0.3in}&&M(p(\vec{k},\alpha_1) \Sigma^-(- \vec{k},\alpha_2)
\to K^-(\vec{0}\,) p(\vec{K},\sigma_p) n( - \vec{K},\sigma_n);
{^3}{\rm P}_1) = i\,C^{\,(p \Sigma^-; {^3}{\rm
P}_1)}_{K^-(pn; {^3}{\rm S}_1)}\,\nonumber\\
\hspace{-0.3in}&&\times\, \frac{[\bar{u}(\vec{K},\sigma_p)
\vec{\gamma} u^c( - \vec{K},\sigma_n)]\cdot
[\bar{u^c}(\vec{k},\alpha_1) \vec{\gamma} \gamma^5u(-
\vec{k},\alpha_2)] }{\displaystyle 1 - \frac{1}{2}\,r^t_{np} a^t_{np}
K^2 - i\,a^t_{np} K}\,f^{\,(p \Sigma^-; {^3}{\rm P}_1)}_{K^-(pn; {^3}{\rm S}_1)}(k_0),
\end{eqnarray}
where $f^{\,(p \Sigma^-; {^3}{\rm P}_1)}_{K^-(pn; {^3}{\rm
S}_1)}(k_0)$ is the amplitude, describing the $p \Sigma^-$
rescattering in the ${^3}{\rm P}_1$ state near threshold of the $K^-(p
n)_{{^3}{\rm S}_1}$ system production and $C^{\,(p \Sigma^-; {^3}{\rm
P}_1)}_{K^-(pn; {^3}{\rm S}_1)}$ is the effective coupling constant of
the transition $(p \Sigma^- )_{{^3}{\rm P}_1} \to K^- (pn)_{{^3}{\rm
S}_1}$.

The effective Lagrangian of the transition $(p \Sigma^-)_{{^3}{\rm
P}_1} \to K^- (pn)_{{^3}{\rm S}_1}$, computed at threshold, reads
\begin{eqnarray}\label{label8.7}
{\cal L}^{(p \Sigma^-; {^3}{\rm P}_1)\to K^-(pn; {^3}{\rm S}_1)}_{\rm
eff}(x)_P = i\,C^{\,(p \Sigma^-; {^3}{\rm P}_1)}_{K^-(pn; {^3}{\rm
S}_1)}\, K^{-\dagger}(x)\, [\bar{n}(x)\vec{\gamma}\, p^c(x)] \cdot
[\bar{p^c}(x) \vec{\gamma}\, \gamma^5 \Sigma^-(x)].
\end{eqnarray}
The effective coupling constant $C^{\,(p \Sigma^-; {^3}{\rm
P}_1)}_{K^-(pn; {^3}{\rm S}_1)}$ is defined by
\begin{eqnarray}\label{label8.8}
\hspace{-0.3in}C^{\,(p \Sigma^-; {^3}{\rm P}_1)}_{K^-(pn; {^3}{\rm
S}_1)} &=& -\,\frac{1}{2\sqrt{2}}\,\frac{(2\alpha - 1)^3\,g^3_{\pi
NN}}{m^2_K - (E_{\Sigma^-} - m_N)^2 + k^2_0}\,\frac{1}{m_{\Sigma} +
m_N + m_K}\nonumber\\
\hspace{-0.3in}&&+ \,\frac{1}{3\sqrt{2}}\,\frac{(2\alpha - 1)\,(3-
2\alpha)^2\,g^3_{\pi NN}}{m^2_K - (E_{\Sigma^-} - m_N)^2 +
k^2_0}\,\frac{1}{m_{\Lambda^0} + m_N + m_K}\nonumber\\ &&-
\,\frac{1}{2\sqrt{2}}\,\frac{(2\alpha - 1)\,g^3_{\pi NN}}{m^2_{\pi} -
(E_N - m_N)^2 + k^2_0}\,\frac{m_N + m_K - m_{\Sigma}}{m^2_N -
(E_{\Sigma^-} - m_K)^2 + k^2_0}\nonumber\\ &&-
\,\frac{1}{\sqrt{2}}\,\frac{(1 - \alpha)\,(2\alpha - 1)\,g^3_{\pi
NN}}{m^2_{\pi} - (E_N - m_N)^2 + k^2_0}\,\frac{1}{m_{\Sigma} + m_N +
m_K}\nonumber\\ &&- \,\frac{1}{\sqrt{2}}\,\frac{(2\alpha -
1)\,g^3_{\pi NN}}{m^2_{\pi} - (E_N - m_N)^2 + k^2_0}\,\frac{m_N + m_K
- m_{\Sigma}}{m^2_N - (E_{\Sigma^-} - m_K)^2 + k^2_0}\nonumber\\
&&-\,\frac{1}{\sqrt{2}}\,\frac{(1 - \alpha)\,(2\alpha - 1)\,g^3_{\pi
NN}}{m^2_{\pi} - (E_N - m_N)^2 + k^2_0}\,\frac{1}{m_{\Sigma} + m_N +
m_K}\nonumber\\ &&+ \,\frac{1}{3\sqrt{2}}\,\frac{\alpha\,(3 -
2\alpha)\,g^3_{\pi NN}}{m^2_{\pi} - (E_N - m_N)^2 +
k^2_0}\,\frac{1}{m_{\Lambda^0} + m_N + m_K}\nonumber\\
&&+\,\frac{1}{6\sqrt{2}}\,\frac{(2 \alpha - 1)\,g^3_{\pi
NN}}{m^2_{\eta} - (E_N - m_N)^2 + k^2_0}\frac{m_N + m_K -
m_{\Sigma}}{m^2_N - (E_{\Sigma^-} - m_K)^2 + k^2_0}\nonumber\\ &&+
\,\frac{1}{3\sqrt{2}}\,\frac{\alpha\,(2 \alpha - 1)\,(3 -
4\alpha)\,g^3_{\pi NN}}{m^2_{\eta} - (E_N - m_N)^2 +
k^2_0}\,\frac{1}{m_{\Sigma} + m_N + m_K}\nonumber\\ &&-
\,\frac{1}{4\sqrt{2}}\,\frac{1}{g_A F_{\pi}}\,\frac{(2\alpha -
1)\,g^2_{\pi NN}}{m^2_K - (E_{\Sigma^-} - m_N)^2 + k^2_0}\nonumber\\
&&- \,\frac{1}{4\sqrt{2}}\,\frac{1}{g_A F_{\pi}}\,\frac{(2\alpha -
1)\,g^2_{\pi NN}}{m^2_{\pi} - (E_N - m_N)^2 + k^2_0}\nonumber\\ &&+
\,\frac{1}{12\sqrt{2}}\,\frac{1}{g_A F_{\pi}}\,\frac{(2\alpha - 1)\,(3
- 4\alpha)\,g^2_{\pi NN}}{m^2_{\eta} - (E_N - m_N)^2 + k^2_0} =
-\,7\times 10^{-7}\,{\rm MeV}^{-3}.
\end{eqnarray}
The amplitude $f^{\,(p\Sigma^-; {^3}{\rm P}_1)}_{K^-(pn; {^3}{\rm
S}_1)}(k_0)$, describing the rescattering of the $p\Sigma^-$ pair in
the ${^3}{\rm P}_1$ state near threshold of the $K^-(p n)_{{^3}{\rm
S}_1}$ system production, is defined by the Feynman diagrams similar
to those depicted in Fig.6. The procedure of the calculation of these
diagrams is expounded in Appendix E. The result of the calculation
reads 
\begin{eqnarray}\label{label8.9}
\Big|f^{\,(p\Sigma^-; {^3}{\rm P}_1)}_{K^-(pn; {^3}{\rm
P}_1)}(k_0)\Big| =\Big|\Big\{ 1 - \frac{C_{p\Sigma^-}({^3}{\rm
P}_1)}{12\pi^2}\,\frac{k^3_0}{E(k_0)}\,\Big[{\ell
n}\Big(\frac{E(k_0) + k_0}{E(k_0) - k_0} \Big) -
i\,\pi\Big]\Big\}^{-1}\Big| \simeq 0.6,
\end{eqnarray}
where $E(k_0) = \sqrt{k^2_0 + m^2_B}$\,\footnote{For simplicity we use
the equal masses of baryons for the calculation of the rescattering of
the $p\Sigma^-$ pair, where $m_B = \sqrt{(2 m_N + m_K)^2 - 4 k^2_0}/2
= 1070\,{\rm MeV}$.} and the effective coupling constant
$C_{p\Sigma^-}({^3}{\rm P}_1)$ is equal to 
\begin{eqnarray}\label{label8.10}
C_{p\Sigma^-}({^3}{\rm P}_1) &=& -\,(1 - \alpha)\,\frac{g^2_{\pi NN}}{2
k^2_0}\,{\ell n}\Big(1 + \frac{4 k^2_0}{~m^2_{\pi}}\Big) + \alpha (3
-4\alpha)\,\frac{g^2_{\pi NN}}{6 k^2_0}\,{\ell n}\Big(1 + \frac{4
k^2_0}{~m^2_{\eta}}\Big) = \nonumber\\ &=& -\,4.0\times 10^{-4}\,{\rm
MeV}^{-2}.
\end{eqnarray}
The rescattering of the $p\Sigma^-$ pair in the ${^3}{\rm P}_1$ state
is defined by the interaction, computed in the one--meson exchange
approximation
\begin{eqnarray}\label{label8.11}
{\cal L}^{(p\Sigma^-;{^3}{\rm P}_1) \to (p \Sigma^-;{^3}{\rm
P}_1)}_{\rm eff}(x) = -\,\frac{1}{4}\,C_{p\Sigma^-}({^3}{\rm
P}_1))\,[\bar{\Sigma}^-(x) \vec{\gamma}\,\gamma^5 p^c(x)]\cdot
[\bar{p^c}(x)\vec{\gamma}\,\gamma^5 \Sigma^-(x)].
\end{eqnarray}
We would like to remind that the interaction (\ref{label8.11}) defines
also the final--state $(p\Sigma^-)_{{^3}{\rm P}_1}$ interaction near
threshold of the reaction $K^- (pn)_{{^3}{\rm P}_1} \to
(p\Sigma^-)_{{^3}{\rm P}_1}$.

\subsection{S--wave amplitude $\tilde{f}^{\,K^-d}_0(0)_{(p\Sigma^-; 
{^3}{\rm P}_1)}$ of $K^-d$ scattering}

The amplitude $\tilde{f}^{\,K^-d}_0(0)_{(p\Sigma^-; {^3}{\rm P}_1)}$
can be computed in a way similar to
$\tilde{f}^{\,K^-d}_0(0)_{(n\Lambda^0; {^3}{\rm P}_1)}$ and
$\tilde{f}^{\,K^-d}_0(0)_{(n\Sigma^0; {^3}{\rm P}_1)}$. The result
reads
\begin{eqnarray}\label{label8.12}
\hspace{-0.5in}&&\tilde{f}^{\,K^-d}_0(0)_{(p\Sigma^-; {^3}{\rm P}_1)}
= 4.6\times 10^{-3}\, \frac{1}{3\pi^2}\,\frac{m^3_{\pi}}{1 +
m_K/m_d}\,[C^{\,(p\Sigma^-; {^3}{\rm P}_1)}_{K^-(pn; {^3}{\rm
S}_1)}]^2\, |f^{(p\Sigma^-; {^3}{\rm P}_1)}_{K^-(pn; {^3}{\rm
S}_1)}(k_0)|^2\nonumber\\
\hspace{-0.5in}&&\times\,\Big[\frac{1}{4}\,\frac{k^2_0}{1 + m_K/2
m_B}\,F\Big(\frac{\Lambda}{m_B},\frac{k_0}{m_B}\Big) +
\,i\,\frac{k^3_0}{2m_N + m_K}\Big] = (0.02 + i\,0.7)\times
10^{-3}\,{\rm fm}.
\end{eqnarray}
Now we proceed to computing the contribution of the reaction
$(n\Sigma^0)_{{^1}{\rm P}_1} \to K^- (pn)_{{^3}{\rm S}_1}$.

\subsection{\bf Reaction $(p \Sigma^-)_{{^1}{\rm P}_1} \to K^-
(pn)_{{^3}{\rm S}_1}$}

The amplitude of the reaction $(p \Sigma^-)_{{^1}{\rm P}_1} \to K^-
(pn)_{{^3}{\rm S}_1}$ is defined by 
\begin{eqnarray}\label{label8.13}
\hspace{-0.3in}&&M(p(\vec{k},\alpha_1) \Sigma^-(- \vec{k},\alpha_2)
\to K^-(\vec{0}\,) p(\vec{K},\sigma_p) n( - \vec{K},\sigma_n);
{^1}{\rm P}_1) = i\,C^{\,(p\Sigma^-; {^1}{\rm P}_1)}_{K^- (pn;
{^3}{\rm S}_1)}\,\nonumber\\
\hspace{-0.3in}&&\times\, \frac{[\bar{u}(\vec{K},\sigma_p)
\vec{\gamma} u^c( - \vec{K},\sigma_n)]\cdot
[\bar{u^c}(\vec{k},\alpha_1) \vec{\gamma} \gamma^5u(-
\vec{k},\alpha_2)] }{\displaystyle 1 - \frac{1}{2}\,r^t_{np} a^t_{np}
K^2 + i\,a^t_{np} K}\,f^{\,(p\Sigma^-; {^1}{\rm P}_1)}_{K^- (pn;
{^3}{\rm S}_1)}(k_0).
\end{eqnarray}
The effective Lagrangian of the transition $(p\Sigma^-)_{{^1}{\rm
P}_1} \to K^- (pn)_{{^3}{\rm S}_1}$ at threshold can be defined by
\begin{eqnarray}\label{label8.14}
{\cal L}^{\,(p\Sigma^-; {^1}{\rm P}_1) \to K^-(pn; {^3}{\rm
S}_1)}_{\rm eff}(x) = i\,C^{\,(p\Sigma^-; {^1}{\rm P}_1)}_{K^- (pn;
{^3}{\rm S}_1)}\, K^{-\dagger}(x)\, [\bar{n}(x)\vec{\gamma}\, p^c(x)]
\cdot [\bar{p^c}(x) \gamma^0\vec{\gamma}\, \gamma^5 \Sigma^-(x)].
\end{eqnarray}
Using (\ref{label8.3}), (\ref{label8.4}) and (\ref{label8.5}) we
compute the effective coupling constant $C^{\,(p\Sigma^-; {^1}{\rm
P}_1)}_{K^- (pn; {^3}{\rm S}_1)}$. It is equal to
\begin{eqnarray}\label{label8.15}
C^{\,(n\Lambda^0; {^1}{\rm P}_1)}_{K^- (pn; {^3}{\rm S}_1)} =
-\,12\times 10^{-7}\,{\rm MeV}^{-3}.
\end{eqnarray}
The Lagrangian (\ref{label8.14}) describes the interaction of the
$p\Sigma^-$ pair in the ${^1}{\rm P}_1$ state with the $np$ pair in
the ${^1}{\rm S}_1$ state through the emission of the $K^-$--meson.

\subsection{ Amplitude of $(p\Sigma^-)_{{^1}{\rm P}_1}$
rescattering}.

The amplitude $f^{\,(p\Sigma^-; {^1}{\rm P}_1)}_{K^-(pn; {^3}{\rm
S}_1)}(k_0)$, describing the rescattering of the $n\Sigma^0$ pair in
the ${^1}{\rm P}_1$ state near threshold of the $K^-p n$ system
production, is given by (see Appendix E)
\begin{eqnarray}\label{label8.16}
\Big|f^{\,(p\Sigma^-; {^1}{\rm P}_1)}_{K^-(pn; {^3}{\rm
S}_1)}(k_0)\Big| = \Big|\Big\{1 - \frac{C_{p\Sigma^-}({^1}{\rm
P}_1)}{24\pi^2}\,\frac{k^3_0}{E(k_0)}\,\Big[{\ell n}\Big(\frac{E(k_0)
+ k_0}{E(k_0) - k_0} \Big) - i\,\pi\Big]\Big\}^{-1}\Big| \simeq 0.8.
\end{eqnarray}
The effective coupling constant $C_{p\Sigma^-}({^1}{\rm P}_1)$ is
equal to
\begin{eqnarray}\label{label8.17}
C_{p\Sigma^-}({^1}{\rm P}_1) = C_{p\Sigma^-}({^3}{\rm P}_1) =
-\,4.0\times 10^{-4}\,{\rm MeV}^{-2}.
\end{eqnarray}
The rescattering of the $p\Sigma^-$ pair in the ${^3}{\rm P}_1$ state
is defined by the interaction
\begin{eqnarray}\label{label8.18}
{\cal L}^{\,(p\Sigma^-; {^1}{\rm P}_1) \to (p\Sigma^-; {^1}{\rm
P}_1)}_{\rm eff}(x) = -\,\frac{1}{4}\,C_{p\Sigma^-}({^1}{\rm
P}_1)\,[\bar{\Sigma}^-(x) \vec{\gamma}\,\gamma^5 p^c(x)]\cdot
[\bar{p^c}(x)\vec{\gamma}\,\gamma^5 \Sigma^-(x)].
\end{eqnarray}
Recall that according to \cite{AI01} the effective coupling constant
$C_{p\Sigma^-}({^1}{\rm P}_1)$ is computed in the one--pseudoscalar
meson exchange approximation.

\subsection{S--wave amplitude
$\tilde{f}^{\,K^-d}_0(0)_{(p\Sigma^-; {^1}{\rm P}_1)}$ of $K^-d$
scattering}

The S--wave amplitude $\tilde{f}^{\,K^-d}_0(0)_{(p\Sigma^-; {^1}{\rm
P}_1)}$ of $K^-d$ scattering near threshold, caused by the reaction
$K^- (pn)_{{^3}{\rm S}_1} \to (p\Sigma^-)_{{^1}{\rm P}_1} \to K^-
(pn)_{{^3}{\rm S}_1}$, is equal to
\begin{eqnarray}\label{label8.19}
\hspace{-0.5in}&&\tilde{f}^{\,K^-d}_0(0)_{(p\Sigma^-; {^1}{\rm P}_1)}
= 4.6\times 10^{-3}\, \frac{1}{6\pi^2}\,\frac{m^3_{\pi}}{1 +
m_K/m_d}\,[C^{\,(p\Sigma^-; {^1}{\rm P}_1)}_{K^-(pn; {^3}{\rm
S}_1)}]^2\, |f^{(p\Sigma^-; {^1}{\rm P}_1)}_{K^-(pn; {^3}{\rm
S}_1)}(k_0)|^2\nonumber\\
\hspace{-0.5in}&&\times\,\Big[\frac{1}{4}\,\frac{k^2_0}{1 + m_K/2
m_B}\,F\Big(\frac{\Lambda}{m_B},\frac{k_0}{m_B}\Big) +
\,i\,\frac{k^3_0}{2m_N + m_K}\Big] = (0.05 + i\,1.8)\times
10^{-3}\,{\rm fm}.
\end{eqnarray}
Now we can compute the contribution of the two--body inelastic channel
$K^- (pn)_{{^3}{\rm S}_1} \to p\Sigma^- \to K^- (pn)_{{^3}{\rm S}_1}$
to the S--wave amplitude $f^{K^-d}_0(0)_{p\Sigma^-}$ of $K^-d$
scattering near threshold and the energy level displacement of the
ground state of kaonic deuterium.

\subsection{S--wave amplitude $f^{K^-d}_0(0)_{p\Sigma^-}$ 
and the energy level displacement}

The S--wave amplitude of $K^-d$ scattering at threshold, saturated by
the inelastic reaction $K^- (pn)_{{^3}{\rm S}_1} \to p \Sigma^- \to
K^- (pn)_{{^3}{\rm S}_1}$ with the $p\Sigma^-$ pair in the ${^3}{\rm
P}_1$ and ${^1}{\rm P}_1$ state, is equal to the sum of the
contributions (\ref{label8.12}) and (\ref{label8.19})
\begin{eqnarray}\label{label8.20}
\tilde{f}^{\,K^-d}_0(0)_{p \Sigma^-} = (0.07 + i\,2.5) \times
10^{-3}\,{\rm fm}.
\end{eqnarray}
The contribution of the decay $A_{Kd} \to p \Sigma^-$ to the energy
level displacement of the ground state of kaonic deuterium amounts to
\begin{eqnarray}\label{label8.21}
-\,\epsilon^{(p\Sigma^-)}_{1s} +
i\,\frac{\Gamma^{(n\Sigma^0)}_{1s}}{2} =
602\,\tilde{f}^{\,K^-d}_0(0)_{n\Sigma^0} = (0.04 + i\,1.5)\,{\rm eV}.
\end{eqnarray}
Hence, the partial width of the decay $A_{Kd} \to p \Lambda^-$ is
equal to $\Gamma^{(p\Sigma^-)}_{1s} = 3.0\,{\rm eV}$.

According to \cite{VV70}, the experimental rate of the production of
the $n\Sigma^0$ pair at threshold of the reaction $K^-d \to p
\Sigma^-$ is equal to $R(K^-d \to p \Sigma^-) = (0.505 \pm
0.036)\,\%$.

Using our estimate of the partial width, $\Gamma^{(p\Sigma^-)}_{1s} =
(3.0\pm 0.6)\,{\rm eV}$, where $\pm 0.6\,{\rm eV}$ is a
theoretical accuracy of the result, and the experimental rate,
$R(K^-d \to n \Sigma^0) = (0.505 \pm 0.036)\,\%$, we can estimate the
expected value of the total width of the energy level of the ground
state of kaonic deuterium
\begin{eqnarray}\label{label8.22}
\Gamma_{1s} = \frac{\Gamma^{(p\Sigma^-)}_{1s} }{(0.505 \pm
0.036)\times 10^{-2}} = (590 \pm 130)\,{\rm eV}.
\end{eqnarray}
This value is compared well with our estimate $\Gamma_{1s} = (570 \pm
130)\,{\rm eV}$ and $\Gamma_{1s} = (700 \pm 200)\,{\rm eV}$, which we
have made in Sections 6 and 7 using the theoretical values of the
partial widths of the decays $A_{Kd} \to n\Lambda^0$ and $A_{Kd} \to
n\Sigma^0$ and the experimental rates of the $n\Lambda^0$ and
$n\sigma^0$ production in the reactions $K^-d \to n \Lambda^0$ and
$K^-d \to n \Sigma^0$.

\section{Comparison with experimental data and the energy level 
displacement}
\setcounter{equation}{0}

The imaginary parts of the amplitudes $\tilde{f}^{\,K^-d}_0(0)_{NY}$
with $NY = n\Lambda^0, n \Sigma^0$ and $p \Sigma^-$ are proportional
to the cross sections for the reactions $K^-d \to n\Lambda^0$, $K^-d
\to n \Sigma^0$ and $K^-d \to p \Sigma^- $ near threshold. According
to the experimental data by Veirs and Burnstein \cite{VV70}, the
production rates of $NY$ pairs in the reactions $K^-d \to n\Lambda^0$,
$K^-d \to n \Sigma^0$ and $K^-d \to p \Sigma^- $ are equal to:
\begin{eqnarray}\label{label9.1}
R(K^-d \to n \Lambda^0) &=& (0.387 \pm 0.041)\,\%,\nonumber\\ R(K^-d
\to n \Sigma^0) &=& (0.337 \pm 0.070)\,\%,\nonumber\\ R(K^-d \to p
\Sigma^-) &=& (0.505 \pm 0.036)\,\%,\nonumber\\ R = \sum_{NY}R(K^-d
\to NY) &=& (1.229 \pm 0.090)\,\%.
\end{eqnarray}
The ratios, independent on a total yield, read
\begin{eqnarray}\label{label9.2}
R(\Lambda^0/\Sigma^0) &=& \frac{R(K^-d \to n \Lambda^0)}{R(K^-d \to n
\Sigma^0)} = (1.15 \pm 0.27),\nonumber\\
R(\Sigma^0/\Sigma^-) &=&
\frac{R(K^-d \to n \Sigma^0)}{ R(K^-d \to p \Sigma^-)} = (0.68 \pm
0.15),\nonumber\\ R(\Lambda^0/\Sigma^-) &=& \frac{R(K^-d \to n
\Lambda^0)}{ R(K^-d \to p \Sigma^-)} = (0.77 \pm 0.10).
\end{eqnarray}
For these ratios we predict the following theoretical values
\begin{eqnarray}\label{label9.3}
R(\Lambda^0/\Sigma^0) &=& \frac{{\cal
I}m\,\tilde{f}^{\,K^-d}_0(0)_{n\Lambda^0}}{{\cal
I}m\,\tilde{f}^{\,K^-d}_0(0)_{n\Sigma^0}} = 1.0 \pm 0.3,\nonumber\\
R(\Sigma^0/\Sigma^-) &=& \frac{{\cal
I}m\,\tilde{f}^{\,K^-d}_0(0)_{n\Sigma^0}}{{\cal
I}m\,\tilde{f}^{\,K^-d}_0(0)_{p\Sigma^-}} = 0.8 \pm 0.2, \nonumber\\
R(\Lambda^0/\Sigma^-) &=& \frac{{\cal
I}m\,\tilde{f}^{\,K^-d}_0(0)_{n\Lambda^0}}{{\cal
I}m\,\tilde{f}^{\,K^-d}_0(0)_{p\Sigma^-}} = 0.8 \pm 0.2.
\end{eqnarray}
The theoretical predictions agree well with the experimental data.

We would like to emphasize that according to the requirement of
isospin invariance the ratio $R(\Sigma^0/\Sigma^-)$ of the cross
sections for the reactions $K^-d \to n\Sigma^0$ and $K^-d \to
p\Sigma^-$ should be equal to
\begin{eqnarray}\label{label9.4}
R(\Sigma^0/\Sigma^-) = 0.5.
\end{eqnarray}
We would like to notice that the strength of the forces responsible
for the transitions $K^-d \to n\Sigma^0$ and $K^-d \to p \Sigma^-$ is
of order of a strength of the forces violating isospin
invariance. Indeed, relative mass differences of the neutron and
proton $(m_n - m_p)/m_N = 0.138\,\%$ and the charged and neutral
$K$--mesons $(m_{\bar{K}^0} - m_{K^-})/m_K = 0.607\,\%$ are of order
of the production rates of the $NY$ pairs near threshold of the
reactions $K^-d \to NY$. Therefore, a departure from the isospin
invariance for the ratio of the cross sections of the reactions $K^-d
\to n\Sigma^0$ and $K^-d \to p \Sigma^-$ should not be a surprise
\cite{VV70}.

The contribution of the inelastic two--body channels $K^-d \to NY$ to
the energy level displacement of the ground state of kaonic deuterium
is  given by
\begin{eqnarray}\label{label9.5}
\hspace{-0.3in}-\,\epsilon^{(n\Lambda^0)}_{1s} + i\,\frac{\Gamma^{(n
\Lambda^0)}_{1s}}{2} &=& 602\,\tilde{f}^{K^-d}_0(0)_{n\Lambda^0} =
(-\,0.10 \pm 0.02) + i\,(1.1 \pm 0.2) \,{\rm eV},\nonumber\\
\hspace{-0.3in}-\,\epsilon^{(n\Sigma^0)}_{1s} + i\,\frac{\Gamma^{(n
\Sigma^0)}_{1s}}{2} &=& 602\,\tilde{f}^{K^-d}_0(0)_{n\Sigma^0} =
(+\,0.03\pm 0.01) + i\,(1.2 \pm 0.3) \,{\rm eV},\nonumber\\
\hspace{-0.3in}-\,\epsilon^{(p\Sigma^-)}_{1s} + i\,\frac{\Gamma^{(p
\Sigma^-)}_{1s}}{2} &=& 602\,\tilde{f}^{K^-d}_0(0)_{p\Sigma^-} =
(+\,0.04\pm 0.01) + i\,(1.5 \pm 0.3) \,{\rm eV}.
\end{eqnarray}
The partial widths $\Gamma^{(NY)}_{1s}$, equal to
\begin{eqnarray}\label{label9.6}
\Gamma^{(n \Lambda^0)}_{1s}&=& (2.2 \pm 0.5)\,{\rm eV},\nonumber\\
\Gamma^{(n \Sigma^0)}_{1s} &=& (2.4 \pm 0.5)\,{\rm eV},\nonumber\\
\Gamma^{(p \Sigma^-)}_{1s} &=& (3.0 \pm 0.6)\,{\rm eV},
\end{eqnarray}
are compared well with theoretical estimates discussed by Reitan
\cite{AR69}
\begin{eqnarray}\label{label9.7}
0.66\,{\rm eV} \le &\left(\begin{array}{c}\Gamma^{(n \Lambda^0)}_{1s}
 \\ \Gamma^{(n \Sigma^0)}_{1s}\end{array}\right)& \le 190\,{\rm
 eV},\nonumber\\ 0.66\,{\rm eV} \le &\Gamma^{(p \Sigma^-)}_{1s}& \le
 3.95\,{\rm eV}.
\end{eqnarray}
The S--wave amplitude $\tilde{f}^{K^-d}_0(0)_{(\rm two-body)}$ of
$K^-d$ scattering near threshold, caused by the two--body inelastic
channels $K^-d \to NY \to K^-d$ with the intermediate $NY =
n\Lambda^0, n \Sigma^0$ and $p \Sigma^-$ states, is equal to
\begin{eqnarray}\label{label9.8}
\tilde{f}^{K^-d}_0(0)_{(\rm two-body)} = (-\,0.08\pm 0.02) + i\,(6.4
\pm 0.8)\times 10^{-3}\,{\rm fm}.
\end{eqnarray}
The energy level displacement of the ground state of kaonic deuterium
caused by the inelastic two--body decays $A_{Kd} \to n\Lambda^0$,
$A_{Kd} \to n \Sigma^0$ and $A_{Kd} \to p \Sigma^-$ is equal to
\begin{eqnarray}\label{label9.9}
-\,\epsilon^{(\rm two-body)}_{1s} + i\,\frac{\Gamma^{(\rm
two-body)}_{1s}}{2} &=& 602\,\tilde{f}^{K^-d}_0(0)_{(\rm two-body)}
=\nonumber\\ &=& (-\,0.05\pm 0.01) + i\,(3.9 \pm 0.5) \,{\rm eV}.
\end{eqnarray}
Using the experimental value of the total production rate $R =
(1.229\pm 0.090)\,\%$ and our theoretical prediction for $\Gamma^{(\rm
two-body)}_{1s}$, given by (\ref{label9.9}), we can estimate the
expected value of the total width of the ground state of kaonic
deuterium
\begin{eqnarray}\label{label9.10}
\Gamma_{1s} &=& \frac{\sum_{NY}\Gamma^{(NY)}_{1s}}{ (1.229 \pm
0.090)\times 10^{-2}} = \frac{\Gamma^{(n\Lambda^0)}_{1s} + \Gamma^{(n
\Sigma^0)}_{1s} + \Gamma^{(p \Sigma^-)}_{1s}}{(1.229 \pm 0.090)\times
10^{-2}} = \nonumber\\ &=& \frac{7.8\pm 1.0}{(1.229 \pm 0.090)\times
10^{-2}} = (630 \pm 100)\,{\rm eV}.
\end{eqnarray}
This value agrees well with the estimates (\ref{label6.37}),
(\ref{label7.26}) and (\ref{label8.22}).

Following the estimate of the total width, based on the theoretical
values of the widths of the decays $A_{Kd} \to n\Lambda^0, n \Sigma^0$
and $A_{Kd} \to p\Sigma^-$ and the experimental rates of the reactions
$K^-d \to n \Lambda^0$, $K^-d \to n \Sigma^0$ and $K^-d \to p
\Sigma^-$, we can estimate the expected contribution of the
three--body decays $A_{K^d} \to NY \pi$ to the shift of the energy
level of the ground state of kaonic deuterium. We get
\begin{eqnarray}\label{label9.11}
\epsilon^{(\rm three-body)}_{1s} = (9 \pm 8)\,{\rm eV}.
\end{eqnarray}
This implies that the contribution of the inelastic channels with
two--body $K^-d \to NY \to K^-d$ and three--body $K^-d \to NY \pi \to
K^-d$ intermediate states to the real part of the S--wave amplitude of
$K^-d$ scattering near threshold is negligible small, and the real
part of the S--wave amplitude of $K^-d$ scattering near threshold is
fully defined by the Ericson--Weise formula for the S--wave scattering
length (\ref{label4.17}). This gives the following value for the shift
of the energy level of the ground state of kaonic deuterium
\begin{eqnarray}\label{label9.12}
\epsilon_{1s} = - 602\,(a^{K^-d}_0)_{\rm EW} + \epsilon^{(\rm
three-body)}_{1s}= (353 \pm 60)\,{\rm eV}.
\end{eqnarray}
Thus, we predict that the S--wave amplitude of $K^-d$ scattering near
threshold is equal to
\begin{eqnarray}\label{label9.13}
f^{K^-d}_0(0) = (-\,0.586 \pm 0.095) + i\,(0.521 \pm 0.075)\,{\rm fm},
\end{eqnarray}
This defines the energy level displacement of the ground state of
kaonic deuterium
\begin{eqnarray}\label{label9.14}
-\,\epsilon_{1s} + i\,\frac{\Gamma_{1s}}{2} = 602\,f^{K^-d}_0(0) =
 (-\,353 \pm 60) + i\,(315 \pm 50)\;{\rm eV}.
\end{eqnarray}
A confirmation of these estimates should go through the calculation of
the contributions of the reactions $K^- (p n)_{{^3}{\rm S}_1} \to NY
\pi$ to the amplitude of low--energy elastic $K^-d$ scattering.

\section{Conclusion}

The quantum field theoretic and relativistic covariant approach,
developed in \cite{IV1}--\cite{IV3} for the description of the energy
level displacement of the ground and excited states of pionic hydrogen
\cite{IV1,IV2} and the energy level displacement of the ground state
of kaonic hydrogen \cite{IV3}, we have applied to the analysis of the
energy level displacement of the ground state of kaonic deuterium and
the S--wave amplitude of elastic $K^-d$ scattering near threshold.

According to \cite{IV1}--\cite{IV3} we have represented the energy
level displacement of the ground state of kaonic deuterium in terms of
the momentum integrals of the amplitude of elastic $K^-d$ scattering
for arbitrary relative momenta of the $K^-d$ pair weighted with the
wave functions of kaonic deuterium in the ground state. The knowledge
of this amplitude should allow to compute the energy level
displacement of the ground state of kaonic deuterium without any
low--energy approximation. As has been shown in \cite{IV1}--\cite{IV3}
the low--energy reduction of our representation of the energy level
displacements of exotic atoms reproduces the well--known DGBT formula
with additional corrections caused by the smearing of wave functions
around the origin. Such a smearing is defined by the relativistic
factors, related to the recoil energies of the nuclei
\cite{IV1}. These corrections are negative and of order 1$\,\%$. They
are universal for all exotic atoms, the existence of which is caused
by Coulombic forces. Since the experimental accuracy of the
measurement of the energy level displacement of the ground state of
pionic hydrogen, reached recently by the PSI Collaboration
\cite{PSI1}, is of order 1$\,\%$, the corrections, obtained in
\cite{IV1}, play an important role for the correct extraction of the
S--wave scattering lengths of $\pi N$ scattering from the experimental
data on the energy level displacement of the ground state of pionic
hydrogen \cite{TE04}.

Since the experiments on the energy level displacement of the ground
state of kaonic deuterium are in the stage of preparation, for the
analyses of the energy level displacement of the ground state of
kaonic deuterium we can neglect the correction of order of $1\,\%$ and
use the DGBT formula.

In the analysis of the energy level displacement of the ground state
of kaonic deuterium using the DGBT formula the main object of the
theoretical investigation is the S--wave amplitude $f^{K^-d}_0(0)$ of
$K^-d$ scattering near threshold. As has been pointed out by Ericson
and Weise \cite{TE88} for the analysis of elastic low--energy $\pi^-d$
scattering, the S--wave scattering length of $\pi^- d$ scattering can
be represented in the form of the superposition of the S--wave
scattering lengths of $\pi^-p$ and $\pi^-n$ scattering, realizing
so--called the {\it impulse approximation}, and the term, describing
elastic $\pi^-pn$ scattering.

Following Ericson and Weise \cite{TE88} and assuming that at threshold
the S--wave amplitude of $K^-d$ scattering is defined by the
superposition of the S--wave amplitudes of $K^-p$ and $K^-n$
scattering, reproducing the {\it impulse approximation}, and the
S--wave amplitude of the three--body to three--body reaction
$K^-(pn)_{{^3}{\rm S}_1} \to K^-(pn)_{{^3}{\rm S}_1}$, where the $np$
pair couples in the ${^3}{\rm S}_1$ state with isospin zero, we have
introduced the wave function of the ground state of kaonic deuterium
and the wave function of the deuteron in the momentum and the particle
number representation in terms of the operators of creation of the
$K^-$--meson, the proton and the neutron. In Appendix A we have shown
that these wave functions describe the bound $K^-d$ state and the
bound $np$ state with quantum numbers of the deuteron.

We have shown that due to such a representation of the wave function
of the deuteron, the S--wave amplitude of $K^-d$ scattering at
threshold can be represented in the Ericson--Weise form. The real part
of the S--wave amplitude of $K^-d$ scattering contains two terms,
defined by the S--wave amplitudes of $K^-p$ and $K^-n$ scattering near
threshold, and the terms coming from the interaction of three--body
scattering $K^-(pn)_{{^3}{\rm S}_1} \to K^-(pn)_{{^3}{\rm S}_1}$. The
imaginary part of the S--wave amplitude of $K^-d$ scattering near
threshold is fully defined by S--wave amplitude of three--body
scattering $K^-(pn)_{{^3}{\rm S}_1} \to K^-(pn)_{{^3}{\rm S}_1}$. The
amplitudes of elastic $K^-p$, $K^-n$ and $K^-(pn)_{{^3}{\rm S}_1}$
scattering are weighted with the wave functions of the deuteron in the
momentum representation.

We would like to accentuate that the Ericson--Weise form of the
S--wave scattering length of $\pi^-d$ scattering has been derived
within a potential model approach. We have proved this form within the
quantum field theoretic approach \cite{AI04} and have derived the
$K^-d$ version of this formula. The main object of the Ericson--Weise
formula, $\langle 1/r_{12}\rangle$, where $1/r_{12}$ is the inverse
distance between the proton and the neutron averaged over all
positions of them \cite{TE88}, we have computed equal to $\langle
1/r_{12}\rangle = 0.69\,m_{\pi}$ for $\pi^-d$ scattering \cite{AI04}
agreeing well with the Ericson--Weise value $\langle 1/r_{12}\rangle =
0.64\,m_{\pi}$ \cite{TE88}.

We have shown that in the case of $K^-d$ scattering the Ericson--Weise
term, caused by elastic three--body $K^-(pn)_{{^3}{\rm S}_1}$
scattering, is defined by the S--wave scattering lengths of $\bar{K}N$
scattering $a^0_0$ and $a^1_0$ with isospin zero and one,
respectively, $a^I_0$ for $I = 0$ and $I = 1$. In our approach the
real part ${\cal R}e\,f^{K^-d}_0(0)$ of the S--wave amplitude of
$K^-d$ scattering near threshold is defined by the Ericson--Weise kind
expression and the contribution of the inelastic channels of
three--body reaction $K^-(pn)_{{^3}{\rm S}_1} \to K^-(pn)_{{^3}{\rm
S}_1}$.

The term of the Ericson--Weise formula, defining the {\it impulse
approximation}, takes the form of the superposition of the S--wave
scattering length $a^{K^-p}_0$ of $K^-p$ scattering and the S--wave
scattering length $a^{K^-n}_0$ of $K^-n$ scattering. The S--wave
scattering length $a^{K^-p}_0$ of $K^-p$ scattering has been computed
in \cite{IV3}. 

For the calculation of the S--wave scattering length $a^{K^-n}_0$ of
elastic $K^-n$ scattering we have followed \cite{IV3}. We have
represented the S--wave amplitude of $K^-n$ scattering near threshold
in the form of the contribution of the $\Sigma^-(1750)$ resonance and
the smooth elastic background. In our approach \cite{IV3} the smooth
elastic background is given by the contribution of defined by the
low--energy interactions, which can be described by the Effective
Chiral Lagrangians (ECL), and the exotic states such as the scalar
mesons $a_0(980)$ and $f_0(980)$, which are the four--quarks states
(or $K\bar{K}$ molecule), the description of which goes beyond the
scope of ECL. Unlike $K^-p$ scattering, where the contribution of the
exotic four--quark states $a_0(980)$ and $f_0(980)$ is very important
for the correct description of the smooth elastic background, the
scalar mesons $a_0(980)$ and $f_0(980)$ do not contribute to the real
part of the S--wave amplitude of $K^-n$ scattering near threshold. As
a result the smooth elastic background is fully determined by the
contribution, described by ECL. We have computed the smooth elastic
background for $K^-n$ scattering near threshold within the soft--kaon
technique\,\footnote{This is equivalent to the leading order in chiral
expansion of ChPT by Gasser and Leutwyler with a non--linear
realization of chiral $U(3)\times U(3)$ symmetry \cite{JG99} realizing
the ECL approach to the description of strong low--energy interactions
of hadrons.} and within the Effective chiral quark model with chiral
$U(3)\times U(3)$ symmetry \cite{AI99}. We have shown that these two
approaches lead to the values of the smooth elastic background of
$K^-n$ scattering which are compared within the accuracy about
10$\,\%$.

This has made the result of the calculation of the smooth elastic
background for $K^-p$ scattering, carried out in \cite{IV3} within the
Effective quark model with chiral $U(3)\times U(3)$ symmetry, more
credible. Recall, that the calculation of the smooth elastic
background for $K^-p$ scattering within the Effective quark model with
chiral $U(3)\times U(3)$ symmetry has been justified by the absence of
the theoretical and experimental information about the coupling
constants of the exotic scalar mesons $a_0(980)$ and $f_0(980)$ with
nucleons. In \cite{IV3} we have computed the smooth elastic background
for $K^-p$ scattering near threshold within the Effective Chiral
Lagrangian approach and fixed the coupling constants of the $SNN$
interactions, where $S = a_0(980)$ and $f_0(980)$.

The imaginary part ${\cal I}m\,f^{K^-d}_0(0)$ of the S--wave amplitude
of the three--body reaction $K^-(pn)_{{^3}{\rm S}_1} \to
K^-(pn)_{{^3}{\rm S}_1}$ is determined by the inelastic channels with
the two--body intermediate states $K^-(pn)_{{^3}{\rm S}_1} \to NY \to
K^-(pn)_{{^3}{\rm S}_1}$, where $NY$ is $n\Lambda^0$, $n\Sigma^0$ and
$p \Sigma^-$, and the three--body intermediate states
$K^-(pn)_{{^3}{\rm S}_1} \to NY\pi \to K^-(pn)_{{^3}{\rm S}_1}$.

We have computed the contributions of the two--body channels
$K^-(pn)_{{^3}{\rm S}_1} \to NY \to K^-(pn)_{{^3}{\rm S}_1}$, where
$NY$ is $n\Lambda^0$, $n\Sigma^0$ and $p \Sigma^-$. The calculation of
the amplitudes of the reactions $K^-(pn)_{{^3}{\rm S}_1} \to NY$ we
have carried out in the one--pseudoscalar and one--scalar meson
exchange approximation. The contribution of scalar meson is computed
in the infinite mass limit that corresponds to a non--linear
realization of chiral $U(3)\times U(3)$ symmetry.

We have shown that the $NY$ pairs in the reactions $K^-(pn)_{{^3}{\rm
S}_1} \to NY$ can be produced in the ${^3}{\rm P}_1$ and ${^1}{\rm
P}_1$ states. Accounting for the rescattering of the $np$ pair in the
${^3}{\rm S}_1$ state and the $NY$ pairs in the ${^3}{\rm P}_1$ and
${^1}{\rm P}_1$ states we have computed the S--wave amplitudes
$f^{K^-d}_0(0)_{NY}$ of $K^-d$ scattering near threshold, caused by
the two--body inelastic channels $K^-(pn)_{{^3}{\rm S}_1} \to NY \to
K^-(pn)_{{^3}{\rm S}_1}$. As has been pointed out in \cite{AI01} the
amplitudes of the rescattering of the $NY$ pairs produced near
threshold of the reactions $K^-(pn)_{{^3}{\rm S}_1} \to NY$ describe
effectively the contribution of the set of resonances with the quantum
numbers of the $NY$ pairs.

Using the DGBT formula for the energy level displacement we have
computed the energy level displacements of the ground state of kaonic
deuterium induced by the two--body inelastic channels
$K^-(pn)_{{^3}{\rm S}_1} \to NY \to K^-(pn)_{{^3}{\rm S}_1}$. Using
the experimental data on the rates of the production of the states
$NY$ is $n\Lambda^0$, $n\Sigma^0$ and $p \Sigma^-$ in the reactions
$K^-d \to NY$, we have estimated the expected value of the total width
of the energy level of the ground state of kaonic deuterium and the
contribution of the three--body inelastic channels to the shift of the
energy level of the ground state of kaonic deuterium. As a result, we
have found that the total energy level displacement of the ground
state of kaonic deuterium should be equal to
$$
-\,\epsilon_{1s} + i\,\frac{\Gamma_{1s}}{2} = 602\,f^{K^-d}_0(0) =
 (-\,353 \pm 60) + i\,(315 \pm 50)\;{\rm eV} 
$$
that corresponds to the S--wave amplitude of $K^-d$ scattering near
threshold\footnote{According to recent calculations of the S--wave
  scattering lengths of $\bar{K}N$ scattering \cite{IV8} $a^0_0 =
  (-1.50 \pm 0.05)\,{\rm fm}$ and $a^1_0 = (0.50 \pm 0.02)\,{\rm fm}$,
  the single scattering contribution to the S--wave amplitude of
  $K^-d$ scattering at threshold, defined by the impulse
  approximation, vanishes and the S--wave scattering length of $K^-d$
  scattering is equal to $f^{K^-d}_0(0) = (-\,0.584 \pm 0.095) +
  i\,(0.521 \pm 0.075)\,{\rm fm}$.}
$$
f^{K^-d}_0(0) = (-\,0.586 \pm 0.095) + i\,(0.521 \pm 0.075)\,{\rm fm}.
$$
The theoretical value of the S--wave amplitude of $K^-d$ scattering
near threshold and the approach, used for the calculation of the
contribution of the inelastic two--body channels, have been justified
by the description of the S--wave amplitude of $\pi^-d$ scattering and
the calculation of the contribution of the inelastic two--body channel
$\pi^-d \to (nn)_{{^3}{\rm P}_1} \to \pi^-d$ in agreement with the
experimental data on the energy level displacement of the ground state
of pionic deuterium \cite{AI04}.

Of course, the complete confirmation of the theoretical prediction for
the S--wave amplitude of $K^-d$ scattering and the energy level
displacement of the ground state of kaonic deuterium obtained above
should go through the calculation of the contribution of the
three--body inelastic channels, which we are planning to carry out in
our forthcoming publication. Indeed, there are seven three--body
inelastic channels $K^-(pn)_{{^3}{\rm S}_1} \to (NY)_{{^3}{\rm
S}_1}\pi \to K^-(pn)_{{^3}{\rm S}_1}$ with $NY\pi = n \Sigma^- \pi^+,
p \Sigma^- \pi^0, n \Sigma^0 \pi^0, p\Sigma^0 \pi^-, n \Sigma^+ \pi^-,
n \Lambda^0 \pi^0$ and $p \Lambda^0 \pi^-$, which would exceed a
reasonable size of this paper.

The theoretical value of the energy level displacement of the ground
state of kaonic deuterium obtained above can be used for the planning
of experiments on the measurement of the energy level displacement of
kaonic deuterium by the DEAR Collaboration at Frascati.

\section*{Acknowledgement}

We are grateful to Torleif Ericson, Alexander Kobushkin and Yaroslav
Berdnikov for numerous fruitful discussions.

\newpage

\section*{Appendix A. Wave function of the $np$ pair in the bound state
with isospin zero and spin one}

In this Appendix we show that the wave function of the deuteron
(\ref{label2.6}) describes the bound $np$ pair in the state with
isospin $I = 0$. In terms of the operators of creation and
annihilation of the proton and the neutron the operators of isospin
read \cite{SS61}
$$
\hat{I}_+ = \sum_{\alpha = \pm 1/2}\int \frac{d^3Q}{(2\pi)^3 2
E(\vec{Q}\,)}\,a^{\dagger}_p(\vec{Q},\alpha)a_n(\vec{Q},\alpha),
$$
$$
\hat{I}_- = \sum_{\alpha = \pm 1/2}\int \frac{d^3Q}{(2\pi)^3 2
E(\vec{Q}\,)}\,a^{\dagger}_n(\vec{Q},\alpha)a_p(\vec{Q},\alpha),
$$
$$
\hat{I}_0 = \frac{1}{2}\sum_{\alpha = \pm 1/2}\int
\frac{d^3Q}{(2\pi)^3 2
E(\vec{Q}\,)}\,[a^{\dagger}_p(\vec{Q},\alpha)a_p(\vec{Q},\alpha) -
a^{\dagger}_n(\vec{Q},\alpha)a_n(\vec{Q},\alpha)],\eqno({\rm A}.1)
$$
where the operator $\hat{I}_+(\hat{I}_-)$ increases (decreases) the
isospin eigenvalue of the state. Using the anti--commutation relations
for the operators of creation and annihilation of the proton and the
neutron \cite{IV1} one can show that the operators ({\rm A}.1) obey
the commutation relation
$$
[\hat{I}_{\pm}, \hat{I}_0] = \mp\,\hat{I}_{\pm} \quad,\quad 
[\hat{I}_+, \hat{I}_-] = 2\,\hat{I}_0.  \eqno({\rm A}.2)
$$
In order to show that the wave functions (\ref{label2.8}) describe the
bound $np$ pair with the isospin zero we have to act by the operator
$\hat{I}_+$ on the wave function (\ref{label2.8}). Acting $\hat{I}_+$
on $|d(\vec{k}_d,\lambda_d = \pm 1)\rangle $ and using the
anti--commutation relations for the operators of creation and
annihilation of the proton and the neutron \cite{IV1} we get
$$
\hat{I}_+|A^{(1s)}_{K d}(\vec{P},\lambda_d = \pm 1)\rangle =
-\,\frac{1}{(2\pi)^3}\int \frac{d^3k_K}{\sqrt{2
E_K(\vec{k}_K)}}\frac{d^3k_d}{\sqrt{2
E_d(\vec{k}_d)}}\,\sqrt{2 E^{(1s)}_A(\vec{k}_K +
\vec{k}_d)}
$$
$$
\times\,\,\delta^{(3)}(\vec{P} - \vec{k}_K -
\vec{k}_d)\,\Phi_{1s}(\vec{k}_K)\,\frac{1}{(2\pi)^3}\int
\frac{d^3k_p}{\sqrt{2 E_p(\vec{k}_p)}}\frac{d^3k_n}{\sqrt{2
E_n(\vec{k}_n)}}\,\sqrt{2 E_d(\vec{k}_p + \vec{k}_n)}
$$
$$
\times\, \delta^{(3)}(\vec{k}_d - \vec{k}_p -
\vec{k}_n)\Phi_{d}\Big(\frac{\vec{k}_p - \vec{k}_n}{2}\Big)\,
a^{\dagger}_p(\vec{k}_n, \pm 1/2)a^{\dagger}_p(\vec{k}_p, \pm
1/2)|0\rangle.\eqno({\rm A}.3)
$$
Permuting the operators 
$$
a^{\dagger}_p(\vec{k}_n, \pm
1/2)a^{\dagger}_p(\vec{k}_p, \pm 1/2) = - a^{\dagger}_p(\vec{k}_p, \pm
1/2)a^{\dagger}_p(\vec{k}_n, \pm 1/2)\eqno({\rm A}.4)
$$ 
and making a change of
variables $\vec{k}_n \to \vec{k}_p$ and $\vec{k}_p \to \vec{k}_n$ we
transcribe the r.h.s. of ({\rm A}.3) to the form
$$
\hat{I}_+|A^{(1s)}_{K d}(\vec{P},\lambda_d = \pm 1)\rangle =
+\,\frac{1}{(2\pi)^3}\int \frac{d^3k_K}{\sqrt{2
E_K(\vec{k}_K)}}\frac{d^3k_d}{\sqrt{2
E_d(\vec{k}_d)}}\,\sqrt{2 E^{(1s)}_A(\vec{k}_K +
\vec{k}_d)}
$$
$$
\times\,\,\delta^{(3)}(\vec{P} - \vec{k}_K -
\vec{k}_d)\,\Phi_{1s}(\vec{k}_K)\,\frac{1}{(2\pi)^3}\int
\frac{d^3k_p}{\sqrt{2 E_p(\vec{k}_p)}}\frac{d^3k_n}{\sqrt{2
E_n(\vec{k}_n)}}\,\sqrt{2 E_d(\vec{k}_p + \vec{k}_n)}
$$
$$
\times\, \delta^{(3)}(\vec{k}_d - \vec{k}_p -
\vec{k}_n)\Phi_{d}\Big(\frac{\vec{k}_p - \vec{k}_n}{2}\Big)\,
a^{\dagger}_p(\vec{k}_n, \pm 1/2)a^{\dagger}_p(\vec{k}_p, \pm
1/2)|0\rangle.\eqno({\rm A}.5)
$$
where we have used the symmetry of the wave function
$\Phi_d(\vec{k}\,) = \Phi_d(-\,\vec{k}\,)$. The relations ({\rm A}.4)
and ({\rm A}.5) testify that
$$
\hat{I}_+|d(\vec{k}_d,\lambda_d = \pm 1)\rangle = 0.\eqno({\rm A}.6)
$$
In analogy one can show that 
$$
\hat{I}_+|A^{(1s)}_{\pi d}(\vec{P},\lambda_d = \pm 1)\rangle = 0.\eqno({\rm A}.7)
$$
Hence, one obtains that
$$
\hat{I}_+|A^{(1s)}_{K d}(\vec{P},\lambda_d = 0)\rangle =
0.\eqno({\rm A}.8)
$$
Applying the same procedure one gets that
$$
\hat{I}_-|A^{(1s)}_{K d}(\vec{P},\lambda_d )\rangle =
\hat{I}_0|A^{(1s)}_{K d}(\vec{P},\lambda_d )\rangle = 0.\eqno({\rm
A}.9)
$$
This testifies that the wave functions (\ref{label2.8}) describe the
bound $np$ pair in the state with isospin zero.

One can also easily show that the wave function (\ref{label2.8})
describes the $np$ pair in the bound state with spin one. The spin
operator of the $np$ pair can be defined as
$$
\hat{\vec{S}} = \hat{\vec{s}}_p\otimes \hat{1}_n +
\hat{1}_p\otimes\hat{\vec{s}}_n.
$$
It is obvious that 
$$
\hat{S}^{+}|A^{(1s)}_{K d}(\vec{P},\lambda_d = + 1 )\rangle =
\hat{S}^{-}|A^{(1s)}_{K d}(\vec{P},\lambda_d = - 1 )\rangle = 0,
$$
$$
\hat{S}^{+}|A^{(1s)}_{K d}(\vec{P},\lambda_d = - 1 )\rangle =
\hat{S}^{-}|A^{(1s)}_{K d}(\vec{P},\lambda_d = + 1 )\rangle
\sqrt{2}\,|A^{(1s)}_{K d}(\vec{P},\lambda_d = 0 )\rangle. \eqno({\rm
A}.11)
$$
In turn one can easily show that
$$
\hat{S}^{+}|A^{(1s)}_{K d}(\vec{P},\lambda_d = 0)\rangle =
\sqrt{2}\,|A^{(1s)}_{K d}(\vec{P},\lambda_d = + 1)\rangle,
$$
$$
\hat{S}^{-}|A^{(1s)}_{K d}(\vec{P},\lambda_d = 0 )\rangle =
\sqrt{2}\,|A^{(1s)}_{K d}(\vec{P},\lambda_d = - 1 )\rangle.
\eqno({\rm A}.12)
$$
Thus, the wave function $|A^{(1s)}_{K d}(\vec{P},\lambda_d)\rangle$
given by (\ref{label2.6}) describes the $np$ pair in the bound
${^3}{\rm S}_1$ state with isospin zero, $I = 0$.

\section*{Appendix B. Smooth elastic background of $K^-n$ scattering 
within Effective quark model with chiral $U(3)\times U(3)$ symmetry}

Using the expression for the external sources $\eta_n(x_2)$ and
$\bar{\eta}_n(x_3)$, given by (\ref{label3.8}), and substituting them
in (\ref{label3.11}) we obtain
$$
M(K^- n \to K^- n) = i\,\frac{1}{4}\,g^2_{\rm
B}\,g^2_K\,\varepsilon^{i\,'j\,'k\,'}\,\varepsilon^{ijk}\int d^4x_1
d^4x_2 d^4x_3\,e^{\textstyle i\,q\,' \cdot x_1 + ip\,'\cdot x_2 -
ip\cdot x_3}
$$
$$
\times\,\bar{u}(p\,',\sigma\,'\,)_a (i\gamma^5)_{a_1b_1}
(C\gamma^{\mu})_{a_2b_2} (\gamma_{\mu}\gamma^5)_{a c_2}
(\gamma_{\nu}\gamma^5)_{c_3 b}(\gamma_{\nu}C)_{a_3b_3}
(i\gamma^5)_{a_4b_4}u(p,\sigma)_b
$$
$$
\hspace{-0.1in}\times\langle 0|{\rm
T}(\bar{u}_{\ell}(x_1)_{a_1}s_{\ell}(x_1)_{b_1}
d_{i\,'}(x_2)_{a_2}d_{j\,'}(x_2)_{b_2}
u_{k\,'}(x_2)_{c_2}\bar{u}_i(x_3)_{c_3}\bar{d}_j(x_3)_{a_3}
\bar{d}_k(x_3)_{b_3} \bar{s}_t(0)_{a_4} u_t(0)_{b_4})|0\rangle_c,
\eqno({\rm B}.1)
$$
where the index $c$ stands for the abbreviation {\it connected}.

Making contractions of the $u$-- and $s$--quark field operators we
reduce the r.h.s of ({\rm B}.1) to the form
$$
M(K^- n \to K^- n) = \frac{1}{4}\,g^2_{\rm
B}\,g^2_K\,\varepsilon^{i i\,'j\,'}\,\varepsilon^{ijk}\int d^4x_1
d^4x_2 d^4x_3\,e^{\textstyle i\,q\,' \cdot x_1 + ip\,'\cdot x_2 -
ip\cdot x_3}
$$
$$
\times\,\bar{u}(p\,',\sigma\,'\,)_a (\gamma^5)_{a_1b_1}
(C\gamma^{\mu})_{a_2b_2} (\gamma_{\mu}\gamma^5)_{a c_2}
(\gamma_{\nu}\gamma^5)_{c_3 b}(\gamma_{\nu}C)_{a_3b_3}
(\gamma^5)_{a_4b_4}u(p,\sigma)_b
$$
$$
\times\,S^{(s)}_F(x_1)_{b_1 a_4}\,S^{(u)}_F(- x_3)_{b_4
c_3}\,S^{(u)}_F(x_2 - x_1)_{c_2 a_1}
$$
$$
\times\,\langle 0|{\rm T}( d_{i\,'}(x_2)_{a_2}d_{j\,'}(x_2)_{b_2}
\bar{d}_j(x_3)_{a_3} \bar{d}_k(x_3)_{b_3})|0\rangle_c, \eqno({\rm
B}.2)
$$
The requirement to deal with only {\it connected} quark diagrams
prohibits the contraction of the $u$--quark field operators
$\bar{u}_{\ell}(x_1)_{a_1}$ and $u_{\ell}(0)_{b_4}$ and
$u_{k\,'}(x_2)_{c_2}$ and $\bar{u}_i(x_3)_{c_3}$. Contracting the
$d$--quark field operators we get
$$
M(K^- n \to K^- n) = \frac{3}{2}\,g^2_{\rm B}\,g^2_K\int d^4x_1
d^4x_2 d^4x_3\,e^{\textstyle i\,q\,' \cdot x_1 + ip\,'\cdot x_2 -
ip\cdot x_3}
$$
$$
\times\,\bar{u}(p\,',\sigma\,'\,)_a (\gamma^5)_{a_1b_1}
(C\gamma^{\mu})_{a_2b_2} (\gamma_{\mu}\gamma^5)_{a c_2}
(\gamma_{\nu}\gamma^5)_{c_3 b}(\gamma_{\nu}C)_{a_3b_3}
(\gamma^5)_{a_4b_4}u(p,\sigma)_b
$$
$$
\times\,S^{(s)}_F(x_1)_{b_1 a_4}\,S^{(u)}_F(- x_3)_{b_4
c_3}\,S^{(u)}_F(x_2 - x_1)_{c_2 a_1}
$$
$$
\times [S^{(d)}_F(x_2 - x_3)_{b_2 a_3}\,S^{(d)}_F(x_2 - x_3)_{a_2 b_3}
+ S^{(d)}_F(x_2 - x_3)_{a_2 a_3}\,S^{(d)}_F(x_2 - x_3)_{b_2
b_3}].\eqno({\rm B}.3)
$$
Summing over the indices we end up with the expression
$$
M(K^- p \to K^- p) = -\,3\,g^2_{\rm B}\,g^2_K\int d^4x_1 d^4x_2
d^4x_3\,e^{\textstyle i\,q\,' \cdot x_1 + ip\,'\cdot x_2 - ip\cdot
x_3}
$$
$$
\times\,\bar{u}(p\,',\sigma\,'\,)\gamma^{\mu}\gamma^5S^{(u)}_F(x_2 -
x_1) \gamma^5 S^{(s)}_F(x_1) \gamma^5 S^{(u)}_F(- x_3)
\gamma_{\nu}\gamma^5 u(p,\sigma)
$$
$$
\times\,{\rm tr}\{\gamma^{\mu} S^{(d)}_F(x_2 - x_3) \gamma^{\nu}S^{(d)}_F(x_3 - x_2)\},\eqno({\rm B}.4)
$$
where we have used the relations $C^T\gamma^T_{\nu} = \gamma_{\nu}C$
and 
$$
CS^{(d)}_F(x_2 - x_3)^T C= - S^{(d)}_F(x_3 - x_2).\eqno({\rm B}.5)
$$
In the momentum representation the r.h.s. of ({\rm B}.5) reads
$$
M(K^- n \to K^- n) = 3\,g^2_{\rm B}\,g^2_K
$$
$$
\times \int \frac{d^4k_1}{(2\pi)^4 i}\,
\bar{u}(p\,',\sigma\,'\,)\gamma^{\mu}\gamma^5\frac{1}{m_u -
\hat{k}_1}\gamma^5 \frac{1}{m_s - \hat{k}_1 - \hat{q}\,'}\gamma^5
\frac{1}{m_u - \hat{k}_1 - \hat{p} + \hat{p}\,'}\gamma^{\nu}\gamma^5
u(p,\sigma)
$$
$$
\times \int \frac{d^4k_2}{(2\pi)^4 i}{\rm tr}\Big\{\gamma^{\mu}
\frac{1}{m_d - \hat{k}_2}\gamma^{\nu} \frac{1}{m_d - \hat{k}_2 -
\hat{k}_1 + \hat{p}\,'}\Big\}.\eqno({\rm B}.7)
$$
The result of the calculation of momentum integrals within the
procedure accepted in the Effective quark model with chiral
$U(3)\times U(3)$ symmetry \cite{AI99}--\cite{AI92} is equal to
$$
M(K^- n \to K^- n) =- (2m - m_s)\,\frac{g^2_B}{8\pi^2}\,\frac{\langle
\bar{q}q\rangle}{F^2_K}\,\frac{m}{m_N}\,
$$
$$
\times\,\frac{m_s + m}{m_s - m}\Big[m^2_s\,{\ell n}\Big(1 +
\frac{\Lambda^2_{\chi}}{m^2_s}\Big) + (m^2_s - 2m^2)\,{\ell
n}\Big(1 + \frac{\Lambda^2_{\chi}}{m^2}\Big)\Big],\eqno({\rm B}.8)
$$
where $\langle \bar{q}q\rangle = - (252.630 {\rm MeV})^3$ is the quark
condensate, $\Lambda_{\chi} = 940\,{\rm MeV}$ is the scale of the
spontaneous breaking of chiral symmetry \cite{AI99,AI92}. The parameter
$A^{K^-n}_B$ is given by
$$
A^{K^-n}_B = \frac{M(K^- n \to K^- n)}{8\pi (m_K + m_N)} = -
\frac{g^2_B}{64\pi^3}\frac{\langle \bar{q}q\rangle}{F^2_K}
\frac{m}{m_N}\frac{2m - m_s}{m_K + m_N}
$$
$$
\times\,\frac{m_s + m}{m_s - m}\Big[m^2_s {\ell n}\Big(1 +
\frac{\Lambda^2_{\chi}}{m^2_s}\Big) + (m^2_s - 2 m^2) {\ell n}\Big(1 +
\frac{\Lambda^2_{\chi}}{m^2}\Big)\Big] = (0.221 \pm 0.022)\,{\rm
fm}. \eqno({\rm B}.9)
$$
A theoretical accuracy of this result is about of 10$\%$
\cite{AI99}--\cite{AI92} and \cite{AI80}.

\section*{Appendix C. Spinorial wave functions of $np$ and $NY$ pairs 
in the ${^3}{\rm S}_1$, ${^3}{\rm P}_1$ and ${^1}{\rm P}_1$ states}

In this Appendix we compute the spinorial wave functions of the pairs
$(np)_{{^3}{\rm S}_1}$, $(NY)_{{^3}{\rm P}_1}$ and $(NY)_{{^1}{\rm
S}_1}$ coupled in the ${^3}{\rm S}_1$, ${^3}{\rm P}_1$ and ${^1}{\rm
P}_1$ states. 

\subsection*{\bf 1. $(np)_{{^3}{\rm S}_1}$ pair}

In the quantum field theoretic approach the $np$ in the ${^3}{\rm S}_1$
state is described by the product of the neutron and proton Dirac
bispinors
$$
\vec{N}_{\sigma_n\sigma_p}({^3}{\rm S}_1) =
\bar{u^c}(-\vec{K},\sigma_n)\vec{\gamma}\,u(\vec{K},\sigma_p).
\eqno({\rm C}.1)
$$
In terms of spinorial wave functions the wave function of the $np$
pair reads \cite{TE88}
$$
\vec{N}_{\sigma_n\sigma_p}({^3}{\rm S}_1)
=\bar{u^c}(-\vec{K},\sigma_n)\vec{\gamma}\,u(\vec{K},\sigma_p) = 2m_N
\, \chi^{\dagger}(\sigma_n)\,i\sigma_2\,\vec{\sigma}\,\chi(\sigma_p),
\eqno({\rm C}.2)
$$
where we have used the relation $(\vec{\sigma}\cdot
\vec{k}\,)\vec{\sigma}(\vec{\sigma}\cdot \vec{k}\,) = -
\vec{k}^{\,2}\vec{\sigma} + 2\,\vec{k}\,(\vec{\sigma}\cdot
\vec{k}\,)$.  For different polarizations of the neutron and the
proton in the $np$ pair the wave function ({\rm C}.2) has the
following components
$$
\vec{N}_{+ 1/2 + 1/2}({^3}{\rm S}_1) = \bar{u^c}(- \vec{K},+ 1/2)
\vec{\gamma}\, u( \vec{K}, + 1/2) = - 2m_N(- \vec{e}_x - i\,\vec{e}_y)
= - 2m_N \sqrt{2}\,\vec{e}_{+1},
$$
$$
\vec{N}_{- 1/2 - 1/2}({^3}{\rm S}_1) = \bar{u^c}(- \vec{K},- 1/2)
\vec{\gamma}\, u( \vec{K}, - 1/2) = - 2m_N(+ \vec{e}_x - i\,\vec{e}_y)
= - 2 m_N \sqrt{2}\,\vec{e}_{-1},
$$
$$
\vec{N}_{+ 1/2 - 1/2}({^3}{\rm S}_1) = \bar{u^c}(- \vec{K},+ 1/2)
\vec{\gamma}\, u( \vec{K}, - 1/2) = - 2m_N \vec{e}_z = - 2m_N
\vec{e}_0,
$$
$$
\vec{N}_{- 1/2 + 1/2}({^3}{\rm S}_1) = \bar{u^c}(- \vec{K},- 1/2)
\vec{\gamma}\, u( \vec{K}, + 1/2)= - 2m_N \vec{e}_z = - 2m_N
\vec{e}_0, \eqno({\rm C}.3)
$$
where $\vec{e}_i\,(i = x,y, z)$ are unit orthogonal vectors of
Cartesian coordinate system and $\vec{e}_{\pm 1} = \mp ( \vec{e}_x \pm
i\,\vec{e}_y)/\sqrt{2}$ and $ \vec{e}_0 = \vec{e}_z$ are cyclic unit
vectors of the spherical coordinate system, corresponding to the
states with $|{^3}{\rm S}_1; 1, + 1\rangle$, $|{^3}{\rm S}_1; 1, -
1\rangle$ and $|{^3}{\rm S}_1; 1, 0 \rangle$ \cite{DV88}, which
describe the spinorial states of the $np$ pair in the reaction $K^-
(pn)_{{^3}{\rm S}_1} \to K^- (pn)_{{^3}{\rm S}_1}$.

Thus, the wave functions $|{^3}{\rm S}_1; 1, \lambda_d \rangle$ with
$\lambda_d = 0, \pm 1$ read
$$
 \vec{N}_{+1/2 +1/2}({^3}{\rm S}_1) = - 2\sqrt{2}\,m_N \,\vec{e}_{+1}
= |{^3}{\rm S}_1; 1, + 1\rangle,
$$
$$
\vec{N}_{-1/2 -1/2}({^3}{\rm S}_1) = - 2\sqrt{2}\,
m_N \,\vec{e}_{-1} = |{^3}{\rm S}_1; 1, - 1\rangle,
$$
$$
 \frac{1}{\sqrt{2}}\,[\vec{N}_{+1/2 -1/2}({^3}{\rm S}_1) + \vec{N}_{-
1/2 +1/2}({^3}{\rm S}_1)] = - 2\sqrt{2}\, m_N \,\vec{e}_0 = |{^3}{\rm
S}_1; 1, 0 \rangle.\eqno({\rm C}.4)
$$
The result of the average over polarizations of the $np$ pair in the
${^3}{\rm S}_1$ state in (\ref{label4.10}) can be obtained as follows
$$
\frac{1}{3}\sum_{(\sigma_p,\sigma_n; {^3}{\rm
S}_1)}[\bar{u}(\vec{K},\sigma_p) \vec{\gamma}\,u^c(-
\vec{K},\sigma_n)]\cdot
[\bar{u^c}(-\vec{Q},\sigma_n)\vec{\gamma}\,u(\vec{Q},\sigma_p)] =
$$
$$
= \frac{1}{3}\,[\langle 1, + 1; {^3}{\rm S}_1|{^3}{\rm S}_1; 1, +
1\rangle + \langle 1, - 1; {^3}{\rm S}_1|{^3}{\rm S}_1; 1, - 1\rangle
+ \langle 1, 0; {^3}{\rm S}_1|{^3}{\rm S}_1; 1, 0 \rangle] =
8m^2_N,\eqno({\rm C}.5)
$$
where we have used the normalization and orthogonality relations for
the unit cyclic vectors \cite{DV88}
$$
\vec{e}^{\;*}_{\pm 1}\cdot \vec{e}_{\pm 1} = \vec{e}^{\;2}_0 = 1,
$$
$$
\vec{e}^{\;*}_{\pm 1}\cdot \vec{e}_{\mp 1} = \vec{e}^{\;*}_{\pm
1}\cdot \vec{e}_0 = \vec{e}_{\pm 1} \cdot \vec{e}_0 = 0 \eqno({\rm
C}.6)
$$
The equation ({\rm C}.5) defines the result of the averaging over
polarizations of the $(np)_{{^3}{\rm S}_1}$ pair for the derivation of
the Ericson--Weise formula of the S--wave scattering length of $K^-d$
scattering (\ref{label4.12}).

\subsection*{\bf 2. $(NY)_{{^3}{\rm P}_1}$ pair}

The wave function of the $NY$ pair in the state ${^3}{\rm P}_1$,
coupled to the $np$ pair in the ${^3}{\rm S}_1$ state near threshold
of the reaction $(NY)_{{^3}{\rm P}_1}\to K^- (p n)_{{^3}{\rm S}_1}$, we
denote as
$$
\vec{N}_{\alpha_1\alpha_2}({^3}{\rm P}_1) =
[\bar{u^c}(\vec{k},\alpha_1) \vec{\gamma}\, \gamma^5u(-
\vec{k},\alpha_2)].\eqno({\rm C}.7)
$$
In terms of spinorial wave functions the wave function ({\rm C}.7)
reads \cite{TE88}
$$
\vec{N}_{\alpha_1\alpha_2}({^3}{\rm P}_1) = \bar{u^c}(\vec{k},\alpha_1) \vec{\gamma}
\,\gamma^5u(- \vec{k},\alpha_2) = -\,2i\,\varphi^T(\alpha_1)\,i
\sigma_2\,(\vec{k}\times \vec{\sigma}\,)\,\varphi(\alpha_2).\eqno({\rm
C}.8)
$$
For different polarizations of the neutrons the wave function of the
$NY$ pair in the ${^3}{\rm P}_1$ state has the following components
$$
\vec{N}_{+ 1/2 + 1/2}({^3}{\rm P}_1) = \bar{u^c}(\vec{k}, + 1/2)
\vec{\gamma} \, \gamma^5u(- \vec{k}, +1/2) = -2 i\,\vec{k}\times
(\vec{e}_x + i\,\vec{e}_y) = i\,2\sqrt{2}\,(\vec{k}\times
\vec{e}_{+1}),
$$
$$
\vec{N}_{- 1/2 - 1/2}({^3}{\rm P}_1) = \bar{u^c}(\vec{k}, - 1/2)
\vec{\gamma} \,\gamma^5u(- \vec{k}, - 1/2) = +2 i\,\vec{k}\times
(\vec{e}_x - i\,\vec{e}_y) = i\,2 \sqrt{2}\,(\vec{k}\times
\vec{e}_{-1}),
$$
$$
\vec{N}_{+ 1/2 - 1/2}({^3}{\rm P}_1) = \bar{u^c}(\vec{k}, + 1/2)
\vec{\gamma} \,\gamma^5u(- \vec{k}, - 1/2) = -2 i\,(\vec{k}\times
\vec{e}_z) = - 2 i\,(\vec{k}\times \vec{e}_0),
$$
$$
\vec{N}_{- 1/2 + 1/2}({^3}{\rm P}_1) = \bar{u^c}(\vec{k}, - 1/2)
\vec{\gamma} \,\gamma^5u(- \vec{k}, + 1/2) = - 2 i\,(\vec{k}\times
\vec{e}_z) = - 2 i\,(\vec{k}\times \vec{e}_0). \eqno({\rm C}.9)
$$
For the subsequent analysis it is convenient to use the expansion of
the momentum $\vec{k}$ into cyclic unit vectors. It reads \cite{DV88}
$$
\vec{k} = - k_{+1}\,\vec{e}_{-1} - k_{-1}\,\vec{e}_{+1} +
k_0\,\vec{e}_0,\eqno({\rm C}.10)
$$ 
where we have denoted \cite{DV88}
$$
 k_{+1} = -\,\frac{k_x + i\,k_y}{\sqrt{2}} =
k\,\sqrt{\frac{4\pi}{3}}\,Y_{1,+ 1}\;,\; k_{-1} = +\,\frac{k_x -
i\,k_y}{\sqrt{2}} = k\,\sqrt{\frac{4\pi}{3}}\,Y_{1,- 1}\;,
$$
$$
k_0 = k_z = k\,\sqrt{\frac{4\pi}{3}}\,Y_{1,0}.\eqno({\rm C}.11)
$$  
Here $Y_{1,M}$ are spherical harmonics in the momentum space,
describing the states with the angular momentum $L = 1$ and a magnetic
quantum number $M = 0, \pm 1$ \cite{DV88}.  Using the relations
\cite{DV88}
$$
\vec{e}_{+1}\times \vec{e}_{-1} = i\,\vec{e}_0\;,\, \vec{e}_0 \times
\vec{e}_{\pm 1} = \mp\,i\,\vec{e}_{\pm 1}\eqno({\rm C}.12)
$$
we obtain
$$
\vec{k}\times \vec{e}_{+1} = +\,i\,k_0\,\vec{e}_{+1} - \,
i\,k_{+1}\,\vec{e}_0,
$$
$$
\vec{k}\times \vec{e}_{-1} = + \,i\,k_0\,\vec{e}_{-1} -
\,i\,k_{-1}\,\vec{e}_0,
$$
$$
\vec{k}\times \vec{e}_0 = +\,i\,k_{+1}\,\vec{e}_{-1} -
\,i\,k_{-1}\,\vec{e}_{+1}.\eqno({\rm C}.13)
$$
This gives the following components of the wave function of the $NY$
pair in the ${^3}{\rm P}_1$ state
$$
\vec{N}_{+ 1/2 + 1/2}({^3}{\rm P}_1) = \bar{u^c}(\vec{k}, + 1/2)
\vec{\gamma} \,\gamma^5u(- \vec{k}, +1/2) =
2\sqrt{2}\,(k_{+1}\,\vec{e}_0 - \,k_0\,\vec{e}_{+1}),
$$
$$
\vec{N}_{- 1/2 - 1/2}({^3}{\rm P}_1) = \bar{u^c}(\vec{k}, - 1/2)
\vec{\gamma} \,\gamma^5u(- \vec{k}, - 1/2) =
2\sqrt{2}\,(k_{-1}\,\vec{e}_0 - \,k_0\,\vec{e}_{-1}),
$$
$$
\vec{N}_{+ 1/2 - 1/2}({^3}{\rm P}_1) = \bar{u^c}(\vec{k}, + 1/2)
\vec{\gamma} \,\gamma^5u(- \vec{k}, - 1/2) = 2(k_{+1}\,\vec{e}_{-1} -
k_{-1}\,\vec{e}_{+1}),
$$
$$
\vec{N}_{- 1/2 + 1/2}({^3}{\rm P}_1) = \bar{u^c}(\vec{k}, - 1/2)
\vec{\gamma} \,\gamma^5u(- \vec{k}, + 1/2) = 2(k_{+1}\,\vec{e}_{-1} -
k_{-1}\,\vec{e}_{+1}). \eqno({\rm C}.14)
$$
The wave functions ({\rm C}.13) have the properties of the wave
functions $|{^3}{\rm P}_1; J, J_z\rangle$ with a total momentum $J =
1$ and $J_z = 0, \pm 1$.

Thus, the wave function of the $NY$ pair in the ${^3}{\rm P}_1$ state
can be defined by
$$
\vec{N}_{+ 1/2 + 1/2}({^3}{\rm P}_1) = \bar{u^c}(\vec{k}, + 1/2)
\vec{\gamma} \,\gamma^5u(- \vec{k}, +1/2) =
2\sqrt{2}\,(k_{+1}\,\vec{e}_0 - \,k_0\,\vec{e}_{+1}) = |{^3}{\rm P}_1;
1, + 1\rangle,
$$
$$
\vec{N}_{- 1/2 - 1/2}({^3}{\rm P}_1) = \bar{u^c}(\vec{k}, - 1/2)
\vec{\gamma} \,\gamma^5u(- \vec{k}, - 1/2) =
2\sqrt{2}\,(k_{-1}\,\vec{e}_0 - \,k_0\,\vec{e}_{-1}) = |{^3}{\rm P}_1;
1, - 1\rangle
$$
$$
\frac{1}{\sqrt{2}}\,[\vec{N}_{+ 1/2 - 1/2}({^3}{\rm P}_1) +
\vec{N}_{- 1/2 + 1/2}({^3}{\rm P}_1)] =
$$
$$
= \frac{1}{\sqrt{2}}\,[\bar{u^c}(\vec{k}, + 1/2) \vec{\gamma}
\,\gamma^5u(- \vec{k}, - 1/2) + \bar{u^c}(\vec{k}, - 1/2) \vec{\gamma}
\,\gamma^5u(- \vec{k}, + 1/2)] =
$$
$$
= 2\sqrt{2}\,(k_{+1}\,\vec{e}_{-1} - k_{-1}\,\vec{e}_{+1}) =
|{^3}{\rm P}_1; 1, 0\rangle.\eqno({\rm C}.15)
$$
The product of the spinorial wave functions of the reaction
$(NY)_{{^3}{\rm P}_1} \to K^- (p n)_{{^3}{\rm S}_1}$, squared, summed
over polarizations of the $(NY)_{{^3}{\rm P}_1}$ pair and averaged
over polarizations of the $(np)_{{^3}{\rm S}_1}$ pair, reads
$$
\frac{1}{3}\sum_{(\sigma_p,\sigma_n; {^3}{\rm S}_1)}
\sum_{(\alpha_2,\alpha_1; {^3}{\rm P}_1)}|[\bar{u}(\vec{K},\sigma_p)
\vec{\gamma}\, u^c( - \vec{K},\sigma_n)]\cdot
[\bar{u^c}(\vec{k},\alpha_1) \vec{\gamma}\,\gamma^5u(-
\vec{k},\alpha_2)]|^2 =
$$
$$
= \frac{64}{3}\,m^2_N\,[(k_{+1}\,\vec{e}_0 - \,k_0\,\vec{e}_{+1})^2 +
(k_{-1}\,\vec{e}_0 - \,k_0\,\vec{e}_{-1})^2 + (k_{+1}\,\vec{e}_{-1} -
k_{-1}\,\vec{e}_{+1})^2\,].\eqno({\rm C}.15)
$$
Using the orthogonality and normalization relations ({\rm C}.6) we get
$$
\frac{1}{3}\sum_{(\sigma_p,\sigma_n; {^3}{\rm S}_1)}
\sum_{(\alpha_2,\alpha_1; {^3}{\rm P}_1)}|[\bar{u}(\vec{K},\sigma_p)
\vec{\gamma}\, u^c( - \vec{K},\sigma_n)]\cdot
[\bar{u^c}(\vec{k},\alpha_1) \vec{\gamma}\,\gamma^5u(-
\vec{k},\alpha_2)]|^2 =
$$
$$
= \frac{128}{3}\,m^2_N\,(k^2_0 + k_{+1}k^*_{+1} + k_{-1}k^*_{-1}) =
\frac{128}{3}\,m^2_N\,\vec{k}^{\;2}.\eqno({\rm C}.16)
$$
This average value defines the S--wave amplitude of the reaction $K^-
(p n)_{{^3}{\rm S}_1} \to K^- (p n)_{{^3}{\rm S}_1}$ saturated by the
intermediate $(NY)_{{^3}{\rm P}_1}$ state.

\subsection*{\bf 3. $(NY)_{{^1}{\rm P}_1}$ pair}.

The wave function of the $NY$ pair in the state ${^1}{\rm P}_1$,
coupled to the $np$ pair in the ${^3}{\rm S}_1$ state near threshold
of the reaction $(NY)_{{^1}{\rm P}_1}\to K^- (p n)_{{^3}{\rm S}_1}$, we
denote as
$$
\vec{N}_{\alpha_1\alpha_2}({^1}{\rm P}_1) =
[\bar{u^c}(\vec{k},\alpha_1)\gamma^0 \vec{\gamma}\, \gamma^5u(-
\vec{k},\alpha_2)].\eqno({\rm C}.17)
$$
In terms of spinorial wave functions the wave function ({\rm C}.17)
reads \cite{TE88}
$$
\vec{N}_{\alpha_1\alpha_2}({^1}{\rm P}_1) =
\bar{u^c}(\vec{k},\alpha_1) \gamma^0 \vec{\gamma} \,\gamma^5u(-
\vec{k},\alpha_2) = 2i\,\vec{k}\,\varphi^T(\alpha_1)\,i
\sigma_2\,\varphi(\alpha_2).\eqno({\rm C}.18)
$$
According to the expansion ({\rm C}.10) we can define three
independent states $|{^1}{\rm P}_1; J, J_z\rangle$ of the
$(NY)_{{^1}{\rm P}_1}$ pair. They read
$$
|{^1}{\rm P}_1; 1, + 1\rangle = \vec{N}^{(+)}_{\alpha_1\alpha_2}({^1}{\rm
P}_1) = - \,2i\,k_{+1}\,\varphi^T(\alpha_1)\,i
\sigma_2\,\varphi(\alpha_2),
$$
$$
|{^1}{\rm P}_1; 1, - 1\rangle =
\vec{N}^{(-)}_{\alpha_1\alpha_2}({^1}{\rm P}_1) = -
\,2i\,k_{-1}\,\varphi^T(\alpha_1)\,i
\sigma_2\,\varphi(\alpha_2), 
$$
$$
|{^1}{\rm P}_1; 1,~~ 0\rangle =
\vec{N}^{(0)}_{\alpha_1\alpha_2}({^1}{\rm P}_1) = +
\,2i\,k_0\,\varphi^T(\alpha_1)\,i
\sigma_2\,\varphi(\alpha_2). \eqno({\rm C}.19)
$$
Now we can calculate the average values of the squared matrix element
of the reaction $(NY)_{{^1}{\rm P}_1} \to K^- (p n)_{{^3}{\rm
S}_1}$. 

The product of the spinorial wave functions of the reaction
 $(NY)_{{^1}{\rm P}_1} \to K^- (pn)_{{^3}{\rm S}_1}$, squared and
 summed over polarizations of the coupled particles and averaged over
 polarizations of the $(np)_{{^3}{\rm S}_1}$ pair, is equal to
$$
\frac{1}{3}\sum_{(\sigma_p,\sigma_n; {^3}{\rm S}_1)} \sum_{\alpha_1 =
\pm ^/2}\sum_{\alpha_2 = \pm ^/2}|[\bar{u}(\vec{K},\sigma_p)
\vec{\gamma}\, u^c( - \vec{K},\sigma_n)]\cdot
[\bar{u^c}(\vec{k},\alpha_1) \gamma^0 \vec{\gamma}\,\gamma^5u(-
\vec{k},\alpha_2)]|^2 =
$$
$$
= \frac{32}{3}\,m^2_N\,\vec{k}^{\,2}\sum_{\alpha_1 = \pm
^/2}\sum_{\alpha_2 = \pm ^/2}|\varphi^T(\alpha_1)\,i
\sigma_2\,\varphi(\alpha_2)|^2 = \frac{64}{3}\,m^2_N\,
\vec{k}^{\,2}. \eqno({\rm C}.20)
$$

\section*{Appendix D. Coupling constants of $PBB$,  $SBB$ and $SPP$
interactions}

\subsection*{Phenomenological $PBB$ interactions}

The Lagrangian of the phenomenological $PBB$ interactions is defined
by \cite{MN79}
$$
{\cal L}_{PBB}(x) = \sqrt{2}\,g\,{\rm tr}\{\{\bar{B},i\gamma^5 B\}P\}
+ \sqrt{2}\,f\,{\rm tr}\{[\bar{B},i\gamma^5B]P\}=
$$
$$
= \sqrt{2}\,(g + f)\,\bar{B}^b_a i\gamma^5 B^a_cP^c_b + \sqrt{2}\,(g
- f)\,\bar{B}^b_a i\gamma^5B^c_b P^a_c,\eqno({\rm D}.1)
$$
where $g$ and $f$ are phenomenological coupling constants,
$\bar{B}^b_a(x)$, $B^b_a(x)$ and $P^a_b(x)$\,($a(b) = 1,2,3$)
are interpolating fields of the octets of light baryons and
pseudoscalar mesons, respectively:
$$
\bar{B}^b_a =\left(\begin{array}{llcl} {\displaystyle
\frac{\bar{\Sigma}^0}{\sqrt{2}} + \frac{\bar{\Lambda}^0}{\sqrt{6}}} &
\hspace{0.3in}\bar{\Sigma}^- & - \bar{\Xi}^- \\
\hspace{0.3in}\bar{\Sigma}^+ &{\displaystyle -
\frac{\bar{\Sigma}^0}{\sqrt{2}} +
\frac{\bar{\Lambda}^0}{\sqrt{6}}} & \bar{\Xi}^0 \\
\hspace{0.3in}\bar{p}& \hspace{0.3in} \bar{n} & {\displaystyle
-\frac{2}{\sqrt{6}}\bar{\Lambda}^0} \\
\end{array}\right),
$$
$$
B^b_a = \left(\begin{array}{llcl} {\displaystyle
\frac{\Sigma^0}{\sqrt{2}} + \frac{\Lambda^0}{\sqrt{6}}} &
\hspace{0.3in}\Sigma^+ & p \\ \hspace{0.3in}\Sigma^- &{\displaystyle -
\frac{\Sigma^0}{\sqrt{2}} + \frac{\Lambda^0}{\sqrt{6}}} & n \\
\hspace{0.15in}- \Xi^- & \hspace{0.3in}\Xi^0 & {\displaystyle
-\frac{2}{\sqrt{6}}\Lambda^0}
\end{array}\right),
$$
$$
P^a_b = \left(\begin{array}{llcl} {\displaystyle
\frac{\pi^0}{\sqrt{2}} + \frac{\eta}{\sqrt{6}}} &
\hspace{0.3in} \pi^+ & K^+\\ \hspace{0.3in} \pi^-
&{\displaystyle - \frac{\pi^0}{\sqrt{2}} +
\frac{\eta}{\sqrt{6}}} & K^0\\
\hspace{0.15in}- K^- & \hspace{0.3in} \bar{K}^0& {\displaystyle
-\frac{2}{\sqrt{6}}\,\eta} \\
\end{array}\right),\eqno({\rm D}.2)
$$
where $\bar{B}^2_1 = \bar{\Sigma}^-$, $B^2_1 = \Sigma^+$, $P^2_1 =
\pi^+$ and so on.  For simplicity we identify the eighth component of
the pseudoscalar octet $\eta(x)$ with the observed pseudoscalar meson
$\eta(550)$ \cite{DG00}.

In terms of the components of octets the Lagrangian ({\rm D}.1) reads
$$
{\cal L}_{PBB}(x) = \sqrt{2}\,(g + f)\,\bar{p}(x)i\gamma^5n(x)\pi^+(x)
+ \sqrt{2}\,(g + f)\,\bar{n}(x)i\gamma^5p(x)\pi^-(x)
$$
$$
 + (g + f)\,[\bar{p}(x)i\gamma^5p(x) -
\bar{n}(x)i\gamma^5n(x)]\,\pi^0(x)
$$
$$
+ 2 f\,[\bar{\Sigma}^0(x)i\gamma^5 \Sigma^-(x) - \bar{\Sigma}^+(x)i
\gamma^5 \Sigma^0(x)]\,\pi^+(x)
$$
$$
 + 2 f\,[\bar{\Sigma}^-(x)i\gamma^5
\Sigma^0(x) - \bar{\Sigma}^0(x)i \gamma^5 \Sigma^+(x)]\,\pi^-(x)
$$
$$
 + 2 f\,[\bar{\Sigma}^+(x)i\gamma^5 \Sigma^+(x) - \bar{\Sigma}^-(x)i
\gamma^5 \Sigma^-(x)]\,\pi^0(x)
$$
$$
+ \frac{2}{\sqrt{3}}\,g\,[\bar{\Sigma}^+(x)i\gamma^5 \Lambda^0(x)
 \pi^+(x) + \bar{\Sigma}^-(x)i\gamma^5 \Lambda^0(x) \pi^- (x) +
 \bar{\Sigma}^0(x)i\gamma^5 \Lambda^0(x) \pi^0(x)]
$$
$$
+ \frac{2}{\sqrt{3}}\,g\,[\bar{\Lambda}^0(x)i\gamma^5 \Sigma^+(x)
\pi^-(x) + \bar{\Lambda}^0(x)i\gamma^5 \Sigma^-(x) \pi^+(x) +
\bar{\Lambda}^0(x)i\gamma^5 \Sigma^0(x) \pi^0(x)]
$$
$$
- \frac{1}{\sqrt{3}}\,(g + 3 f)\,\bar{p}(x)i\gamma^5 \Lambda^0(x)
  K^+(x) + \frac{1}{\sqrt{3}}\,(g + 3 f)\,\bar{\Lambda}^0(x)i \gamma^5
  p(x)K^-(x)
$$
$$
- \frac{1}{\sqrt{3}}\,(g + 3 f)\,\bar{n}(x)i\gamma^5 \Lambda^0(x)
  K^0(x) - \frac{1}{\sqrt{3}}\,(g + 3 f)\,\bar{\Lambda}^0(x)i \gamma^5
  n(x) \bar{K}^0(x)
$$
$$
- (g - f)\,\bar{\Sigma}^0(x)i \gamma^5 p(x) K^-(x) + (g -
  f)\,\bar{p}(x)i \gamma^5 \Sigma^0(x) K^+(x)
$$
$$
- (g - f)\,\bar{\Sigma}^0(x)i \gamma^5 n(x) \bar{K}^0(x) - (g -
  f)\,\bar{n}(x)i \gamma^5 \Sigma^0(x) K^0(x)
$$
$$
+ \sqrt{2}\, (g - f)\,\bar{\Sigma}^+(x)i \gamma^5 p(x) \bar{K}^0(x) +
  \sqrt{2}\, (g - f)\,\bar{p}(x)i \gamma^5 \Sigma^+(x) K^0(x)
$$
$$
- \sqrt{2}\, (g - f)\,\bar{\Sigma}^-(x)i \gamma^5 n(x) K^-(x) +
  \sqrt{2}\, (g - f)\,\bar{n}(x)i \gamma^5 \Sigma^-(x) K^+(x)
$$
$$
+ \frac{1}{\sqrt{3}}\,(3 f - g)\,[\bar{p}(x)i\gamma^5p(x) +
\bar{n}(x)i\gamma^5n(x)]\,\eta(x)
$$
$$
+ \frac{2}{\sqrt{3}}\,g\,[\bar{\Sigma}^+(x)i\gamma^5 \Sigma^+(x) +
\bar{\Sigma}^-(x)i\gamma^5 \Sigma^-(x) + \bar{\Sigma}^0(x)i\gamma^5
\Sigma^0(x) - \bar{\Lambda}^0(x)i\gamma^5 \Lambda^0(x)]\,\eta(x) +
\ldots\,\eqno({\rm D}.3)
$$
Following Nagels {\it et al.} \cite{MN79} and denoting  $g + f = g_{\pi
NN}$ and $\alpha = g/(g + f)$ we get
$$
{\cal L}_{PBB}(x) = \sqrt{2}\,g_{\pi
NN}\,\bar{p}(x)i\gamma^5n(x)\pi^+(x)
+ \sqrt{2}\,g_{\pi
NN}\,\bar{n}(x)i\gamma^5p(x)\pi^-(x)
$$
$$
 + g_{\pi NN}\,[\bar{p}(x)i\gamma^5p(x) -
\bar{n}(x)i\gamma^5n(x)]\,\pi^0(x)
$$
$$
+ 2\,(1 - \alpha)\,g_{\pi NN}\,(\bar{\Sigma}^0(x)i\gamma^5 \Sigma^-(x)
- \bar{\Sigma}^+(x)i \gamma^5 \Sigma^0(x))\,\pi^+(x)
$$
$$
 + 2\,(1 - \alpha)\,g_{\pi NN}\,[\bar{\Sigma}^-(x)i\gamma^5
\Sigma^0(x) - \bar{\Sigma}^0(x)i \gamma^5 \Sigma^+(x)]\,\pi^-(x)
$$
$$
 + 2\,(1 - \alpha)\,g_{\pi NN}\,[\bar{\Sigma}^+(x)i\gamma^5 \Sigma^+(x) - \bar{\Sigma}^-(x)i
\gamma^5 \Sigma^-(x)]\,\pi^0(x)
$$
$$
+ \frac{2}{\sqrt{3}}\,\alpha\,g_{\pi NN}\,[\bar{\Sigma}^+(x)i\gamma^5
 \Lambda^0(x) \pi^+(x) + \bar{\Sigma}^-(x)i\gamma^5 \Lambda^0(x) \pi^-
 (x) + \bar{\Sigma}^0(x)i\gamma^5 \Lambda^0(x) \pi^0(x)]
$$
$$
+ \frac{2}{\sqrt{3}}\,\alpha\,g_{\pi NN}\,[\bar{\Lambda}^0(x)i\gamma^5
\Sigma^+(x) \pi^-(x) + \bar{\Lambda}^0(x)i\gamma^5 \Sigma^-(x)
\pi^+(x) + \bar{\Lambda}^0(x)i\gamma^5 \Sigma^0(x) \pi^0(x)]
$$
$$
- \frac{1}{\sqrt{3}}\,(3 - 2\alpha)\,g_{\pi NN}\,\bar{p}(x)i\gamma^5
  \Lambda^0(x) K^+(x) + \frac{1}{\sqrt{3}}\,(3 - 2\alpha)\,g_{\pi
  NN}\,\bar{\Lambda}^0(x)i \gamma^5 p(x)K^-(x)
$$
$$
- \frac{1}{\sqrt{3}}\,(3 - 2\alpha)\,g_{\pi NN}\,\bar{n}(x)i\gamma^5
  \Lambda^0(x) K^0(x) - \frac{1}{\sqrt{3}}\,(3 - 2\alpha)\,g_{\pi
  NN}\,\bar{\Lambda}^0(x)i \gamma^5 n(x) \bar{K}^0(x)
$$
$$
- (2\alpha - 1)\,g_{\pi NN}\,\bar{\Sigma}^0(x)i \gamma^5 p(x) K^-(x) +
  (2\alpha - 1)\,g_{\pi NN}\,\bar{p}(x)i \gamma^5 \Sigma^0(x) K^+(x)
$$
$$
- (2\alpha - 1)\,g_{\pi NN}\,\bar{\Sigma}^0(x)i \gamma^5 n(x)
  \bar{K}^0(x) - (2\alpha - 1)\,g_{\pi NN}\,\bar{n}(x)i \gamma^5
  \Sigma^0(x) K^0(x)
$$
$$
+ \sqrt{2}\,(2\alpha - 1)\,g_{\pi NN}\,\bar{\Sigma}^+(x)i \gamma^5
  p(x) \bar{K}^0(x) + \sqrt{2}\,(2\alpha - 1)\,g_{\pi NN}\,\bar{p}(x)i
  \gamma^5 \Sigma^+(x) K^0(x)
$$
$$
- \sqrt{2}\, (2\alpha - 1)\,g_{\pi NN}\,\bar{\Sigma}^-(x)i \gamma^5
  n(x) K^-(x) + \sqrt{2}\,(2\alpha - 1)\,g_{\pi NN}\,\bar{n}(x)i
  \gamma^5 \Sigma^-(x) K^+(x)
$$
$$
+ \frac{1}{\sqrt{3}}\,(3 - 4 \alpha)\,g_{\pi NN}\,
[\bar{p}(x)i\gamma^5p(x) + \bar{n}(x)i\gamma^5n(x)]\,\eta(x)
$$
$$
+ \frac{2}{\sqrt{3}}\,\alpha\,g_{\pi NN}\,[\bar{\Sigma}^+(x)i\gamma^5
\Sigma^+(x) + \bar{\Sigma}^-(x)i\gamma^5 \Sigma^-(x) +
\bar{\Sigma}^0(x)i\gamma^5 \Sigma^0(x) - \bar{\Lambda}^0(x)i\gamma^5
\Lambda^0(x)]\,\eta(x)
$$
$$
 + \ldots\,.\eqno({\rm D}.4)
$$
For numerical calculation we set $g_{\pi NN} = 13.21$ \cite{PSI2} and
$\alpha = 0.635$ \cite{MS89} (see also \cite{AI01}).

\subsection*{Phenomenological $SBB$ interactions}

The Lagrangian of the phenomenological $SBB$ interactions can be
defined as
$$
{\cal L}_{SBB}(x) = \sqrt{2}\,g_S\,{\rm tr}\{\{\bar{B},B\}S\}
+ \sqrt{2}\,f_S\,{\rm tr}\{[\bar{B},B]S\}=
$$
$$
= \sqrt{2}\,(g_S + f_S)\,\bar{B}^b_a B^a_cS^c_b + \sqrt{2}\,(g_S -
f_S)\,\bar{B}^b_a B^c_b S^a_c,\eqno({\rm D}.5)
$$
where $S^a_b$ is a nonet of scalar mesons defined by
$$
S^a_b = \left(\begin{array}{llcl} {\displaystyle
\frac{\delta^0}{\sqrt{2}} + \frac{\sigma}{\sqrt{2}}} &
\hspace{0.3in} \delta^+ & \kappa^+\\ \hspace{0.3in} \delta^-
&{\displaystyle - \frac{\delta^0}{\sqrt{2}} +
\frac{\sigma}{\sqrt{2}}} & \kappa^0\\
\hspace{0.15in}- \kappa^- & \hspace{0.3in} \bar{\kappa}^0&
{\displaystyle \sigma_s} \\
\end{array}\right),\eqno({\rm D}.6)
$$
where the meson field $\sigma(x)$ and $\sigma_s(x)$ are isoscalar
scalar fields with quark structure $(\bar{u}u + \bar{d}d)/\sqrt{2}$
and $\bar{s}s$, respectively. The interaction of the baryons with the
$\sigma$--meson reads
$$
{\cal L}_{S BB}(x) = \sqrt{2}\,\frac{g_{\pi NN}}{g_A}\,\bar{p}(x)
n(x)\delta^+(x) + \sqrt{2}\,\frac{g_{\pi
NN}}{g_A}\,\bar{n}(x)p(x)\delta^-(x)
$$
$$
 + \frac{g_{\pi NN}}{g_A}\,[\bar{p}(x)p(x) -
\bar{n}(x)n(x)]\,\delta^0(x)
$$
$$
+ 2\,(1 - \alpha)\,\frac{g_{\pi NN}}{g_A}\,(\bar{\Sigma}^0(x)
\Sigma^-(x) - \bar{\Sigma}^+(x)\Sigma^0(x))\,\delta^+(x)
$$
$$
 + 2\,(1 - \alpha)\,\frac{g_{\pi NN}}{g_A}\,[\bar{\Sigma}^-(x)
\Sigma^0(x) - \bar{\Sigma}^0(x)\Sigma^+(x)]\,\delta^-(x)
$$
$$
 + 2\,(1 - \alpha)\,\frac{g_{\pi
NN}}{g_A}\,[\bar{\Sigma}^+(x)\Sigma^+(x) -
\bar{\Sigma}^-(x)\Sigma^-(x)]\,\delta^0(x)
$$
$$
+ \frac{2}{\sqrt{3}}\,\alpha\,\frac{g_{\pi
NN}}{g_A}\,[\bar{\Sigma}^+(x) \Lambda^0(x) \delta^+(x) +
\bar{\Sigma}^-(x) \Lambda^0(x) \delta^- (x) +
\bar{\Sigma}^0(x)\Lambda^0(x) \delta^0(x)]
$$
$$
+ \frac{2}{\sqrt{3}}\,\alpha\,\frac{g_{\pi
NN}}{g_A}\,[\bar{\Lambda}^0(x)\Sigma^+(x) \delta^-(x) +
\bar{\Lambda}^0(x) \Sigma^-(x) \delta^+(x) +
\bar{\Lambda}^0(x)\Sigma^0(x) \delta^0(x)]
$$
$$
- \frac{1}{\sqrt{3}}\,(3 - 2\alpha)\,\frac{g_{\pi
NN}}{g_A}\,\bar{p}(x)\Lambda^0(x) \kappa^+(x) + \frac{1}{\sqrt{3}}\,(3
- 2\alpha)\,\frac{g_{\pi NN}}{g_A}\,\bar{\Lambda}^0(x) p(x) \kappa^-(x)
$$
$$
- \frac{1}{\sqrt{3}}\,(3 - 2\alpha)\,\frac{g_{\pi
NN}}{g_A}\,\bar{n}(x) \Lambda^0(x) \kappa^0(x) -
\frac{1}{\sqrt{3}}\,(3 - 2\alpha)\,\frac{g_{\pi
NN}}{g_A}\,\bar{\Lambda}^0(x)n(x) \bar{\kappa}^0(x)
$$
$$
- (2\alpha - 1)\,\frac{g_{\pi NN}}{g_A}\,\bar{\Sigma}^0(x) p(x)
\kappa^-(x) + (2\alpha - 1)\,\frac{g_{\pi NN}}{g_A}\,\bar{p}(x)
\Sigma^0(x) \kappa^+(x)
$$
$$
- (2\alpha - 1)\,\frac{g_{\pi NN}}{g_A}\,\bar{\Sigma}^0(x) n(x)
  \bar{\kappa}^0(x) - (2\alpha - 1)\,\frac{g_{\pi
  NN}}{g_A}\,\bar{n}(x)\Sigma^0(x) \kappa^0(x)
$$
$$
+ \sqrt{2}\,(2\alpha - 1)\,\frac{g_{\pi NN}}{g_A}\,\bar{\Sigma}^+(x)
p(x) \bar{\kappa}^0(x) + \sqrt{2}\,(2\alpha - 1)\,\frac{g_{\pi
NN}}{g_A}\,\bar{p}(x) \Sigma^+(x) \kappa^0(x)
$$
$$
- \sqrt{2}\, (2\alpha - 1)\,\frac{g_{\pi NN}}{g_A}\,\bar{\Sigma}^-(x)
n(x) \kappa^-(x) + \sqrt{2}\,(2\alpha - 1)\,\frac{g_{\pi
NN}}{g_A}\,\bar{n}(x) \Sigma^-(x) \kappa^+(x)
$$
$$
+ \frac{g_{\pi NN}}{g_A}\,[\bar{p}(x)p(x) + \bar{n}(x)n(x)]\,\sigma(x)
$$
$$
 + 2 \alpha\,\frac{g_{\pi NN}}{g_A}\,[\bar{\Sigma}^+(x)\Sigma^+(x) +
\bar{\Sigma}^-(x) \Sigma^-(x) + \bar{\Sigma}^0(x) \Sigma^0(x)
+\frac{1}{3}\,\bar{\Lambda}^0(x)\Lambda^0(x)]\,\sigma(x) + \ldots
,\eqno({\rm D}.7)
$$
where the coupling constants are fixed as $g_S + f_S = g_{\pi NN}/g_A$
according to \cite{HP73,IV90}.

\subsection*{Phenomenological $SPP$ interactions}

A phenomenological $SPP$ interaction we describe by the Lagrangian 
$$
{\cal L}_{SPP}(x) = \sqrt{2}\,g_0\,{\rm tr}\{S(x)P(x)P(x)\} =
\sqrt{2}\,g_0\,S^b_a(x)P^c_b(x)P^a_c(x),\eqno({\rm
D}.8)
$$
where $g_0$ is a phenomenological coupling constant which we fix below.

In terms of the components of multiplets the Lagrangian ${\cal
L}_{SPP}(x)$ reads
$$
{\cal L}_{SPP}(x) = 
$$
$$
= \,g_0\,\sigma(x)\,\Big(2\pi^+(x)\pi^-(x) + \pi^0(x)\pi^0(x) - K^+(x)
K^-(x) + K^0(x)\bar{K}^0(x) + \frac{1}{3}\,\eta^2(x)\Big)
$$
$$
+ \,g_0\,\kappa^-(x)\,\Big(\sqrt{2}\,\pi^+(x) K^0(x) + \pi^0(x) K^+(x) -
\frac{1}{\sqrt{3}}\,\eta(x)K^+(x)\Big)
$$
$$
+ \,g_0\,\bar{\kappa}^0(x)\,\Big(\sqrt{2}\,\pi^-(x)K^+(x) - \pi^0(x)
K^0(x) - \frac{1}{\sqrt{3}}\,\eta(x)K^0(x)\Big)
$$
$$
+ \,g_0\,\kappa^+(x)\,\Big(\sqrt{2}\,\pi^-(x) \bar{K}^0(x) - \pi^0(x)
K^-(x) + \frac{1}{\sqrt{3}}\,\eta(x)K^-(x)\Big)
$$
$$
+ \,g_0\,\kappa^0(x)\,\Big(-\,\sqrt{2}\,\pi^+(x)K^-(x) - \pi^0(x)
\bar{K}^0(x) - \frac{1}{\sqrt{3}}\,\eta(x)\bar{K}^0(x)\Big)
$$
$$
+\,g_0\,\delta^+(x)\,\Big(\frac{2}{\sqrt{3}}\,\pi^-(x)\eta(x) -
\sqrt{2}\,K^0(x)K^-(x)\Big)
$$
$$
 + g_0\,\delta^-(x)\,\Big(\frac{2}{\sqrt{3}}\,\pi^+(x)\eta(x) +
\sqrt{2}\,K^+(x)\bar{K}^0(x)\Big)
$$
$$
+\,g_0\,\delta^0(x)\,\Big(\frac{2}{\sqrt{3}}\,\pi^0(x)\eta(x) - K^+(x)K^-(x) - K^0(x)\bar{K}^0(x)\Big)
$$
$$
+ \,g_0\,\sigma_s(x)\,\Big(- \sqrt{2}\,K^-(x)K^+(x) +
\sqrt{2}\,\bar{K}^0(x)K^0(x) +
\frac{2\sqrt{2}}{3}\,\eta^2(x)\Big).\eqno({\rm D}.9)
$$
According to the $\sigma$--model we set $g_0 = m^2_S/2F_{\pi}$, where
$m_S$ is a mass of scalar nonet which we tend finally to infinity $m_S
\to \infty$. This corresponds to a non--linear realization of chiral
$U(3)\times U(3)$ symmetry \cite{SG69}. 

\section*{Appendix E. Amplitude of  reaction 
$(n \Lambda^0)_{{^3}{\rm P}_1} \to (n \Lambda^0)_{{^3}{\rm P}_1}$}

Summing up the diagrams depicted in Fig.6 we obtain \cite{AI01}
$$
[\bar{u}(\vec{K},\sigma_p)\gamma_{i} u^c( - \vec{K},\sigma_n)]\cdot
[\bar{u^c}(\vec{k},\alpha_1) \gamma_{j} \gamma^5u(- \vec{k},\alpha_2)]
$$
$$
\to [\bar{u}(\vec{K},\sigma_p)\gamma_{i} u^c( -
\vec{K},\sigma_n)]\cdot [\bar{u^c}(\vec{k},\alpha_1) \gamma_{j}
\gamma^5u(- \vec{k},\alpha_2)] \Big(D^{-1}_{(n\Lambda^0)_{{^3}{\rm
P}_1}}(k_{n\Lambda^0})\Big)^{ji},\eqno({\rm E}.1)
$$
where $D_{(n\Lambda^0)_{{^3}{\rm P}_1}}(k_{n\Lambda^0})^{ij}$ is equal to
$$
D_{(n\Lambda^0)_{{^3}{\rm P}_1}}(k_{n\Lambda^0})^{ij} = 
$$
$$
= g^{ji} + \frac{C_{n\Lambda^0}}{64\pi^2}\int \frac{d^4k}{\pi^2
i}\,{\rm tr}\,\Bigg\{\gamma^i\gamma^5 \frac{1}{\displaystyle m_B -
\hat{k} - \frac{1}{2}\,\hat{P} -i\,0}\gamma^j\gamma^5
\frac{1}{\displaystyle m_B - \hat{k} + \frac{1}{2}\,\hat{P}
-i\,0}\Bigg\}.\eqno({\rm E}.2)
$$
In the center of mass frame of the $n\Lambda^0$ pair $P =
(2\sqrt{k^2_0 + m^2_B},\vec{0}\,) = (2 E_0,\vec{0}\,)$. For simplicity
we have used the averaged mass $m_B = (m_{\Lambda^0} + m_N)/2 =
1030\,{\rm MeV}$.

In the center of mass frame of the $n\Lambda^0$ pair the amplitude
$D_{(n\Lambda^0)_{{^3}{\rm P}_1}}(k_{n\Lambda^0})^{ij}$ is defined by
$$
D_{(n\Lambda^0)_{{^3}{\rm P}_1}}(k_{n\Lambda^0})^{ij} = g^{ji} -
\frac{C_{n\Lambda^0}}{64\pi^2}\int \frac{d^4k}{\pi^2
i}\,\frac{1}{\displaystyle \Big(k_0 + \frac{1}{2}\,P_0\Big)^2 -
E^2_{\vec{k}} + i\,0}\,\frac{1}{\displaystyle \Big(k_0 -
\frac{1}{2}\,P_0\Big)^2 - E^2_{\vec{k}} + i\,0}
$$
$$
\times\,{\rm tr}\,\Big\{\gamma^i \Big[m_B +\Big(k_0 +
\frac{1}{2}\,P_0\Big)\gamma^0 - \vec{k}\cdot
\vec{\gamma}\Big]\gamma^j\Big[m_B - \Big(k_0 -
\frac{1}{2}\,P_0\Big)\gamma^0 + \vec{k}\cdot
\vec{\gamma}\Big]\Big\},\eqno({\rm E}.3)
$$
where we have denoted $E_{\vec{k}} = \sqrt{\vec{k}^{\;2} +
m^2_B}$. Integrating over directions of $\vec{k}$ and computing the
trace over Dirac matrices we get
$$
D_{(n\Lambda^0)_{{^3}{\rm P}_1}}(k_{n\Lambda^0})^{ij} = 
$$
$$
= g^{ji} - g^{ji}\,\frac{C_{n\Lambda^0}}{64\pi^2}\int
\frac{d^4k}{\pi^2 i}\,\frac{\displaystyle m^2_B -
\frac{1}{3}\,\vec{k}^{\;2} - \frac{1}{4}\,P^2_0 + k^2_0}{\displaystyle
\Big[\Big(k_0 + \frac{1}{2}\,P_0\Big)^2 - E^2_{\vec{k}} +
i\,0\Big]\Big[\Big(k_0 - \frac{1}{2}\,P_0\Big)^2 - E^2_{\vec{k}} +
i\,0\Big]} =
$$
$$
= D(k_{n\Lambda^0})_{(n\Lambda^0)_{{^3}{\rm P}_1}}\,g^{ji}, \eqno({\rm E}.4)
$$
where we have denoted
$$
D(k_{n\Lambda^0})_{(n\Lambda^0)_{{^3}{\rm P}_1}} = 1 -
\frac{C_{n\Lambda^0}}{16\pi^2}\int \frac{d^4k}{\pi^2
i}\,\frac{\displaystyle m^2_B - \frac{1}{3}\,\vec{k}^{\;2} -
\frac{1}{4}\,P^2_0 + k^2_0}{\displaystyle \Big[\Big(k_0 +
\frac{1}{2}\,P_0\Big)^2 - E^2_{\vec{k}} + i\,0\Big]\Big[\Big(k_0 -
\frac{1}{2}\,P_0\Big)^2 - E^2_{\vec{k}} + i\,0\Big]}.\eqno({\rm E}.5)
$$
The integrand of ({\rm E}.5) has four poles at
$$
(k_0)_1 = -\,\frac{1}{2}\,P_0 + E_{\vec{k}} - i\,0 \quad,\quad (k_0)_2
= -\,\frac{1}{2}\,P_0 - E_{\vec{k}} + i\,0,
$$
$$
(k_0)_3 = +\,\frac{1}{2}\,P_0 + E_{\vec{k}} - i\,0 \quad,\quad (k_0)_4
= +\,\frac{1}{2}\,P_0 - E_{\vec{k}} + i\,0.\eqno({\rm E}.6)
$$
Integrating over $k_0$, folding the contour of integration to the upper
semiplane, we get
$$
D(k_{n\Lambda^0})_{(n\Lambda^0)_{{^3}{\rm P}_1}} = 1 +
\frac{C_{n\Lambda^0}}{16\pi^3}\int \frac{d^3k}{E_{\vec{k}}}
$$
$$
\times\,\Bigg\{\frac{\displaystyle m^2_B - \frac{1}{3}\,\vec{k}^{\;2}
- \frac{1}{4}\,P^2_0 + \Big(\frac{1}{2}\,P_0 + E_{\vec{k}}
\Big)^2}{\displaystyle (P_0 + E_{\vec{k}})^2 - E^2_{\vec{k}}} +
\frac{\displaystyle m^2_B - \frac{1}{3}\,\vec{k}^{\;2} -
\frac{1}{4}\,P^2_0 + \Big(\frac{1}{2}\,P_0 - E_{\vec{k}}
\Big)^2}{\displaystyle (P_0 - E_{\vec{k}})^2 -
E^2_{\vec{k}}}\Bigg\} =
$$
$$
= 1 + \frac{C_{n\Lambda^0}}{16\pi^3}\int
 \frac{d^3k}{E_{\vec{k}}}\Bigg\{\frac{\displaystyle m^2_B -
 \frac{1}{3}\,\vec{k}^{\;2} + P_0E_{\vec{k}} + E^2_{\vec{k}}
 }{\displaystyle (P_0 + E_{\vec{k}})^2 - E^2_{\vec{k}}} +
 \frac{\displaystyle m^2_B - \frac{1}{3}\,\vec{k}^{\;2} -
 P_0E_{\vec{k}} + E^2_{\vec{k}}}{\displaystyle (P_0 - E_{\vec{k}})^2 -
 E^2_{\vec{k}}}\Bigg\} =
$$
$$
= 1 + \frac{C_{n\Lambda^0}}{64\pi^3}\,\frac{1}{E_0}\int
 \frac{d^3k}{E_{\vec{k}}}\Bigg\{\frac{\displaystyle m^2_B -
 \frac{1}{3}\,\vec{k}^{\;2} + 2E_0E_{\vec{k}} + E^2_{\vec{k}}
 }{\displaystyle E_0 + E_{\vec{k}} } + \frac{\displaystyle m^2_B -
 \frac{1}{3}\,\vec{k}^{\;2} - 2E_0E_{\vec{k}} +
 E^2_{\vec{k}}}{\displaystyle E_0 - E_{\vec{k}}}\Bigg\} =
$$
$$
= 1 - \frac{C_{n\Lambda^0}}{24\pi^3}\int
 \frac{d^3k}{E_{\vec{k}}}\,\frac{\vec{k}^{\;2}}{E^2_0 -
 E^2_{\vec{k}}}.  \eqno({\rm E}.7)
$$
Subtracting trivially divergent integrals we obtain the regularized
and renormalized amplitude of $n\Lambda^0$ rescattering
$$
D(k_{n\Lambda^0})_{(n\Lambda^0)_{{^3}{\rm P}_1}} = 1 -
 \frac{C_{n\Lambda^0}}{6\pi^2}\,k^4_0\int^{\infty}_0
 \frac{dk}{\sqrt{k^2 + m^2_B}}\,\frac{1}{k^2_0 - k^2}.  \eqno({\rm
 E}.8)
$$
The integral over $k$ can be computed analytically. The result reads
$$
\int^{\infty}_0 \frac{dk}{\sqrt{k^2 + m^2_B}}\,\frac{1}{k^2_0 - k^2} =
 \frac{1}{2k_0 E_0}\,\Big[{\ell n}\Big(\frac{E_0 + k_0}{E_0 -
 k_0}\Big) - i\,\pi\Big]. \eqno({\rm E}.9)
$$
This gives
$$
D(k_{n\Lambda^0})_{(n\Lambda^0)_{{^3}{\rm P}_1}} = 1 -
 \frac{C_{n\Lambda^0}}{12\pi^2}\,\frac{k^3_0}{E_0}\,\Big[{\ell
 n}\Big(\frac{E_0 + k_0}{E_0 - k_0}\Big) - i\,\pi\Big]. \eqno({\rm
 E}.10)
$$
Thus, the amplitude $f^{\,n\Lambda^0}_{K^-pn}(k_0)$ accounting for the
rescattering of the $n\Lambda^0$ pair in the ${^3}{\rm P}_1$ state, is
equal to
$$
f^{\,n\Lambda^0}_{K^-pn}(k_0)_{{^3}{\rm P}_1} = \frac{1}{\displaystyle 1
- \frac{C_{n\Lambda^0}}{12\pi^2}\,\frac{k^3_0}{E_0}
\,\Big[{\ell n}\Big(\frac{E_0 + k_0}{E_0 - k_0} \Big) -
i\,\pi\Big]}. \eqno({\rm E}.11)
$$
For the reaction $K^- (p n)_{{^3}{\rm S}_1} \to (n
\Lambda^0)_{{^3}{\rm P}_1}$ the amplitude ({\rm E}.11) describes also
the final state interaction of the $n\Lambda^0$ pair in the ${^3}{\rm
P}_1$ state. 

\section*{Appendix F. Amplitude of  reaction 
$(n \Lambda^0)_{{^1}{\rm P}_1} \to (n \Lambda^0)_{{^1}{\rm P}_1}$}

Summing up the diagrams depicted in Fig.6 we obtain \cite{AI01}
$$
[\bar{u}(\vec{K},\sigma_p)\gamma_{i} u^c( - \vec{K},\sigma_n)]\cdot
[\bar{u^c}(\vec{k},\alpha_1)\gamma^0 \gamma_{j} \gamma^5u(-
\vec{k},\alpha_2)]
$$
$$
\to [\bar{u}(\vec{K},\sigma_p)\gamma_{i} u^c( -
\vec{K},\sigma_n)]\cdot [\bar{u^c}(\vec{k},\alpha_1)\gamma^0
\gamma_{j} \gamma^5u(- \vec{k},\alpha_2)]
\Big(D^{-1}_{(n\Lambda^0)_{{^1}{\rm
P}_1}}(k_{n\Lambda^0})\Big)^{ji},\eqno({\rm F}.1)
$$
where $D_{(n\Lambda^0)_{{^1}{\rm P}_1}}(k_{n\Lambda^0})^{ij}$ is equal
to
$$
D_{(n\Lambda^0)_{{^1}{\rm P}_1}}(k_{n\Lambda^0})^{ij} = 
$$
$$
= g^{ji} + \frac{C_{n\Lambda^0}({^1}{\rm P}_1)}{64\pi^2}\int
\frac{d^4k}{\pi^2 i}\,{\rm tr}\,\Bigg\{\gamma^0\gamma^i\gamma^5
\frac{1}{\displaystyle m_B - \hat{k} - \frac{1}{2}\,\hat{P}
-i\,0}\gamma^0\gamma^j\gamma^5 \frac{1}{\displaystyle m_B - \hat{k} +
\frac{1}{2}\,\hat{P} -i\,0}\Bigg\}.\eqno({\rm F}.2)
$$
In the center of mass frame of the $n\Lambda^0$ pair $P =
(2\sqrt{k^2_0 + m^2_B},\vec{0}\,) = (2 E_0,\vec{0}\,)$. For simplicity
we have used the averaged mass $m_B = (m_{\Lambda^0} + m_N)/2 =
1030\,{\rm MeV}$.

In the center of mass frame of the $n\Lambda^0$ pair the amplitude
$D_{(n\Lambda^0)_{{^1}{\rm P}_1}}(k_{n\Lambda^0})^{ij}$ is defined by
$$
D_{(n\Lambda^0)_{{^1}{\rm P}_1}}(k_{n\Lambda^0})^{ij} = $$
$$
=
g^{ji} + \frac{C_{n\Lambda^0}({^1}{\rm P}_1)}{64\pi^2}\int
\frac{d^4k}{\pi^2 i}\,\frac{1}{\displaystyle \Big(k_0 +
  \frac{1}{2}\,P_0\Big)^2 - E^2_{\vec{k}} +
  i\,0}\,\frac{1}{\displaystyle \Big(k_0 - \frac{1}{2}\,P_0\Big)^2 -
  E^2_{\vec{k}} + i\,0} $$
$$
\times\,{\rm tr}\,\Big\{\gamma^0\gamma^i
\Big[m_B +\Big(k_0 + \frac{1}{2}\,P_0\Big)\gamma^0 - \vec{k}\cdot
\vec{\gamma}\Big]\gamma^0 \gamma^j\Big[m_B - \Big(k_0 -
\frac{1}{2}\,P_0\Big)\gamma^0 + \vec{k}\cdot
\vec{\gamma}\Big]\Big\},\eqno({\rm F}.3) $$
where we have denoted
$E_{\vec{k}} = \sqrt{\vec{k}^{\;2} + m^2_B}$. Integrating over
directions of $\vec{k}$ and computing the trace over Dirac matrices we
get $$
D_{(n\Lambda^0)_{{^1}{\rm P}_1}}(k_{n\Lambda^0})^{ij} = $$
$$
=
g^{ji} - g^{ji}\,\frac{C_{n\Lambda^0}({^1}{\rm P}_1)}{64\pi^2}\int
\frac{d^4k}{\pi^2 i}\,\frac{\displaystyle m^2_B +
  \frac{1}{3}\,\vec{k}^{\;2} - \frac{1}{4}\,P^2_0 +
  k^2_0}{\displaystyle \Big[\Big(k_0 + \frac{1}{2}\,P_0\Big)^2 -
  E^2_{\vec{k}} + i\,0\Big]\Big[\Big(k_0 - \frac{1}{2}\,P_0\Big)^2 -
  E^2_{\vec{k}} + i\,0\Big]} = $$
$$
=
D(k_{n\Lambda^0})_{(n\Lambda^0)_{{^1}{\rm P}_1}}\,g^{ji}, \eqno({\rm
  F}.4) $$
where we have denoted $$
D(k_{n\Lambda^0})_{(n\Lambda^0)_{{^3}{\rm P}_1}} = $$
$$
= 1 -
\frac{C_{n\Lambda^0}({^1}{\rm P}_1)}{16\pi^2}\int \frac{d^4k}{\pi^2
  i}\,\frac{\displaystyle m^2_B + \frac{1}{3}\,\vec{k}^{\;2} -
  \frac{1}{4}\,P^2_0 + k^2_0}{\displaystyle \Big[\Big(k_0 +
  \frac{1}{2}\,P_0\Big)^2 - E^2_{\vec{k}} + i\,0\Big]\Big[\Big(k_0 -
  \frac{1}{2}\,P_0\Big)^2 - E^2_{\vec{k}} + i\,0\Big]}.\eqno({\rm
  F}.5) $$
The integrand of ({\rm F}.5) has four poles at $$
(k_0)_1 =
-\,\frac{1}{2}\,P_0 + E_{\vec{k}} - i\,0 \quad,\quad (k_0)_2 =
-\,\frac{1}{2}\,P_0 - E_{\vec{k}} + i\,0, $$
$$
(k_0)_3 =
+\,\frac{1}{2}\,P_0 + E_{\vec{k}} - i\,0 \quad,\quad (k_0)_4 =
+\,\frac{1}{2}\,P_0 - E_{\vec{k}} + i\,0.\eqno({\rm F}.6) $$
Integrating over $k_0$, folding the contour of integration to the
upper semiplane, we get $$
D(k_{n\Lambda^0})_{(n\Lambda^0)_{{^3}{\rm
      P}_1}} = 1 + \frac{C_{n\Lambda^0}({^1}{\rm P}_1)}{16\pi^3}\int
\frac{d^3k}{E_{\vec{k}}} $$
$$
\times\,\Bigg\{\frac{\displaystyle
  m^2_B + \frac{1}{3}\,\vec{k}^{\;2} - \frac{1}{4}\,P^2_0 +
  \Big(\frac{1}{2}\,P_0 + E_{\vec{k}} \Big)^2}{\displaystyle (P_0 +
  E_{\vec{k}})^2 - E^2_{\vec{k}}} + \frac{\displaystyle m^2_B +
  \frac{1}{3}\,\vec{k}^{\;2} - \frac{1}{4}\,P^2_0 +
  \Big(\frac{1}{2}\,P_0 - E_{\vec{k}} \Big)^2}{\displaystyle (P_0 -
  E_{\vec{k}})^2 - E^2_{\vec{k}}}\Bigg\} = $$
$$
= 1 +
\frac{C_{n\Lambda^0}({^1}{\rm P}_1)}{16\pi^3}\int
\frac{d^3k}{E_{\vec{k}}}\Bigg\{\frac{\displaystyle m^2_B +
  \frac{1}{3}\,\vec{k}^{\;2} + P_0E_{\vec{k}} + E^2_{\vec{k}}
}{\displaystyle (P_0 + E_{\vec{k}})^2 - E^2_{\vec{k}}} +
\frac{\displaystyle m^2_B + \frac{1}{3}\,\vec{k}^{\;2} -
  P_0E_{\vec{k}} + E^2_{\vec{k}}}{\displaystyle (P_0 - E_{\vec{k}})^2
  - E^2_{\vec{k}}}\Bigg\} = $$
$$
= 1 + \frac{C_{n\Lambda^0}({^1}{\rm
    P}_1)}{64\pi^3}\,\frac{1}{E_0}\int
\frac{d^3k}{E_{\vec{k}}}\Bigg\{\frac{\displaystyle m^2_B +
  \frac{1}{3}\,\vec{k}^{\;2} + 2E_0E_{\vec{k}} + E^2_{\vec{k}}
}{\displaystyle E_0 + E_{\vec{k}} } + \frac{\displaystyle m^2_B +
  \frac{1}{3}\,\vec{k}^{\;2} - 2E_0E_{\vec{k}} +
  E^2_{\vec{k}}}{\displaystyle E_0 - E_{\vec{k}}}\Bigg\} = $$
$$
= 1 -
\frac{C_{n\Lambda^0}({^1}{\rm P}_1)}{48\pi^3}\int
\frac{d^3k}{E_{\vec{k}}}\,\frac{\vec{k}^{\;2}}{E^2_0 - E^2_{\vec{k}}}.
\eqno({\rm F}.7) $$
Subtracting trivially divergent integrals we
obtain the regularized and renormalized amplitude of $n\Lambda^0$
rescattering $$
D(k_{n\Lambda^0})_{(n\Lambda^0)_{{^3}{\rm P}_1}} = 1 -
\frac{C_{n\Lambda^0}({^1}{\rm P}_1)}{12\pi^2}\,k^4_0\int^{\infty}_0
\frac{dk}{\sqrt{k^2 + m^2_B}}\,\frac{1}{k^2_0 - k^2}.  \eqno({\rm
  F}.8) $$
The integral over $k$ can be computed analytically. The
result reads $$
\int^{\infty}_0 \frac{dk}{\sqrt{k^2 +
    m^2_B}}\,\frac{1}{k^2_0 - k^2} = \frac{1}{2k_0 E_0}\,\Big[{\ell
  n}\Big(\frac{E_0 + k_0}{E_0 - k_0}\Big) - i\,\pi\Big]. \eqno({\rm
  F}.9) $$
This gives $$
D(k_{n\Lambda^0})_{(n\Lambda^0)_{{^1}{\rm
      P}_1}} = 1 - \frac{C_{n\Lambda^0}({^1}{\rm
    P}_1)}{24\pi^2}\,\frac{k^3_0}{E_0}\,\Big[{\ell n}\Big(\frac{E_0 +
  k_0}{E_0 - k_0}\Big) - i\,\pi\Big]. \eqno({\rm F}.10) $$
Thus, the
amplitude $f^{\,n\Lambda^0}_{K^-pn}(k_0)$ accounting for the
rescattering of the $n\Lambda^0$ pair in the ${^3}{\rm P}_1$ state, is
equal to $$
f^{\,n\Lambda^0}_{K^-pn}(k_0)_{{^1}{\rm P}_1} =
\frac{1}{\displaystyle 1 - \frac{C_{n\Lambda^0}({^1}{\rm
      P}_1)}{24\pi^2}\,\frac{k^3_0}{E_0} \,\Big[\Big(\frac{E_0 +
    k_0}{E_0 - k_0} \Big) - i\,\pi\Big]}. \eqno({\rm F}.11) $$
For the
reaction $K^- (p n)_{{^3}{\rm S}_1} \to (n \Lambda^0)_{{^3}{\rm P}_1}$
the amplitude ({\rm F}.11) describes also the final state interaction
of the $n\Lambda^0$ pair in the ${^3}{\rm P}_1$ state.

\end{document}